\documentclass[a4paper,twoside]{article}
\usepackage{geometry}
\usepackage{bbm}
\usepackage{graphicx}
\usepackage{oldgerm}
\usepackage{amsmath}
\usepackage{amssymb}
\usepackage{graphics}
\usepackage{pstricks}
\usepackage{here} 

\numberwithin{figure}{section}
\numberwithin{equation}{section}

\newcommand{\mathsym}[1]{{}}

\newcommand{\bra}[1]{\ensuremath{\left\langle #1\right|}}
\newcommand{\ket}[1]{\ensuremath{\left|#1\right\rangle}}

\begin{document}

\begin{titlepage}

\begin{flushright}
\includegraphics[width=.5\textwidth]{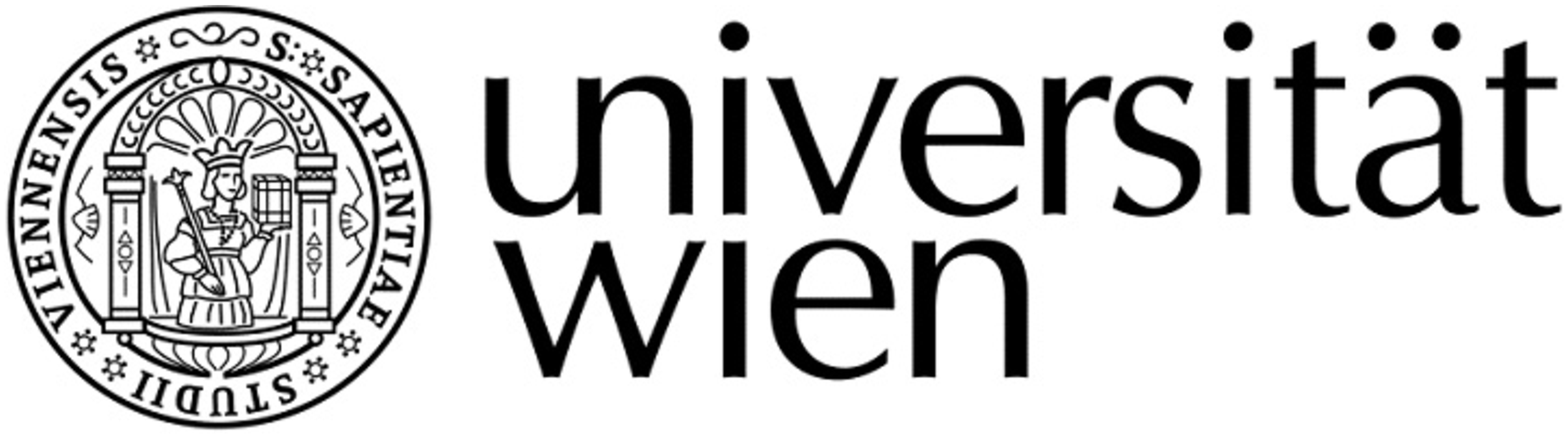}
\end{flushright}

\vspace{2cm}

\begin{center}
\Huge\textbf{\fontfamily{phv}\selectfont DIPLOMARBEIT}\\
\vspace{2cm}
\large
\fontfamily{phv}\selectfont{Titel der Diplomarbeit}\\
\vspace{0.5cm}
\Large\textbf{\fontfamily{phv}\selectfont Geometrical aspects of qudits concerning Bell inequalities}\\
\vspace{2cm}
\large {\fontfamily{phv}\selectfont{angestrebter akademischer Grad}}\\
\vspace{0.5cm}

\Large\textbf{\fontfamily{phv}\selectfont Magister der Naturwissenschaften (Mag.rer.nat.)}\\
\end{center}
\vspace{2cm}

\linespread{1.3}
\large \fontfamily{phv}\selectfont{
\begin{flushleft}
 \begin{tabbing}
    Verfasser: \hspace{3.5cm} \= Christoph Spengler \\
    Matrikel-Nummer: \> 0549723 \\
    Studienrichtung: \> Physik \\
    Betreuerin: \> Univ. Doz. Dr. Beatrix C. Hiesmayr
  \end{tabbing}
 \end{flushleft}}

\vspace{3cm}

\noindent Wien, am 17.11.2008
\end{titlepage}
\newpage
\hfill
\vfill
\textcolor{white}.
\newpage

\large

\vspace*{4cm}

\begin{center}
\Large{\textbf{Abstract}}
\end{center}
\noindent
The aim of this thesis is to investigate quantum entanglement and quantum nonlocality of bipartite finite-dimensional systems (bipartite qudits). Entanglement is one of the most fascinating non-classical features of quantum theory, and besides its impact on our view of the world, it can be exploited for applications such as quantum cryptography and quantum computing. This circumstance has led to a growing interest and profound investigations in this area. Although entanglement and nonlocality are ordinarily regarded as one and the same, under close consideration this cannot be taken for granted. The reason for this is that entanglement is defined by the mathematical structure of a quantum state in a composite Hilbert space, whereas nonlocality signifies that the statistical behaviour of a system cannot be described by a local realistic theory. For the latter it is essential that the correlation probabilities of such theories obey so-called Bell inequalities, which are violated for certain quantum states. The main focus of this thesis is on the comparison of both properties with the objective of understanding their relation. In terms of the analysis of entanglement, recent methods for the detection are presented and discussed. Because of the fact that the correlation probabilities in general depend on the measurement settings it is necessary to optimise these in order to reveal nonlocality. This problem is solved for a particular Bell inequality (CGMLP) by means of a self-developed numerical search algorithm. These methods are then applied to density matrices of a subspace spanned by the projectors of maximally entangled two-qudit states. This set of states has not only interesting properties with respect to our investigations, but also serves to visualise and analyse the state space geometrically. 

\newpage
\hfill
\vfill
\newpage

\vspace*{4cm}

\begin{center}
\Large{\textbf{Kurzfassung}}
\end{center}
\noindent
Diese Diplomarbeit setzt sich mit Verschr\"ankung und Nichtlokalit\"at in bipartiten endlich-dimensionalen Systemen (bipartite Qudits) auseinander. Die Verschr\"ankung ist eines der faszinierendsten nichtklassischen Ph\"anomene der Quantentheorie, und neben ihrer Bedeutung f\"ur unser Weltbild findet sie Anwendung in der Quanten- kryptographie und der Quanteninformatik. Diese Tatsache hat zu wachsendem Interesse und ausgiebiger Forschung auf diesem Gebiet gef\"uhrt. Verschr\"ankung und Nichtlokalit\"at werden f\"ur gew\"ohnlich als ein und dasselbe angesehen. Jedoch ist dies unter genauer Betrachtung nicht als selbstverst\"andlich hinzunehmen, was daran liegt, dass Verschr\"ankung durch die mathematische Struktur eines Zustands in einem zusammengesetzten Hilbertraum definiert ist. Nichtlokalit\"at hingegen besagt, dass das statistische Verhalten eines Systems nicht durch eine lokal-realistische Theorie beschrieben werden kann. F\"ur letzteres ist wesentlich, dass die Korrelationswahrscheinlichkeiten solcher Theorien sogenannte Bell-Ungleichungen erf\"ullen, welche jedoch durch bestimmte Quantenzust\"ande verletzt werden. Diese Diplom- arbeit dient insbesondere dazu, beide Eigenschaften miteinander zu vergleichen. F\"ur die Untersuchung der Verschr\"ankung werden aktuelle Separabilit\"atskriterien vorgestellt und diskutiert. Aufgrund der Tatsache, dass die Korrelationswahrscheinlichkeiten im Allgemeinen von der Messsituation abh\"angen, ist es notwendig diese zu optimieren, um Nichtlokalit\"at nachzuweisen. Dieses Problem wird f\"ur eine be- stimmte Bell-Ungleichung (CGLMP) durch einen selbstentwickelten numerischen Suchalgorithmus gel\"ost. Die besprochenen Methoden werden dann auf Dichte- \mbox{matrizen} eines Unterraums, aufgespannt durch Projektoren von maximal verschr\"ank- ten Zwei-Qudit-Zust\"anden, angewandt. Diese Menge von Zust\"anden hat nicht nur interessante Eigenschaften im Bezug auf unsere Untersuchungen, sondern dient auch dazu, den Zustandsraum zu visualisieren und geometrisch zu analysieren.

\newpage
\hfill
\vfill
\newpage

\textcolor{white}{-}
\large
\tableofcontents

\newpage

\large

\section*{Introduction}
\addcontentsline{toc}{section}{Introduction} 
Modern physical theories often contradict human intuition and this especially applies to quantum physics. One of the most sensational quantum phenomena was recognised by Einstein, Podolsky and Rosen in 1935. In their seminal work \cite{EPR} they presented a physical situation in which quantum theory seems to violate the principles of relativity. Besides the fact that in such a situation best possible knowledge of the whole does not include best possible knowledge of its parts, particles seem to somehow influence each other at the point of measurement, even if these are space-like separated. The theory here predicts a new type of correlations which cannot be understood on the basis of a local realistic description. These correlations have become known as EPR correlations\footnote{or simply quantum correlations}. At that time, their conclusion was that quantum theory must be incomplete. It was thought that this peculiarity can be eliminated through a more fundamental theory based on local hidden variables. Such a formulation has never been found. Furthermore, the existence of such a theory can be revised experimentally via so-called Bell inequality tests. Experimental evidence of EPR correlations in 1982 by Aspect et al. (see \cite{AspectEPR}) has led to a growing interest in this subject and a new research field named quantum information has emerged. Many useful applications such as quantum teleportation, quantum cryptography and algorithms for quantum computers have been proposed. Entanglement and nonlocality are now being accepted and investigated by various physicists.\\
\\
Irrespective of intensive research within the last decades, there still remain many open problems from a theoretical point of view and one of the aims of this thesis is to outline several fundamental ones. In this thesis we investigate bipartite qudit systems, i.e. two d-dimensional systems, in order to gain deeper insight into the subject. The motivation for studying such systems arises from the fact that those are the most simple extensions of the two-qubit system to higher dimensions (Another simple extension can be realised via multipartite systems, where the number of qubits is increased). At this point we might anticipate that investigations become more and more unfeasible with increasing dimensions. Systems with continuous dimensions and/or a great number of parties are almost impossible to analyse with current mathematical tools.\\
\\
The thesis is organised as follows: In Chapter 1, we first provide an introduction to the mathematical framework necessary to study entanglement, followed by a closer examination of how it can be detected, quantified and classified. In Chapter 2, the focus is on the existence of local realistic theories. We discuss Bell inequalities in details, particularly, the CGLMP inequality. As it will be seen, in order to reveal the nonlocality of a quantum state the measurement setup, which can be expressed in the form of a Bell operator $\mathcal{B}$, has to be optimised. In general, this is a very demanding subject and we contribute to the solution of this problem by presenting a self-developed numerical search algorithm. We end this chapter with further remarks on the comparison of entanglement and nonlocality. The purpose of Chapter 3 is to expose the properties of quantum states in a geometric context by applying entanglement detection criteria and our algorithm. The analysis of the state space of bipartite qubits first leads to a tetrahedron and motivated by its attributes we construct its extension for higher dimensions: the magic simplex.
\newpage
\hfill
\vfill
\newpage

\section{Entanglement}
\subsection{Basics}
\label{basics}
\subsubsection{Definition}
\label{entdef}
In this section we introduce the mathematical definition of entanglement using the most common notation of quantum physics - the \textbf{Dirac notation}, which utilises bra and ket vectors. As we do not go beyond bipartite systems, we restrict all mathematical definitions to these in order to simplify our considerations. Consequently, all state vectors are elements of a composite Hilbert space $\mathcal{H}^{AB}= \mathcal{H}^{A} \otimes \mathcal{H}^{B}$, where $\mathcal{H}^{A}$ and $\mathcal{H}^{B}$ are the Hilbert spaces of the two subsystems $A$ and $B$. All operators acting on $\mathcal{H}^{AB}$ will be elements of the  
corresponding Hilbert-Schmidt space $\mathcal{H}^{AB}_{HS}$, that is the set of bounded linear operators. We will utilise the Hilbert-Schmidt inner product defined by $\left\langle A|B\right\rangle_{HS}=Tr(A^{\dagger} B)$, which induces the Hilbert-Schmidt norm $\left\|A\right\|_{HS}=\sqrt{Tr(A^{\dagger}A)}$. Usually the operations on the subsystems $\mathcal{H}^{A}$ and $\mathcal{H}^{B}$ are attributed to the fictive persons Alice and Bob, respectively. In 1935, Schr\"odinger recognised that EPR correlations are related to states $\ket{\Psi^{AB}} \in \mathcal{H}^{AB}$ that cannot be written as tensor products of state vectors $\ket{{\Psi}^{A}} \in \mathcal{H}^{A}$ and $\ket{{\Psi}^{B}} \in \mathcal{H}^{B}$. Consider an arbitrary state $\ket{\Psi^{AB}} \in \mathcal{H}^{AB}$. By choosing a basis $\left\{\ket{i^{A}}\right\}$ of $\mathcal{H}^{A}$ and $\left\{\ket{j^{B}}\right\}$ of $\mathcal{H}^{B}$ any state of $\mathcal{H}^{AB}$ can be written in the form
\begin{align}
\label{anystate}
\ket{{\Psi}^{AB}}=\sum_{i,j} c_{ij} \ket{{i}^{A}} \otimes \ket{{j}^{B}} \ , 
\end{align}
with $c_{ij} \in \mathbb{C} $ and normalisation $\sum_{i,j} c_{ij}^{*}c_{ij}=1$. We call states that can be written in the form
\begin{align}
\label{sepdef}
\ket{{\Psi}^{AB}}=\ket{{\Psi}^{A}} \otimes \ket{{\Psi}^{B}} 
\end{align}
separable. There is a class of states that cannot be written that way and we denote them as non-separable or entangled. A separable state can contain only classical correlations while an entangled state can contain quantum/EPR correlations. 
The definition implies that a state (of a subsystem) in general cannot be described by a state vector. It follows that open quantum systems need to be described by density matrices. Those also enable us to take into account decoherence and imperfect state preparations in experiments. For the density matrix formalism however, a generalised definition of separability for mixed states is required. If we suppose that a product state, regardless of whether it is pure or mixed remains separable under local operations and classical communication (see \S \ref{LOCC}) we can infer that a state is separable iff it can be written in the form
\begin{align}
\label{sepdefdens}
\rho ^{AB}=\sum_{i} p_{i}  \rho _{i}^{A} \otimes \rho _{i}^{B} \ ,
\end{align}
with $p_{i}\geq 0$ and $\sum_{i} p_{i} =1$.  
\subsubsection{The two-qubit system}
\label{qubits}
We start with the two-qubit system which has the minimal number of degrees of freedom essential for entanglement. The qubit stands for a system with a two dimensional Hilbert space $\mathcal{H}=\mathbb{C}^2$ with an orthonormal basis denoted by $\left\{ \ket{0}, \ket{1}\right\}$ and is therefore the quantum mechanical counterpart of a classical bit. Nevertheless, it should emphasized that there are major differences between the classical bit and the quantum bit. While a classical bit can either have the value 0 or 1 a qubit can in principle store an infinite amount of information because of the infinitely many superpositions of $\ket{0}$ and $\ket{1}$. However, this information is unusable since we cannot distinguish between two non-orthogonal states with a single measurement. Hence, we can only work with superpositions during information processing (quantum computations), the outcome however should be in an eigenstate of the measured observable. It should be mentioned that qubit systems are more than just simple examples of low dimensional quantum systems. This is because of their relevance in quantum optics (polarisation of photons in horizontal $\ket{H}$ or vertical $\ket{V}$ direction) and experiments with spin $\frac{1}{2}$ particles (spin up $\ket{\uparrow}$ or spin down $\ket{\downarrow}$). As we have a two dimensional Hilbert space we need an operator basis with four elements to declare an operator. In many cases it is useful to work with the Pauli operator basis $\left\{\mathbbm{1},\sigma_{1} , \sigma_{2}, \sigma_{3}\right\}$\footnote{$\sigma_{1}=\begin{pmatrix} 0 & 1 \\ 1 & 0 \end{pmatrix}=\ket{0}\bra{1}+\ket{1}\bra{0}$, \ $\sigma_{2}=\begin{pmatrix} 0 & -i \\ i & 0 \end{pmatrix}=-i\ket{0}\bra{1}+i\ket{1}\bra{0}$, \ $\sigma_{3}=\begin{pmatrix} 1 & 0 \\ 0 & -1 \end{pmatrix}=\ket{0}\bra{0}-\ket{1}\bra{1}$}, that is an orthogonal basis according to the Hilbert-Schmidt inner product $\langle\sigma_k|\sigma_l\rangle_{HS}=2\delta_{kl}$, $\langle\mathbbm{1}|\sigma_l\rangle_{HS}=0$, $\langle\mathbbm{1}|\mathbbm{1}\rangle_{HS}=2$. Any operator can be written in the form
\begin{align}
\label{POB}
O= a_0 \mathbbm{1} + \sum_{i=1}^{3}{a_{i}\sigma_{i}} \ ,
\end{align}
with all $a_{i} \in \mathbb{C}$.
In the case of a density matrix $O=\rho$ we write
\begin{align}
\rho= \frac{1}{2} \left( b_0 \mathbbm{1} + \vec{b}\cdot  \vec{\sigma} \right) \ ,
\end{align}
which is equivalent to ($\ref{POB}$) and has advantages with respect to the following restrictions
\begin{align}
b_0&=1 \ , \\
\vec{b} & \in \mathbb{R}^3 \ , \\
\|\vec{b} \| & \leq 1 \ .
\end{align}
The first restriction follows from the condition $tr(\rho)=1$. The second is a necessary condition for $\rho$ to be hermitian. Since $\rho$ also has to be positive semi-definite we get the third restriction by computing $det(\rho)=\frac{1}{4}(1-\|\vec{b} \| ^2 )$. ($det(\rho)\geq0$ is a necessary condition for non-negativity, in our case it is also sufficient because of $tr(\rho)=1$ it is not possible for $\rho$ to have two negative eigenvalues).
As shown from this, the quantum state of a single qubit can be fully described by a three dimensional real vector $\vec{b}$ that lies within a three dimensional sphere with radius 1. Vector $\vec{b}$ is called the Bloch vector and the sphere, Bloch sphere.
If the vector lies on the sphere, the state is pure; if it lies inside, the state is mixed. This follows immediately from $det(\rho)=0$ for $\|\vec{b} \| = 1$, meaning one eigenvalue has to be $0$. This implies that the other eigenvalue has to be $1$, and consequently the state is pure.\\
In the following, we discuss systems of two qubits. To describe such systems we need a four-dimensional Hilbert space $\mathcal{H}^{AB}=\mathbb{C}^2 \otimes \mathbb{C}^2$ and we can choose, for example, an orthonormal basis of type $\{\ket{0^A}\otimes\ket{0^B}, \ket{0^A}\otimes\ket{1^B},\ket{1^A}\otimes\ket{0^B},\ket{1^A}\otimes\ket{1^B}\}$. If there is no likelihood of confusion we can get rid of the indices A and B and the tensor products $\left\{\ket{00},\ket{01},\ket{10},\ket{11}\right\}$. Before we discuss the Pauli operator basis for bipartite qubit systems in detail, we introduce the most well-known entangled states, namely the Bell states
\begin{align}
\ket{\Psi ^\pm}&=\frac{1}{\sqrt{2}}\left(\ket{01} \pm \ket{10}\right) \ , \\
\ket{\Phi ^\pm}&=\frac{1}{\sqrt{2}}\left(\ket{00} \pm \ket{11}\right) \ .
\end{align}
The reader may convince himself that these states are indeed entangled, to be more precise, these states are maximally entangled as will be seen in our subsequent discussion of entanglement measures in \S\ref{EntMeas}. One (perhaps unexpected) feature is that they are all equivalent in terms of local unitaries. The application of a unilateral Pauli matrix $\left\{\sigma_1\otimes1, \sigma_2\otimes1,\sigma_3\otimes1\right\}$ onto a certain Bell state yields another Bell state (up to a non-relevant global phase)
\begin{align}
\sigma_1\otimes1 \  &: \  \ket{\Psi ^\pm} \leftrightarrow \ket{\Phi ^\pm} \ , \\
\sigma_2\otimes1 \  &: \  \ket{\Psi ^\pm} \leftrightarrow \ket{\Psi ^\mp} \ , \\
\sigma_2\otimes1 \  &: \  \ket{\Phi ^\pm} \leftrightarrow \ket{\Phi ^\mp} \ , \\
\sigma_3\otimes1 \  &: \  \ket{\Psi ^\pm} \leftrightarrow \ket{\Phi ^\mp} \ .
\end{align}
In the next section we systemize the above mentioned attribute to get the generalised Bell states in higher dimensional systems. It is certain that properties and applications of Bell states could be further discussed. However, we continue with the analysis of our bipartite qubit system by examining the operators acting on the Hilbert space.
As indicated before we now introduce the Pauli operator basis for the four dimensional Hilbert space. Any operator can be written in the form
\begin{align}
\label{4DPOB}
O=\alpha \mathbbm{1} \otimes \mathbbm{1} +\sum_{i=1}^3a_i\sigma_i\otimes\mathbbm{1} + \sum_{i=1}^3 b_i \mathbbm{1}\otimes \sigma_i + \sum_{i,j=1}^3 c_{ij}\sigma_i\otimes\sigma_j \ ,
\end{align}
with $\alpha,a_i,b_i,c_{ij} \in \mathbb{C}$. Due to the fact that Pauli matrices are hermitian, it is apparent that $\alpha,a_i,b_i,c_{ij} \in \mathbb{R}$ for all hermitian operators (for example observables or density matrices). Once again we change the notation slightly
\begin{align}
\label{generaldensity}
\rho=c\left( \mathbbm{1}\otimes\mathbbm{1} + \vec{r}\cdot\vec{\sigma}\otimes\mathbbm{1} + \mathbbm{1}\otimes \vec{s} \cdot \vec{\sigma} + \sum_{n,m=1}^{3}t_{nm}\sigma_n\otimes \sigma_m \right) \ ,
\end{align}
and specify the constraints for density matrices 
\begin{align}
 c&=\frac{1}{4} \ , \\
r_i,s_i,t_{nm} & \in \mathbb{R} \ , \\
 \left\|\vec{r} \right\|^2 +  \left\|\vec{s} \right\|^2 + \sum_{n,m=1}^{3}t_{nm}^2 & \leq 3 \ , \\
Tr\left(\ket{\Psi}\bra{\Psi}  \rho\right) & \geq 0 \hspace{1cm} \forall \ket{\Psi} \in \mathcal{H}^{AB} \ .
\end{align}
The first restriction has to be fulfilled because $Tr(\rho)$ has to be 1. Hermicity of $\rho$ requires restriction two. The condition $Tr(\rho^2)\leq1$ implies the third one, while the fourth condition is the non-negativity condition.
As a conclusion, we can say that for two qubits any density matrix $\rho \in \mathcal{H}^{AB}_{HS}$ can be fully specified by two real vectors $\vec{r}$ and $\vec{s}$ and a real 3x3 matrix with elements $t_{nm}$. The vectors $\vec{r}$ and $\vec{s}$ determine the local characteristics of $\rho$ related to the systems A and B, which can be seen by computing the reduced density matrices
\begin{align}
\rho_A=Tr_{B}\rho=\frac{1}{2}\left(\mathbbm{1}+\vec{r}\cdot\vec{\sigma}\right) \ , \\
\rho_B=Tr_{A}\rho=\frac{1}{2}\left(\mathbbm{1}+\vec{s}\cdot\vec{\sigma}\right) \ .
\end{align}
It can be seen that there is no dependence on the parameters $t_{nm}$, which can be regarded as correlation parameters since they reflect correlations of classical or EPR type. For this reason, the matrix with components $t_{nm}$ is sometimes referred to as a correlation matrix. For a given density matrix $\rho$ all parameters can be obtained by the use of the associated algebra
\begin{align}
\label{algebra}
r_{i}&=Tr\left(\sigma_i\otimes\mathbbm{1}\cdot\rho\right) \ , \\
s_{i}&=Tr\left(\mathbbm{1}\otimes\sigma_i\cdot\rho\right) \ , \\
t_{nm}&=Tr\left(\sigma_n\otimes\sigma_m\cdot\rho\right) \ .
\end{align}

\subsubsection{Bipartite qudit systems}
\label{Biqudits}
The investigation of quantum systems with few dimensions with the aim of gaining insight into the fundamental properties of quantum theory has been one of the most seminal concepts of quantum information and has led to interesting observations and countless applications of entanglement. The two-qubit system has allowed us to study entanglement in the absence of mathematical complexity caused by high dimensionality. However, within recent years, the focus has been on entanglement in quantum systems with more degrees of freedom. We now go beyond the familiar two-qubit system and concentrate on the entanglement of bipartite systems with arbitrary dimensions.\\
The four Bell states have been useful for many quantum algorithms and seminal experiments. There is a very insightful way to generalise those maximally entangled Bell states onto Hilbert spaces $\mathcal{H}^{AB}=\mathbb{C}^d \otimes \mathbb{C}^d$ with any desired $d \geq2$. The analogue of the $\ket{\Phi^+}$ state in $\mathbb{C}^d \otimes \mathbb{C}^d$ is a state of the form
\begin{align}
\ket{\Omega_{0,0}}=\frac{1}{\sqrt{d}}\sum_{s=0}^{d-1}\ket{s^A}\otimes\ket{s^B} \ ,
\end{align}
with an arbitrary orthonormal basis $\left\{\ket{s^A}\right\}$ of $\mathcal{H^A}$ and $\left\{\ket{s^B}\right\}$ of $\mathcal{H^B}$. We briefly recall that Bell states are equivalent in terms of local unitaries. Consider the application of any local unitary transformation $U_{A} \otimes U_{B}$ onto $\ket{\Omega_{0,0}}$
\begin{align}
U_{A} \otimes U_{B}\ket{\Omega_{0,0}}=\frac{1}{\sqrt{d}}\sum_{s=0}^{d-1}U_A\ket{s^A}\otimes U_B\ket{s^B} \ .
\end{align}
Since the transformations $\ket{s^{'A}}=U_A\ket{s^A}$ and $\ket{s^{'B}}=U_B\ket{s^B}$ are basis transformations giving the orthonormal basis $\left\{\ket{s^{'A}}\right\}$ of $\mathcal{H^A}$ and $\left\{\ket{s^{'B}}\right\}$ of $\mathcal{H^B}$, the resulting state is once again a Bell state
\begin{align}
\frac{1}{\sqrt{d}}\sum_{s=0}^{d-1}\ket{s^{'A}}\otimes\ket{s^{'B}} \ .
\end{align}
If there are $d^2-1$ particular local unitaries that produce additional $d^2-1$ mutually orthonormal Bell states starting from $\ket{\Omega_{0,0}}$, we end up with an orthonormal basis of $d^2$ Bell states. Local unitaries with such properties are the \textbf{Weyl operators}, defined by the action
\begin{align}
\label{weylaction}
W_{k,l}\ket{s}&=w^{k(s-l)}\ket{(s-l)\mbox{mod}\ d} \ , \\
w&=e^{i2\pi/d} \ ,
\end{align}
with $k$, $l$ $\in \left\{0,...,d-1\right\}$. The complete transformation on $\mathcal{H}^{AB}$ is given by $W_{k,l}\otimes\mathbbm{1}$, producing $d^2$ orthonormal \textbf{generalised Bell states}
\begin{align}
\label{GBS}
\ket{\Omega_{k,l}}=\left(W_{k,l}\otimes\mathbbm{1}\right)\ket{\Omega_{0,0}} \ .
\end{align}
We give the explicit verification of orthonormality
\begin{align}
\langle{\Omega_{m,n}}|\Omega_{k,l}\rangle&= \frac{1}{d}\sum_{r,s=0}^{d-1}\langle (r-n)\mbox{mod}\ d|\otimes\bra{r} w^{-m(r-n)}w^{k(s-l)}|(s-l)\mbox{mod}\ d\rangle\otimes\ket{s} \nonumber \\
&=\frac{1}{d}\sum_{r,s=0}^{d-1}w^{k(s-l)-m(r-n)}\langle (r-n)\mbox{mod}\ d|(s-l)\mbox{mod}\ d\rangle\underbrace{\langle r| s\rangle}_{=\delta_{rs}} \nonumber \\
&=\frac{1}{d}\sum_{s=0}^{d-1}w^{k(s-l)-m(s-n)}\underbrace{\langle (s-n)\mbox{mod}\ d|(s-l)\mbox{mod}\ d\rangle}_{\delta_{nl}} \nonumber \\
&=\delta_{nl} \frac{1}{d}\sum_{s=0}^{d-1}w^{k(s-l)-m(s-l)} \nonumber \\
&=\delta_{nl} \frac{1}{d}\sum_{s=0}^{d-1}w^{(k-m)(s-l)} \nonumber \\
&=\delta_{nl}\delta_{km} \ .
\label{orthproof2}
\end{align}
Weyl operators obey the Weyl relations
\begin{align}
W_{j,l}W_{k,m}=w^{kl}W_{j+k,l+m} \ , \\
W^{\dagger}_{k,l}=W^{-1}_{k,l}=w^{kl}W_{-k,-l} \ .
\end{align}
Initially, the Weyl operators were rather contrived for the quantization of classical kinematics instead of the construction of a basis of orthonormal Bell states for Hilbert spaces $\mathcal{H}^{AB}=\mathbb{C}^d \otimes \mathbb{C}^d$. In \S \ref{MagicSimplex} we discuss how this has to be understood and how this concept can help us understand the symmetries and equivalences of quantum states. Since we are now able to construct a basis of Bell states, we continue with seeking a practical operator basis for $\mathcal{H}^{AB}_{HS}$. For qubits, the Pauli operator basis has led to a simple presentation of density matrices via Bloch vectors. For qudits, we once again expand operators in the form
\begin{align}
\label{Blochstyle}
O= a_0 \mathbbm{1} + \sum_{i=1}^{d^2-1}{a_{i}\Gamma_{i}} \ ,
\end{align}
with the $d\times d$ matrices $\left\{\mathbbm{1},\Gamma_{1},...,\Gamma_{d^2-1}\right\}$ forming an orthogonal operator basis $\mathcal{H}^{AB}_{HS}$. If we impose the operators $\left\{\Gamma_{i}\right\}$ to be traceless we can fix the parameter $a_0=\frac{1}{d}$ for density matrices because of the constraint $Tr\rho=1$, just as we did for qubits
\begin{align}
\rho= \frac{1}{d} \mathbbm{1} + \sum_{i=1}^{d^2-1}{a_{i}\Gamma_{i}} \ .
\end{align}
It remains open to show which $\left\{\Gamma_{i}\right\}$ are beneficial for computations and parameterisations. There are various candidates and we want to discuss two of them, namely the \textbf{generalised Gell-Mann matrix basis} and the \textbf{Weyl operator basis}. Both coincide with the Pauli operator basis for dimension two.
We start with the definition of the generalised Gell-Mann (GGM) matrices. For every dimension $d$ we have $d^2-1$ matrices divided into three groups:
\begin{enumerate}
	\item $\frac{d(d-1)}{2}$ symmetric GGM matrices
	\begin{align}
	\Lambda_s^{jk}=\ket{j}\bra{k}+\ket{k}\bra{j} \hspace{2cm}  0\leq j < k\leq d-1  
	\end{align}
	
	\item $\frac{d(d-1)}{2}$ antisymmetric GGM matrices
	\begin{align}\label{GGM1}
	\Lambda_a^{jk}=-i\ket{j}\bra{k}+i\ket{k}\bra{j} \hspace{1.3cm}  0\leq j < k\leq d-1   
	\end{align}
	
	\item $(d-1)$ diagonal GGM matrices
	\begin{align}\label{GGM2}
	\Lambda^{l}=\sqrt{\frac{2}{(l+1)(l+2)}}\left(\sum_{j=0}^l\ket{j}\bra{j}-\left(l+1\right)\ket{l+1}\bra{l+1}\right) \  \  \  \  \  0\leq l\leq d-2 
	\end{align}
\end{enumerate}
The definitions imply that they are all hermitian and traceless. For proof of orthogonality please refer to \cite{BK-Bloch}. Due to hermicity of the GGM matrices all expansion coefficients have to be reals $a_i \in \mathbb{R}$ for hermitian operators, furthermore for density matrices we can again assign a real $d^2-1$ dimensional Bloch vector
\begin{align}
\rho = \frac{1}{d} \mathbbm{1} + \vec{b_{\Lambda}} \cdot \vec{\Lambda} \ ,
\end{align}
with Bloch vector $\vec{b_{\Lambda}}=\left(\left\{b_s^{jk}\right\},\left\{b_a^{jk}\right\},\left\{b^{l}\right\}\right)$ and $\vec{\Lambda}=\left(\left\{\Lambda_s^{jk}\right\},\left\{\Lambda_a^{jk}\right\},\left\{\Lambda^{l}\right\}\right)$, with restrictions for $j,k,l$ given in the definitions of the GGM matrices.
The vector lies within the \textbf{Bloch hypersphere} which precisely means $\|\vec{b_{\Lambda}} \| \leq \sqrt{(d-1)/2d}$, originating from the constraint $Tr(\rho^2)\leq 1$ for density matrices.
While for dimension two, all vectors within the sphere result in positive semi-definite operators $\rho$, whereas in higher dimensions this is not the case. This means that there are areas within the sphere that are restricted for density matrices because of resulting $\rho<0$. Unfortunately, until now no general expression or parametrisation has been found to avoid the holes within the sphere. Hence, this criterion has to be checked separately.\\
The alternative, namely the \textbf{Weyl operator basis} is given by the matrices introduced in ($\ref{weylaction}$) leading to
\begin{align}
W_{kl}=\sum_{s=0}^{d-1}e^{\frac{2\pi i }{d}sk}\ket{s}\bra{(s+l)\ \mbox{mod}\ d}\  \hspace{1cm} \  \  k,l \in \left\{0,...,d-1\right\} \ .
\label{wob}
\end{align}
Once again, we have the unity $W_{00}=\mathbbm{1}$ and $d^2-1$ additional matrices forming a basis of mutually orthogonal operators. The proof of orthogonality is very similar to ($\ref{orthproof2}$) and can be found in \cite{BK-Bloch}. Hence, in the Weyl operator basis the Bloch vector expression of any density matrix is
\begin{align}
\rho = \frac{1}{d} \mathbbm{1} + \vec{b_{W}} \cdot \vec{W} \ ,
\end{align}
with a $d^2-1$ dimensional Bloch vector $\vec{b_{W}}$ and a vector $\vec{W}$ containing all operators $W_{kl}$ except $W_{00}$. It is clear that one has to take into account the arrangement of the vector components. The main discrepancy between $\vec{b_W}$ and the foregoing Bloch vectors is that the components of $\vec{b_{W}}$ can be complex and they have to satisfy $b_{-k\; -l}=e^{\frac{2\pi i}{d} kl} b^*_{kl}$ (to be understood modulo d) for hermicity of $\rho$. This is implied when comparing the definition (\ref{wob}) with hermicity $\rho^{\dagger}=\rho$. As before, the constraint $Tr(\rho^2)\leq 1$ enforces the vector $\vec{b_W}$ to lie within a Bloch hypersphere $\|\vec{b_W}\| \leq \sqrt{(d-1)}/d$. Equal to the vectors $\vec{b_{\Lambda}}$, not all vectors $\vec{b_W}$ within this sphere are permitted, since some lead to matrices with negative eigenvalues. An expression in terms of the components $b_{kl}$ has not yet been established.\\
Finally we want to extend our Bloch-type operator expansion (\ref{Blochstyle}) to bipartite qudit systems.
We can do this by generalising the expression (\ref{4DPOB}) to
\begin{align}
\label{biqudit}
O=\alpha \mathbbm{1} \otimes \mathbbm{1} +\sum_{i=1}^{d^2-1}a_i\Gamma_i\otimes\mathbbm{1} + \sum_{i=1}^{d^2-1} b_i \mathbbm{1}\otimes \Gamma_i + \sum_{i,j=1}^{d^2-1} c_{ij}\Gamma_i\otimes\Gamma_j \ .
\end{align}
\subsection{Detection of entanglement}
In \S \ref{entdef} we have introduced the definition of separability and entanglement for pure and mixed states. Even though the distinction is well defined, in practice it is difficult to either find a separable decomposition or to prove that such a decomposition does not exist. For this reason, it is preferable to find operational criteria. As it will be shown, all known methods are either unfeasible or not sufficient to solve the problem completely.
\subsubsection{Reduced density matrices of pure states}
For all bipartite pure states, a necessary and sufficient criterion for separability arises through the form of the reduced density matrices. It is always possible to change the indexing of expression (\ref{anystate}), using one index running from $1$ to $d_A \cdot d_B$ (dimension of $\mathcal{H}_A$ times dimension of $\mathcal{H}_B$) instead of two indices running from $1$ to $d_A$ and $1$ to $d_B$, respectively,
\begin{align}
\ket{{\Psi}^{AB}}=\sum_{i=1}^{d_A \cdot d_B} a_{i} \ket{{i}^{A}} \otimes \ket{{i}^{B}}  \ ,
\end{align}
with elements $\ket{{i}^{A}} \in \mathcal{H}^A$ and $\ket{{i}^{B}} \in \mathcal{H}^B$ of some orthonormal bases. For a given state $\ket{{\Psi}^{AB}}$ one can minimise the number of nonzero coefficients $a_i$ by the use of proper orthonormal vectors $\{\ket{\widetilde{i}^{A}}\}$ and $\{\ket{\widetilde{i}^{B}}\}$ (see \cite{Audretsch}). If we change the order of the numeration starting with nonzero coefficients $a_i\neq0$ for $i\in \left\{1,..,r\right\}$ followed by $a_i=0$ for $i\in \left\{r+1,..,d_A \cdot d_B\right\}$ we end up with the \textbf{Schmidt decomposition}
\begin{align}
\label{schmidt}
\ket{{\Psi}^{AB}}=\sum_{i=1}^{r} \sqrt{\lambda_{i}} \ket{\widetilde{i}^{A}} \otimes \ket{\widetilde{i}^{B}} \  \  \  \   \   \   \   a_i=\sqrt{\lambda_{i}} \ .
\end{align}
with $r\leq \mbox{min}\left\{d_A,d_B\right\}$ and $\left\{\lambda_i\right\}$ known as the \textbf{Schmidt coefficients}. By computing the partial traces we realise that $\left\{\lambda_i\right\}$ are the eigenvalues of the reduced density matrices
\begin{align}
\rho_A=Tr_B \left(\ket{{\Psi}^{AB}} \bra{{\Psi}^{AB}}\right)=\sum_{i=1}^r \lambda_i \ket{\widetilde{i}^{A}} \bra{\widetilde{i}^{A}}  \ , \\
\rho_B=Tr_A \left(\ket{{\Psi}^{AB}} \bra{{\Psi}^{AB}}\right)=\sum_{i=1}^r \lambda_i \ket{\widetilde{i}^{B}} \bra{\widetilde{i}^{B}} \ .
\end{align}
Obviously, the resulting density matrices are diagonal and have rank $r$. Consequently, r is called the \textbf{Schmidt rank}. By comparing (\ref{sepdef}) with (\ref{schmidt}) we infer Schmidt rank $r=1$ for all separable states and $r>1$ for all entangled states. Since the only density matrices with rank 1 are pure states (they have only one eigenvalue $\lambda_1=1$) we come to the conclusion that all separable states result in pure reduced density matrices, while entangled states give mixed ones. Note that for practical purposes it is not necessary to construct the Schmidt decomposition of $\ket{{\Psi}^{AB}}$ before tracing out a system because the rank will always be equal due to basis independence of the trace. Hence, we have found a powerful tool to distinguish pure separable states from pure entangled ones since it is trivial to examine the mixedness of a density matrix. To summarise, a necessary and sufficient condition for separability of pure states is given by
\begin{align}
\ket{{\Psi}^{AB}} \mbox{ is separable } \hspace{0.5cm}  \Leftrightarrow  \hspace{0.5cm} Tr(\rho_{A/B}^2)=1 \ . 
\end{align}
\subsubsection{Detection via positive maps}
\label{CPmaps}
As we know, in general a quantum state can only be fully described by a density matrix. Here, for a given state it is more difficult to determine whether it is separable or entangled. We begin with a rather theoretical consideration of necessary and sufficient conditions which was presented by the Horodeckis in 1996 \cite{HorodeckiSEP}.\\
Consider an operator $\Omega \in \mathcal{H}^1_{HS}$ and a linear map $\Lambda$ : $\mathcal{H}^1_{HS} \rightarrow \mathcal{H}^2_{HS}$. We say the map $\Lambda$ is \textbf{positive (P)} if it maps any positive operator in $\mathcal{H}^1_{HS}$ into the set of positive operators in $\mathcal{H}^2_{HS}$, in terms of mathematics if
$Tr(\Omega \cdot P_1)\geq0$ implies $Tr(\Lambda(\Omega) \cdot P_2 )\geq0$ for all projectors $P_1\in\mathcal{H}^1_{HS}$ and $P_2 \in \mathcal{H}^2_{HS}$. Now, consider the extended linear map $\left[\Lambda \otimes \mathbbm{1}_n\right]$ : $\mathcal{H}^1_{HS}\otimes\mathcal{M}_n\rightarrow \mathcal{H}^2_{HS}\otimes\mathcal{M}_n$. Here $\mathcal{M}_n$ stands for the set of the complex matrices n$\times$n and $\mathbbm{1}_n$ is the appertaining identity. The map $\Lambda$ is said to be \textbf{completely positive (CP)} if its extension maps any positive operator $\Sigma \in \mathcal{H}^1_{HS}\otimes\mathcal{M}_n$ into the set of positive operators $\mathcal{H}^2_{HS}\otimes\mathcal{M}_n$, i.e. $Tr(\Sigma \cdot P_1)\geq0$ implies $Tr(\left[\Lambda \otimes \mathbbm{1}_n\right](\Sigma) \cdot P_2 )\geq0$ for all projectors $P_1\in\mathcal{H}^1_{HS}\otimes\mathcal{M}_n$ and $P_2 \in \mathcal{H}^2_{HS}\otimes\mathcal{M}_n$ and all $n \in \mathbb{N}$. It might be a bit surprising that positive maps are not necessarily completely positive. All possible physical transformations are CP maps because the transformation of a density matrix has to result in another valid density matrix. For the solution of our separability problem we recognise the following. Consider a separable state $\rho ^{AB}=\sum_{i} p_{i}  \rho _{i}^{A} \otimes \rho _{i}^{B}$ and a positive map $\Lambda_A$ (not certainly CP) inducing $\left[\Lambda_A \otimes \mathbbm{1}_B \right]$. The mapping of $\rho ^{AB}$ gives
\begin{align}
\left[\Lambda_A \otimes \mathbbm{1}_B\right]\left(\sum_{i} p_{i}  \rho _{i}^{A} \otimes \rho _{i}^{B}\right)=\sum_{i} p_{i}  \left(\Lambda_A\left(\rho _{i}^{A}\right)\right) \otimes \rho _{i}^{B} \ .
\end{align}
According to our assumption $\Lambda_A$ being a positive map $\Lambda_A\left(\rho _{i}^{A}\right)$ only has non-negative eigenvalues and the same obviously holds for $\rho _{i}^{B}$. All in all, we have a sum of positive operators weighted with $p_i \geq 0$ yielding a positive operator.
Hence for positive maps $\Lambda_A$ the inequality $\left[\Lambda_A \otimes \mathbbm{1}_B\right]\left(\rho^{AB}\right) \geq 0$ is a necessary condition for $\rho^{AB}$ to be separable. Furthermore, it can be shown (see \cite{HorodeckiSEP}) that for every inseparable state $\rho$ there exists a positive map $\Lambda_{\rho}$ so that $\left[\Lambda_{\rho} \otimes \mathbbm{1}\right] \left(\rho\right) < 0$. The existence of such a map $\Lambda_{\rho}$ is an impressive matter from a theoretical point of view, but for a given state $\rho$ there is no recipe for how to construct such a mapping. Nevertheless, there are several approved positive maps for the detection of entanglement. One of them is the \textbf{reduction map}
\begin{align}
\Lambda_{red}\left(\rho\right)=\mathbbm{1} Tr(\rho) - \rho \ .
\end{align}
The positivity of the map can be proven by
\begin{align}
\bra{\Psi}\mathbbm{1} Tr(\rho) - \rho\ket{\Psi} =Tr(\rho) \underbrace{\left\langle \Psi | \Psi\right\rangle}_{1}  - \bra{\Psi}\rho\ket{\Psi}\geq 0 \mbox{ \  \  \   } \forall \ket{\Psi} \in \left\{\ket{\Psi}| \left\langle \Psi | \Psi\right\rangle =1\right\} \ . \nonumber
\end{align}
$Tr(\rho)$ is the sum of all eigenvalues (all positive or zero for positive operators) and $\bra{\Psi}\rho\ket{\Psi}$ can maximally yield the largest eigenvalue, thus the map is positive.\\
Therefore the \textbf{reduction criterion} \cite{HorodeckiRED} signifies that a separable state has to fulfill the inequalities
\begin{align}
\rho_A \otimes \mathbbm{1} - \rho_{AB} \geq 0 \mbox{ \  \ and   \   } \mathbbm{1} \otimes \rho_B - \rho_{AB} \geq 0 \ ,
\end{align}
with the reduced density matrices $\rho_A$ and $\rho_B$.\\
Another method based on positive maps is called \textbf{positive partial transpose (PPT) criterion} \cite{PeresPPT}. For this criterion the well known transposition is used to determine entanglement. For a given operator $A$, the transposition is defined by
\begin{align}
A&=\sum_{i,j}a_{ij}\ket{i} \bra{j} \ ,\\
\Lambda_T(A)=A^T&=\sum_{i,j}a_{ij}\ket{j} \bra{i} \ .
\end{align}
It is obvious that this map is positive
\begin{align}
\mbox{A is positive}\Rightarrow \bra{\Psi}A\ket{\Psi}&=\sum_{i,j}a_{ij}\overbrace{\langle  \Psi \ket{i}}^{\Psi_i^*} \overbrace{\bra{j} \Psi \rangle}^{\Psi_j}\geq0 \mbox{ \  \  \  \  } \forall \ket{\Psi} \in \left\{\ket{\Psi}: \left\langle \Psi | \Psi\right\rangle =1\right\} \ .
\label{equation23}
\end{align}
Choose the vector $\ket{\Psi'}=\ket{\Psi^*}$ and compute
\begin{align}
\bra{\Psi'}A^T\ket{\Psi'}=\bra{\Psi^*}A^T\ket{\Psi^*}=\sum_{i,j}a_{ij}\overbrace{\langle\Psi^*\ket{j}}^{\Psi_j} \overbrace{\bra{i} \Psi^* \rangle}^{\Psi_i^*}\geq0 \ .
\label{equation24}
\end{align}
due to (\ref{equation23}) this is always true for all $\ket{\Psi'}$, thus transposition is a positive map.
As a result, we have found the positive partial transpose criterion. For a given state $\rho_{AB}$ the inequalities
\begin{align}
\rho_{AB}^{T_A}&=\sum_{ijkl}\bra{ij}\rho_{AB}\ket{kl}\cdot \ket{k}\bra{i}\otimes \ket{j}\bra{l} \geq 0\\
\mbox{and \  \  } \rho_{AB}^{T_B}&=\sum_{ijkl}\bra{ij}\rho_{AB}\ket{kl}\cdot \ket{i}\bra{k} \otimes \ket{l}\bra{j} \geq 0
\end{align}
are necessary conditions for separability. The PPT criterion is stronger than the reduction criterion in all cases (see \cite{HorodeckiRED}). It should be mentioned that in the case of a two-qubit system $\mathcal{H}^{AB}=\mathbb{C}^2 \otimes \mathbb{C}^2$ or a qubit-qutrit system $\mathcal{H}^{AB}=\mathbb{C}^2 \otimes \mathbb{C}^3$ it is also a sufficient criterion (see \cite{HorodeckiSEP}). Since PPT is not a sufficient criterion for higher dimensional systems, there exist entangled states with positive partial transpose. As shown in \S \ref{boundentanglement} this feature leads to a phenomenon called \textbf{bound entanglement}.
\begin{figure}[htp]
\centering
\setlength{\unitlength}{1bp}%
  \begin{picture}(368.56, 135.46)(0,0)
  \put(0,0){\includegraphics{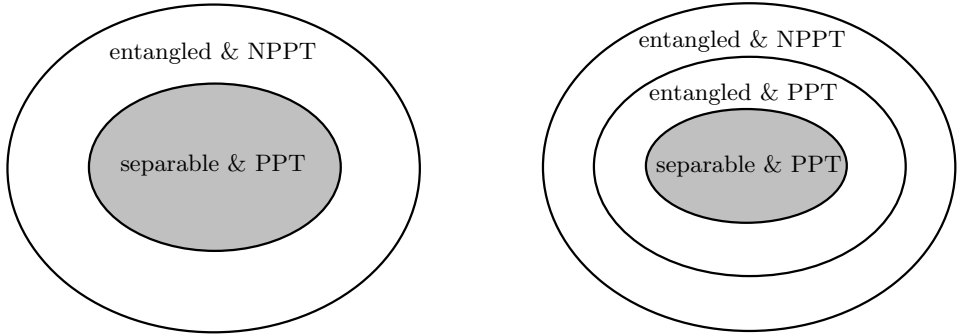}}
  \put(83.24,110.85){\fontsize{8.54}{10.24}\selectfont \makebox(0,0)[]{entangled \& NPPT\strut}}
  \put(83.24,68.44){\fontsize{8.54}{10.24}\selectfont \makebox(0,0)[]{separable \& PPT\strut}}
  \put(285.19,68.23){\fontsize{8.54}{10.24}\selectfont \makebox(0,0)[]{separable \& PPT\strut}}
  \put(282.67,115.26){\fontsize{8.54}{10.24}\selectfont \makebox(0,0)[]{entangled \& NPPT\strut}}
  \put(283.17,95.48){\fontsize{8.54}{10.24}\selectfont \makebox(0,0)[]{entangled \& PPT\strut}}
  \end{picture}%
  \caption{Schematic illustration of the set of quantum states of a two-qubit system (left) and higher dimensional systems (right)}
\end{figure}
\subsubsection{Linear contractions criteria}
Additional operational criteria can be constructed from contraction mappings \cite{HorodeckiCONTR}. The concept behind those criteria also issues from the extension of maps, but yields criteria that are independent from the previously discussed ones. Consider an operator $\Omega \in \mathcal{H}^1_{HS}$ and a linear map $\Lambda$ : $\mathcal{H}^1_{HS} \rightarrow \mathcal{H}^2_{HS}$. The map $\Lambda$ is a \textbf{contraction} iff it does not increase the Hilbert-Schmidt norm, i.e. $\left\|\Lambda\left(\Omega\right)\right\|_{HS}\leq\left\|\Omega\right\|_{HS}$ holds for all $\Omega$. Once again we define the extension of $\Lambda$ analogous to the previous section by $\left[\Lambda \otimes \mathbbm{1}_n\right]$ : $\mathcal{H}^1_{HS}\otimes\mathcal{M}_n\rightarrow \mathcal{H}^2_{HS}\otimes\mathcal{M}_n$. A map $\Lambda$ is a \textbf{complete contraction} iff it is Hilbert-Schmidt norm non-increasing for all possible extensions $n \in \mathbb{N}$, $\left\|\left[\Lambda \otimes \mathbbm{1}_n\right](\Sigma)\right\|_{HS} \leq \left\|\Sigma\right\|_{HS}$ for all $\Sigma \in \mathcal{H}^1_{HS}\otimes\mathcal{M}_n$. Now, for a separable state $\rho ^{AB}=\sum_{i} p_{i}  \rho _{i}^{A} \otimes \rho _{i}^{B}$ and a contraction $\Lambda_A$ we have
\begin{align}
\left\|\left[\Lambda_A \otimes \mathbbm{1}_B\right]\left(\rho^{AB}\right)\right\|_{HS}&=
\left\|\left[\Lambda_A \otimes \mathbbm{1}_B\right]\left(\sum_{i} p_{i}  \rho _{i}^{A} \otimes \rho _{i}^{B}\right)\right\|_{HS} \nonumber \\
& \leq \sum_{i} p_{i}  \underbrace{\left\|\Lambda_A\left(\rho _{i}^{A}\right)\right\|_{HS}}_{\leq \left\|\rho_i^A\right\|_{HS}\leq 1} \underbrace{\left\|\rho _{i}^{B}\right\|_{HS}}_{\leq 1} \nonumber \\
& \leq \sum_{i} p_{i}  = 1  \ .
\end{align}
We have found a necessary condition for separability:
\begin{align}
\left\|\left[\Lambda_A \otimes \mathbbm{1}_B\right]\left(\rho^{AB}\right)\right\|_{HS} \leq 1 \ .
\end{align}
Thus, for the detection of entanglement, the relevant maps are the contractions that are not complete contractions. For instance, transposition is such a map. In general there is no common method to construct the whole set of non-complete contractions. One noteworthy criterion based on the idea of contraction surely is the so called \textbf{Matrix realignment criterion}, which can detect PPT entanglement in some cases. The basic principle is to use a map $\mathcal{R}$ : $\mathcal{H}^{AB}_{HS} \rightarrow \mathcal{H}^{2}_{HS}$ that is a contraction for all separable states within the composite Hilbert space $\mathcal{H}^{AB}_{HS}$. While the contraction $\left[\Lambda_A \otimes \mathbbm{1}_B\right]$ only affects a subspace, the matrix realignment map $\mathcal{R}$ affects the whole space $\mathcal{H}^{AB}_{HS}$ in the following way
\begin{align}
\mathcal{R}(\rho^{AB})=\sum_{ijkl}\bra{ij}\rho^{AB}\ket{kl}\cdot \ket{i} \bra{j} \otimes \ket{k}\bra{l} \ .
\end{align}
As we can see realignment interchanges the basis vectors $\left\{\ket{j}\right\}$ of $\mathcal{H}^A$ applied on the left-hand side with $\left\{\ket{k}\right\}$ of $\mathcal{H}^B$ applied on the right-hand side. This map satisfies
\begin{align}
\left\|\mathcal{R}(\rho^{AB})\right\|_{HS} \leq 1 
\end{align}
for all separable states $\rho ^{AB}=\sum_{i} p_{i}  \rho _{i}^{A} \otimes \rho _{i}^{B}$ (see \cite{chenwu}). All states that violate this inequality are necessarily entangled. Some entangled states that slip through the PPT criterion can be detected via the realignment criterion and examples can be found in \cite{rudolph} and \cite{chenwu}.
\subsubsection{Entanglement witnesses}
\label{witnesses}
The last method for the detection of entanglement we discuss are \textbf{entanglement witnesses}. They originate from the Hahn-Banach theorem in convex analysis stating that a convex set in a vector space (in our case this is the Hilbert-Schmidt space $\mathcal{H}_{HS}^{AB}$) can be fully described by hyperplanes. In other words, there always exists a hyperplane that separates the convex set from the complement. Based on the fact that a hyperplane can always be expressed by a normal vector $W$ (in our case the vector is an element of $\mathcal{H}_{HS}^{AB}$ and therefore a matrix) and the convexity of the set of separable states, we are able to comprehend a theorem stated by the Horodeckis $\cite{HorodeckiSEP}$ and B.M. Terhal $\cite{TerhalSEP}$: A density matrix $\rho_{AB} \in \mathcal{H}_{HS}^{AB}$ is entangled iff there exists a hermitian operator $W \in \mathcal{H}_{HS}^{AB}$ with the properties
\begin{align}
Tr(W\rho_{AB}) & < 0 \ , \\
Tr(W\sigma_{AB}) & \geq 0 \ ,
\end{align}
for all separable density matrices $\sigma_{AB}$. A hermitian operator $W$ accomplishing those requirements is termed an \textbf{entanglement witness}. As a consequence, $W$ has to have at least one negative eigenvalue and due to linearity of the trace, nonnegative expectation values on the subset of product states
\begin{align}
\label{poswitness}
\bra{\Psi^A}\otimes\bra{\Psi^B} W \ket{\Psi^A} \otimes \ket{\Psi^B} \geq 0 \mbox{ \  \  \  \  } \forall \ket{\Psi^A} \otimes \ket{\Psi^B} \in \mathcal{H}^{AB} \ .
\end{align}
We can rank witnesses $W$ by comparing the sets of entangled states they detect, that is $D_W=\left\{\rho | Tr( W \rho) < 0 \right\}$. An entanglement witness $W_1$ is finer than $W_2$ iff $D_{W_2} \subseteq D_{W_1}$. It is optimal if there is no other entanglement witness which is finer. Hence, an optimal witness $W_{optimal}$ defines a tangent plane to the set of separable states which means that there must be at least one separable $\sigma_{AB}$ with $Tr(W_{optimal}\sigma_{AB})=0$.
\begin{figure}[htp]
\centering
  \setlength{\unitlength}{1bp}%
  \begin{picture}(207.46, 141.14)(0,0)
  \put(0,0){\includegraphics{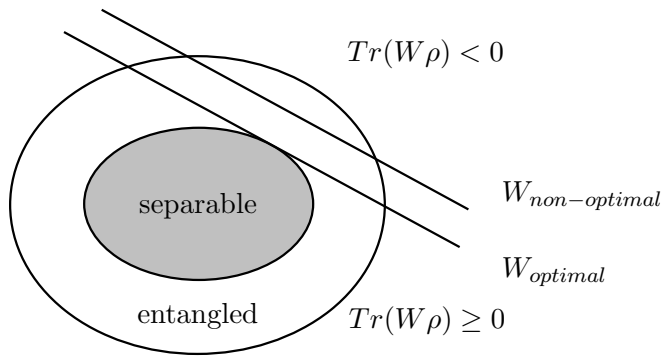}}
  \put(76.55,20.16){\fontsize{11.38}{13.66}\selectfont \makebox(0,0)[]{entangled\strut}}
  \put(76.55,61.63){\fontsize{11.38}{13.66}\selectfont \makebox(0,0)[]{separable\strut}}
  \put(162.36,119.02){\fontsize{11.38}{13.66}\selectfont \makebox(0,0)[]{$Tr(W\rho)<0$\strut}}
  \put(162.84,18.22){\fontsize{11.38}{13.66}\selectfont \makebox(0,0)[]{$Tr(W\rho)\geq0$\strut}}
  \put(191.08,65.94){\fontsize{11.38}{13.66}\selectfont \makebox(0,0)[l]{$W_{non-optimal}$\strut}}
  \put(191.08,37.21){\fontsize{11.38}{13.66}\selectfont \makebox(0,0)[l]{$W_{optimal}$\strut}}
  \end{picture}%
  \caption{Schematic illustration of an optimal and an non-optimal entanglement witness}
\end{figure}\\
Further information for the optimisation of entanglement witnesses can be found in \cite{LewensteinOPT}.
It is evident that a single witness does not detect all entangled states. We only know that for a given entangled state $\rho$ there must be an appertaining entanglement witness $W_{\rho}$, an algorithm for the construction however has not yet been established. For this reason entanglement witnesses cannot be seen as a satisfying solution to the separability problem. Furthermore, it is also not clear how many witnesses are necessary to describe the whole set of separable states. Note that if the set is not a polytope then an infinite number of witnesses is required. The geometry of the set depends on the dimension of the considered space or subspace of $\mathcal{H}^{AB}_{HS}$.
\subsection{Entanglement measures}\label{EntMeas}
The quantification of entanglement is another open problem. This is beyond the separability problem, since the purpose is not only to ascertain if a state is entangled or not, but also to quantify the amount of entanglement within it. This amount should capture the essential features that we associate with entanglement. Before we mention some possible measures we state some requirements on the attributes of a suggestive measure. Take into account that we restrict ourselves to bipartite systems $\mathcal{H}^{AB}= \mathbb{C}^{d} \otimes \mathbb{C}^{d}$ with arbitrary dimension $d$.
\subsubsection{The postulates}
In the first place, an \textbf{entanglement measure} $E(\rho)$ is a mapping of density matrices into the set of positive real numbers
\begin{align}
\rho \rightarrow E(\rho) \in \mathbb{R^+} \ .
\end{align}
Normalisation is not necessary but reasonable and we therefore set
\begin{align}
E(\ket{\Omega_{k,l}}\bra{\Omega_{k,l}})=log_2(d)
\end{align}
for the Bell states $\ket{\Omega_{k,l}}$. This should also be the highest value reachable, because we anticipate them to be maximally entangled due to maximal mixedness of their reduced density matrices $\rho_A=\rho_B=\frac{1}{d}\mathbbm{1}$, which is unique for pure states
\begin{align}
0 \leq E(\rho) \leq log_2(d) \ .
\end{align}
By definition, the outcome of $E(\rho)$ must be zero for all separable states $\rho_{sep}$
\begin{align}
E(\rho_{sep}) = 0 \ .
\end{align}
The most important requirement is \textbf{monotonicity under LOCC}. Entanglement cannot be created by local operations and classical communication ( \S \ref{LOCC} ), hence the contained amount has to be non-increasing under such transformations
\begin{align}
E\left(\Lambda_{LOCC}(\rho)\right) \leq E\left(\rho\right) \ .
\end{align}
This implies that in the special case of local unitaries the outcome of $E(\rho)$ will be invariant \cite{VidalLOCC} 
\begin{align}
E\left(U_A \otimes U_B \rho U_A^{\dagger} \otimes U_B^{\dagger} \right) = E\left(\rho\right) \ .
\end{align}
These postulates are the only ones necessarily required and universally accepted. Some people tend to put further restrictions on entanglement measures. Some of them seem very natural and others can bring mathematical simplicity.
One of them is \textbf{convexity}
\begin{align}
E\left(\sum_i p_i \rho_i\right) \leq \sum_i p_i E\left(\rho_i\right) \ .
\end{align}
Convexity seems to be plausible as we cannot increase entanglement by mixing states and mixtures of entangled states can result in separable states (see \S \ref{Geometry}). \textbf{Asymptotic continuity} is another optional restriction
\begin{align}
\lim_{n\rightarrow \infty}\left\|\rho_n-\sigma_n\right\|_{HS} \rightarrow 0\Rightarrow \lim_{n\rightarrow \infty}  \left|E\left(\rho_n\right)-E\left(\sigma_n\right)\right|\rightarrow0 \ .
\end{align}
The interesting thing is that along with the next limitation called \textbf{additivity} we can obtain a unique measure for pure states \cite{VidalLOCC}. The additivity is separated in three types. We first consider \textbf{partial additivity}
\begin{align}
E(\rho^{\otimes n })= n E(\rho) \ ,
\end{align}
which is the weakest addivity criterion. It only reveals that the entanglement content grows linearly with the number of pairs. We may expect that it is reasonable to assume \textbf{full additivity} (following formula with an equal sign). However, it turned out that this would exclude some appreciated measure candidates. The weakened version is the \textbf{subadditivity } 
\begin{align}
E(\rho \otimes \sigma ) \leq  E(\rho) + E(\sigma) \ .
\end{align}
A more detailed analysis of the postulates and thermodynamical analogies can be found in \cite{Durstberger}, \cite{HorodeckiMEAS} and \cite{Martin}. As mentioned before, the additional constraints of partial additivity and asymptotic continuity, inevitably require the measure to coincide with the \textbf{von Neumann entropy} $S_{vN}$ of the reduced density matrix $\rho^{A/B}=Tr_{B/A}\left(\rho^{AB}\right)$ for a pure state $\rho^{AB}$
\begin{align}
\label{EVN}
E_{vN}(\rho^{AB}) = S_{vN}\left(\rho^{A/B}\right)=-Tr\left(\rho^{A}log_2\left(\rho^{A}\right)\right)=-Tr\left(\rho^{B}log_2\left(\rho^{B}\right)\right) \ .
\end{align}
This is the \textbf{uniqueness theorem of entanglement measures} \cite{DonaldUNITHEOREM}.
\subsubsection{Measures based on distance}
The most intuitive measures are the ones based on distance. The concept is to regard the distance $\mathcal{D}$ of a state $\rho \in \mathcal{H}^{AB}_{HS}$ to the closest separable state $\sigma \in \mathcal{S} \subset \mathcal{H}^{AB}_{HS}$ as the contained amount of entanglement
\begin{align}
E(\rho) = \inf_{\sigma\in\mathcal{S}}\mathcal{D}\left(\rho,\sigma\right) \ .
\end{align}
It is apparent that those measures guarantee $E(\rho)=0$ for separable states. There are several types of possible distance functions $\mathcal{D}$. They do not have to fulfill all criteria for a metric in a mathematical sense, but should meet the requirements of an entanglement measure. Let us consider the \textbf{relative entropy of entanglement} which was introduced by Vedral et al. \cite{VedralRELENT}
\begin{align}
E_R(\rho) = \inf_{\sigma\in\mathcal{S}}Tr\left(\rho\left(log_2\rho-log_2\sigma\right)\right) \ .
\end{align}
The distance used here does not meet the requirements of a metric because it is not symmetric and does not satisfy the triangle inequality. This is acceptable since relative entropy satisfies all criteria of an entanglement measure including asymptotic continuity and partial additivity (see \cite{VedralENTMEAS}). Another distance is induced by the Hilbert-Schmidt norm and was investigated in \cite{WitteTrucks}. The resulting measure is called \textbf{Hilbert-Schmidt measure} or \textbf{Hilbert-Schmidt entanglement}
\begin{align}
E_{HS}(\rho) = \inf_{\sigma\in\mathcal{S}}\left\|\rho-\sigma\right\|_{HS}^2 \ .
\end{align}
This measure has a different scaling than most of the other candidates because no logarithm is taken. For this reason, it is not common to normalise this function in the foregoing way. Up to now it has not been proven that monotonicity under LOCC is accomplished, meaning it is not clear if it is a good measure of entanglement (see \cite{VedralENTMEAS},\cite{Ozawa}).
\subsubsection{Convex roof measures}
The idea of convex roof measures is to use entanglement measures for pure states $E_{pure}$ and generalize them to mixed ones in the following way
\begin{align}
E(\rho) = \inf_{\rho=\sum_i p_i \ket{\Psi_i}\bra{\Psi_i}}\sum_i p_i E_{pure}\left(\ket{\Psi_i}\bra{\Psi_i}\right) \ ,
\end{align}
where the infimum is taken over all possible decompositions of $\rho=\sum_i p_i \ket{\Psi_i}\bra{\Psi_i}$ with $\sum_i p_i=1$ and $p_i\geq0$. The decomposition yielding the infimum is then said to be the \textbf{optimal decomposition} or \textbf{optimal ensemble} of $\rho$. It can be shown that if the utilised measure for pure states $E_{pure}$ is monotonous under LOCC then the induced convex roof measure has this property too (see \cite{Horodecki-summary}). The first measure of this kind was \textbf{entanglement of formation} $E_F$ with the von Neumann entropy of the reduced density matrices as a measure for pure states $E_{pure}=E_{vN}$ introduced in (\ref{EVN})
\begin{align}
E_{F}(\rho) = \inf_{\rho=\sum_i p_i \ket{\Psi_i}\bra{\Psi_i}}\sum_i p_i E_{vN}\left(\ket{\Psi_i}\bra{\Psi_i}\right) \ .
\end{align}
When this measure was introduced by Bennett et al.\cite{BennettEOF}, the notion was to quantify the amount of pure Bell states that is needed per copy to construct $\rho$. In other words, $N$ copies of $\rho$ can be prepared out of a minimum of $E_{F}(\rho) \cdot N$ Bell states and $(1-E_{F}(\rho)) \cdot N$ separable states. $E_{F}(\rho)$ can then be regarded as the contained amount of entanglement. The naming "entanglement of formation" is referable to this concept.
\subsection{Bound entanglement}
\label{boundentanglement}
In the previous sections we frequently used the term \textbf{local operations and classical communication (LOCC)}. Nevertheless, we neither explained its meaning nor given an adequate mathematical definition. This section will serve as an introduction to this issue. After introducing the class of \textbf{quantum operations}, we consider the restrictions for the class of LOCC. Consequences for the purification of entanglement will be briefly discussed, enabling us to justify the term \textbf{bound entanglement}.  
\subsubsection{Quantum operations}
\label{operations}
We investigate the fundamental quantum operations (see \cite{DonaldUNITHEOREM}, \cite{Clarisse}). Our objects of interest are the maps that transform a given state $\rho \in \mathcal{H}^{1}_{HS}$ into another $\rho' \in \mathcal{H}^{2}_{HS}$
\begin{align}
\Lambda:\mathcal{H}^{1}_{HS}& \rightarrow\mathcal{H}^{2}_{HS} \ , \\
\rho'&=\Lambda(\rho) \ .
\end{align}
As far as we know, any $\Lambda$ is a combination of four elementary linear maps\footnote{It is true that the first two types of transformations are not "real physical" operations, but they are the conversion of experimental operating principles into mathematical diction.}:
\begin{itemize}
\item{\textbf{Adding an uncorrelated ancilla} $\sigma\in\mathcal{H}^{2}_{HS}$ to the original quantum system in the state $\rho \in \mathcal{H}^{1}_{HS}$
\begin{align}
&\Lambda_1:\mathcal{H}^{1}_{HS} \rightarrow\mathcal{H}^{1}_{HS}\otimes\mathcal{H}^{2}_{HS} \ ,\\
&\Lambda_1(\rho)=\rho \otimes \sigma \ .
\end{align}}
\item{\textbf{Tracing out part of the system} in the state $\rho\in\mathcal{H}^{1}_{HS}\otimes\mathcal{H}^{2}_{HS}$
\begin{align}
&\Lambda_2:\mathcal{H}^{1}_{HS}\otimes\mathcal{H}^{2}_{HS} \rightarrow\mathcal{H}^{1}_{HS} \ , \\
&\Lambda_2(\rho) =Tr_2\rho \ .
\end{align}
Here $Tr_2$ is the partial trace over the Hilbert-Schmidt space $\mathcal{H}^{2}_{HS}$.}
\item{\textbf{Unitary transformations} of a state $\rho \in \mathcal{H}^{1}_{HS}$
\begin{align}
&\Lambda_3:\mathcal{H}^{1}_{HS}\otimes \rightarrow\mathcal{H}^{1}_{HS} \ ,\\
&\Lambda_3(\rho)=U \rho U^{\dagger} \ ,
\end{align}}
with a unitary operator $U\in\mathcal{H}^{1}_{HS}$.
\item{\textbf{Measurement of an observable $A \in \mathcal{H}^{1}_{HS}$}
\begin{align}
&\Lambda_4:\mathcal{H}^{1}_{HS}\rightarrow\mathcal{H}^{1}_{HS} \ , \\
&\Lambda_4(\rho) =\frac{\sum_i M_i \rho M_i^{\dagger}}{Tr\left(\sum_i M_i \rho M_i^{\dagger}\right)} \ .
\end{align}
Here $\left\{M_i\right\}$ is the set of the spectral projectors associated with the eigenvectors of the observable $A$. It is useful to classify two types of measurements: \textbf{Non-selective measurements}, where we work with all outcomes of the measurement and \textbf{selective measurements}, where we filter outcomes. Non-selective measurements then yield $\sum_i M_i^{\dagger} M_i = \mathbbm{1}$, and selective measurements $\sum_i M_i^{\dagger} M_i \leq \mathbbm{1}$.}
\end{itemize}
The point of the matter is, that the four maps themselves and their compositions are all completely positive. According to Choi's theorem (see \cite{Choi}) any map of this type can be expressed in the form
\begin{align}
&\Lambda:\mathcal{H}^{1}_{HS}\rightarrow\mathcal{H}^{2}_{HS} \ ,\\
&\Lambda(\rho)=\frac{\sum_i V_i \rho V_i^{\dagger}}{Tr\left(\sum_i V_i \rho V_i^{\dagger}\right)} \ ,
\end{align}
with $\rho \in\mathcal{H}^{1}_{HS}$ (Hilbert-Schmidt space of dimension $n \times n$), $\Lambda(\rho)\in\mathcal{H}^{2}_{HS}$ (Hilbert-Schmidt space of dimension $m \times m$) and complex matrices $\left\{V_i\right\}$ of dimension $m \times n$. This expression is the \textbf{Kraus representation} of $\Lambda$ and the matrices $\left\{V_i\right\}$ are the famous \textbf{Kraus operators} which obey $\sum_i V_i^{\dagger} V_i \leq \mathbbm{1}$.
\newpage
\subsubsection{Class of LOCC}
\label{LOCC}
The concept of \textbf{"local operations and classical communication"} stems from quantum communication theory and appears in quantum teleportation, quantum cryptography and distillation protocols. Consider a source and two distant parties A and B, commonly called Alice and Bob. Under realistic circumstances they are able to perform arbitrary quantum operations acting on the particular local Hilbert space $\mathcal{H}^A$ and $\mathcal{H}^B$ and to communicate via classical information. They can neither exchange their quantum systems nor perform "global" transformations involving the entire composite Hilbert space $\mathcal{H}^{AB}=\mathcal{H}^A \otimes \mathcal{H}^B$.
\begin{figure}[htp]
\centering
  \setlength{\unitlength}{1bp}%
  \begin{picture}(316.64, 91.93)(0,0)
  \put(0,0){\includegraphics{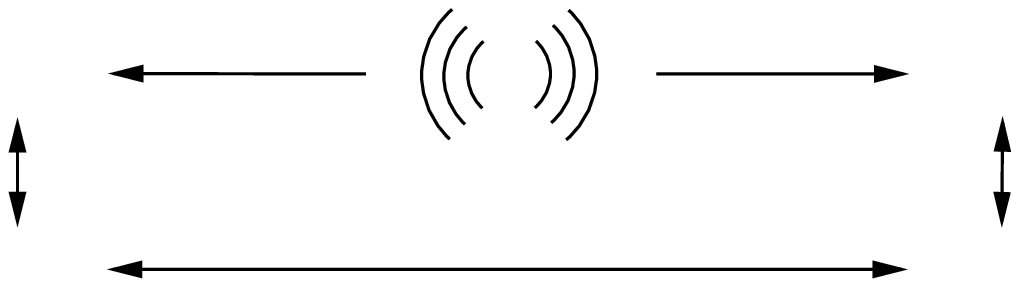}}
  \put(159.73,39.15){\fontsize{8.54}{10.24}\selectfont \makebox(0,0)[]{source\strut}}
  \put(17.95,12.02){\fontsize{11.38}{13.66}\selectfont \makebox(0,0)[]{Alice\strut}}
  \put(301.52,11.34){\fontsize{11.38}{13.66}\selectfont \makebox(0,0)[]{Bob\strut}}
  \put(89.46,74.18){\fontsize{8.54}{10.24}\selectfont \makebox(0,0)[]{quantum channel\strut}}
  \put(230.41,74.31){\fontsize{8.54}{10.24}\selectfont \makebox(0,0)[]{quantum channel\strut}}
  \put(159.10,17.27){\fontsize{8.54}{10.24}\selectfont \makebox(0,0)[]{classical communication channel\strut}}
  \put(17.95,67.92){\fontsize{11.38}{13.66}\selectfont \makebox(0,0)[]{$\mathcal{H}^A$\strut}}
  \put(301.52,67.57){\fontsize{11.38}{13.66}\selectfont \makebox(0,0)[]{$\mathcal{H}^B$\strut}}
  \end{picture}%
  \caption{Schematic illustration of the LOCC situation}
\end{figure}\\
The transcription of this concept into restrictions for the quantum operations \S \ref{operations} implies that a LOCC operation is a map of the form
\begin{align}
\label{LOCCmath}
\Lambda_{AB}\left(\rho\right)=\frac{\sum_i (A_i \otimes B_i) \rho (A_i^{\dagger} \otimes B_i^{\dagger})}{Tr\left[\sum_i (A_i \otimes B_i) \rho (A_i^{\dagger} \otimes B_i^{\dagger})\right]} \ ,
\end{align}
with product Kraus operators $\left\{A_i \otimes B_i\right\}$ where $A_i$ acts on Alice's Hilbert space $\mathcal{H}^A$ and $B_i$ acts on Bob's Hilbert space $\mathcal{H}^B$. The product Kraus operators reflect that the quantum operations only act locally, while the bilateral dependence of $\left(A_i \otimes B_i\right)$ on $i$ reflect that both parties can arrange their actions (they can be classically correlated). While the form of (\ref{LOCCmath}) is comprehensible, the constraints for the operators $\left\{A_i \otimes B_i\right\}$ and their relations can be quite complex. Those depend on the considered communication class, which can be "no communication", "one-way communication" or "two-way communication" (see \cite{Horodecki-summary},\cite{DonaldUNITHEOREM},\cite{Clarisse}). The form of (\ref{LOCCmath}) induces the definition of separable density matrices (\ref{sepdefdens}).
\subsubsection{Distillation and bound entanglement}
\label{distbound}
Outstanding procedures like quantum cryptography or quantum teleportation require pure Bell states. In practice however, we cannot completely neutralise the interaction with the environment that causes decoherence. In the case of the two distant parties Alice and Bob, the linking quantum channel is therefore said to be noisy. We expect that the occurring state $\rho^{AB}$ is no longer a pure Bell state, originally emitted from the source, but a mixture.
We characterise the state $\rho^{AB}$ by its \textbf{fidelity} $F$
\begin{align}
F=Tr\left(\ket{\Omega_{k,l}}\bra{\Omega_{k,l}}\rho^{AB}\right) \ ,
\end{align}
which can be regarded as the remaining content of $\ket{\Omega_{k,l}} \bra{\Omega_{k,l}}$ in $\rho^{AB}$. In order to reconstruct an almost pure Bell state with any desired fidelity close to $1$ via LOCC operations we have to execute a so-called \textbf{distillation protocol}. The first protocols for qubits were introduced by Bennett et al. \cite{BennettDIST}. These are the recurrence, hashing and breeding protocols. The hashing and breeding protocols were later generalised for qudits by Vollbrecht et al. \cite{VollbrechtDIST}. These will not be discussed in detail, but it should be pointed out that they all have one thing in common: They use several copies of $\rho^{AB}$ and local filtering to obtain a smaller number of nearly maximally entangled pure states. For every protocol there is a lower bound for the fidelity $F_{LB}$ in order to work successfully. It is a challenging task to find more universal (ones with a lower $F_{LB}$) and faster protocols. Irrespective of this intention, there will always be entangled states that cannot be distilled. The conventional term therefore is \textbf{bound entanglement} (see \cite{HorodeckiBOUND}). We have already mentioned that in higher dimensional systems there are entangled states which cannot be detected via the PPT criterion. Let us consider how such a state behaves under a LOCC transformation. According to the assumption $\rho\in\mathcal{H}_{HS}^{AB}$ being an entangled density matrix with positive partial transpose we claim
\begin{align}
\bra{\Psi}\rho\ket{\Psi} \geq 0 \mbox{ \ } \forall \ket{\Psi} \in \mathcal{H}^{AB} \ , \\
\bra{\Psi}\rho^{T_B}\ket{\Psi} \geq 0 \mbox{ \ } \forall \ket{\Psi} \in \mathcal{H}^{AB} \ . \label{PPTB}
\end{align}
A LOCC transformation of $\rho$ then yields
\begin{align}
\widetilde{\rho}=\frac{\sum_i{(A_i \otimes B_i) \rho (A_i^{\dagger} \otimes B_i^{\dagger})}}{Tr\left[\sum_i (A_i \otimes B_i) \rho (A_i^{\dagger} \otimes B_i^{\dagger})\right]} \ .
\end{align}
We know that transposition changes the sequence of operators according to $(LMN)^T=N^TM^TL^T$ and in our case hermitian conjugation is simply the combination of transposition and complex conjugation $L^{\dagger}=\left(L^{*}\right)^T$. Partial transposition on $\mathcal{H}_{HS}^B$ then results in
\begin{align}
\widetilde{\rho}^{\mbox{ }  T_B}=N\sum_i {(A_i \otimes B_i^*) \rho^{\mbox{ }  T_B} (A_i^{\dagger} \otimes (B_i^*)^{\dagger})} \ ,
\end{align}
wherein $N$ is a positive real number $N=1/{Tr\left[\sum_i (A_i \otimes B_i) \rho (A_i^{\dagger} \otimes B_i^{\dagger})\right]}>0$. Let us investigate positivity
\begin{align}
\bra{\Psi}\widetilde{\rho}^{ \mbox{ } T_B}\ket{\Psi}&=N\sum_i \overbrace{\bra{\Psi}(A_i \otimes B_i^*)}^{\bra{\Phi_i}} \rho^{T_B} \overbrace{(A_i^{\dagger} \otimes (B_i^*)^{\dagger})\ket{\Psi}}^{\ket{\Phi_i}} \nonumber \\
&=N\sum_i {\underbrace{\bra{\Phi_i} \rho^{  T_B} \ket{\Phi_i}}_{\geq 0 \mbox{ }\forall \mbox{ }  \ket{\Phi_i}\in\mathcal{H}^{AB} (\ref{PPTB})}} \nonumber \\
&\geq 0 \hspace{1cm} \forall \ket{\Psi}\in\mathcal{H}^{AB} \ .
\end{align}
This proves that $\widetilde{\rho}^{\mbox{ }  T_B}$ is a positive operator. The conclusion is that a LOCC transformation cannot transform a PPT density matrix into a density matrix that is non-positive after partial transposition \textbf{(NPPT)}. We have established that a pure maximally entangled state (NPPT) cannot be distilled from a PPT state, even though it contains some entanglement. This justifies the term \textbf{bound entanglement}. The existence of NPPT bound entanglement is controversial and a subject of recent research.
\newpage
\section{Bell inequalities}
\subsection{Local realism versus quantum mechanics}
Since the verification of (special) relativity principles most physicists have believed that any fundamental theory is in compliance with local realism.\\
\textbf{Realism} is the assumption of the existence of definite values for all possible observables. That is, at each point of time these values genuinely exist, whether we measure them or not. "Ideal" measurements with no or marginal disturbance could therefore reproduce these pre-existing values. As a direct consequence, realistic theories are non-contextual.\\
\textbf{Locality} reflects the key consequence of relativity, that is all interactions between distant objects are limited to the speed of light. Space-like separated objects are therefore independent of one another.\\
\textbf{Local realism} is the unification of locality and realism.\\
As obvious as these assumptions may seem, in 1935 Einstein, Podolsky and Rosen showed that quantum mechanics rejects these principles (see \cite{EPR}). At this period of time, their conclusion was that quantum mechanics must be incomplete, meaning there must be elements of reality that do not appear in the theory. An introduction of such elements however, should restore local realism. Due to their hiddenness they have been called \textbf{local hidden variables}, regardless of whether they are hidden in principle or just in experiments at that time. During this period, evidence had not been found yet, demonstrating that all \textbf{local hidden variable theories (LHVT)} are incompatible with the predictions of quantum mechanics in particular experiments. 
In 1964, John Bell derived that the correlation expectation values of local hidden variable theories fulfill inequalities that can be violated by quantum mechanics (see \cite{BellORIGINAL}). Moreover, these \textbf{Bell inequalities} enable us to experimentally revise the validity of \textbf{either} quantum mechanics \textbf{or} a local realistic theory. Before going into further details, it is essential to briefly reconsider why and under which circumstances quantum mechanics is said to be non-local. Nonlocality can be ascribed to entanglement and the measurement problem. This can be best understood by considering the Bohm version of the EPR situation (see \cite{BohmEPR}). Here, two spin-$\frac{1}{2}$ particles, for example electrons or protons, interact and their spins are in a maximally entangled singlet state  $\ket{\Psi^-}=\frac{1}{\sqrt{2}}\left(\ket{\uparrow\downarrow}-\ket{\downarrow\uparrow}\right)$ afterwards. Then both particles fly off in different directions freely while the spins remain in the singlet state. Subsequently $\sigma_z$ spin measurements are performed on both particles.
\begin{figure}[htp]
\centering  
   \setlength{\unitlength}{1bp}%
  \begin{picture}(392.75, 79.78)(0,0)
  \put(0,0){\includegraphics{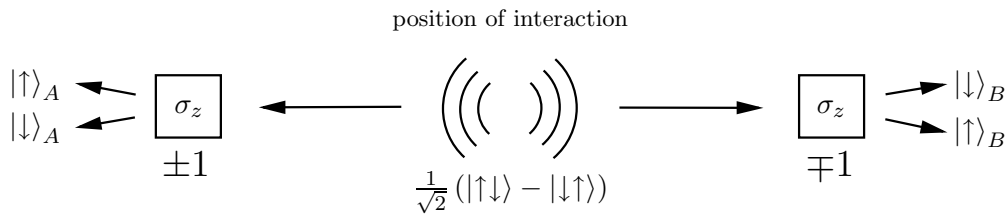}}
  \put(195.60,75.86){\fontsize{8.54}{10.24}\selectfont \makebox(0,0)[]{position of interaction\strut}}
  \put(74.53,43.89){\fontsize{11.38}{13.66}\selectfont \makebox(0,0)[]{$\sigma_z$\strut}}
  \put(316.60,43.89){\fontsize{11.38}{13.66}\selectfont \makebox(0,0)[]{$\sigma_z$\strut}}
  \put(73.25,20.81){\fontsize{14.23}{17.07}\selectfont \makebox(0,0)[]{$\pm 1$\strut}}
  \put(316.07,20.72){\fontsize{14.23}{17.07}\selectfont \makebox(0,0)[]{$\mp 1$\strut}}
  \put(195.74,11.34){\fontsize{11.38}{13.66}\selectfont \makebox(0,0)[]{$\frac{1}{\sqrt{2}}\left(\ket{\uparrow\downarrow}-\ket{\downarrow\uparrow}\right)$\strut}}
  \put(18.27,52.13){\fontsize{11.38}{13.66}\selectfont \makebox(0,0)[]{$\ket{\uparrow}_A$ \strut}}
  \put(374.48,51.78){\fontsize{11.38}{13.66}\selectfont \makebox(0,0)[]{$\ket{\downarrow}_B$ \strut}}
  \put(374.48,34.17){\fontsize{11.38}{13.66}\selectfont \makebox(0,0)[]{$\ket{\uparrow}_B$ \strut}}
  \put(18.27,35.90){\fontsize{11.38}{13.66}\selectfont \makebox(0,0)[]{$\ket{\downarrow}_A$ \strut}}
  \end{picture}%
   \caption{Schematic illustration of the Bohm-EPR situation}
\end{figure}\\
According to the measurement postulate of quantum mechanics the spin will no longer be a superposition of up $\ket{\uparrow}$ and down $\ket{\downarrow}$ though either one \textbf{or} the other. The reduction of the state vector to one of the eigenvectors of the observable $\sigma_z$ of one particle forces the other particle's spin $\sigma_z$ to be anti-correlated. Here appears a new type of simultaneousness. The problem here is that if one measures $\sigma_z$ on one side, the particle's spin on the other side is affected instantaneously, irrespective of how far they are apart. If the wave function is regarded as a real physical object this means that causality is violated, since one particle affects the other with superluminal velocity. When we say the measurements A and B are performed simultaneously then according to relativity there are different inertial frames of reference, in which A happens before B and vice versa. Fact is, quantum mechanics predicts that the spins are always perfectly anticorrelated. This phenomenon can be interpreted in different ways. When we accept the wave function as a real existing physical object we have to reject locality, since the reduction of the wave function is a non-local process. In the conventional version of the Copenhagen interpretation where the wave function is not a physical object of reality but rather a mathematical tool, realism is rejected in order to preserve locality\footnote{"There is no quantum world. There is only an abstract physical description. It is wrong to think that the task of physics is to find out how nature is. Physics concerns what we can say about nature." - Niels Bohr}. In the Bohm interpretation locality is given up to maintain realism. What can be concluded from those interpretations is that, if nature really behaves quantum mechanically, then it is not local \textbf{and} realistic at the same time.
\subsection{A Bell inequality for two-qubit systems: CHSH}
As previously mentioned, local realistic theories are in contradiction to quantum mechanics in certain experiments. In order to show this, we derive a \textbf{Bell inequality} which holds for any local realistic theory but can be violated by quantum mechanics. This will be the famous \textbf{Clauser-Horne-Shimony-Holt inequality (CHSH)}, which is a modification of the original Bell inequality permitting experimental revisal.
\subsubsection{Derivation of the CHSH inequality}
We want to describe the Bohm-EPR situation with a local realistic theory. In order to do this, we introduce the parameter $\lambda$, which represents a hidden variable or a set of those. The spin of a particle then depends on the measured direction represented by a vector $\vec{a}$ and the parameter $\lambda$. We know that a spin measurement on a particle has only two possible outcomes. Without loss of generality we assign the values $+1$ and $-1$ to them. In a local realistic theory the observables of distant parties are then given by
\begin{align}
A(\vec{a},\lambda)=\pm1 \ ,
\end{align}
with measurement direction $\vec{a}$ on Alice's side, and
\begin{align}
B(\vec{b},\lambda)=\pm1 \ ,
\end{align}
with measurement direction $\vec{b}$ on Bob's side. Locality requires that the outcome of A does not depend on $\vec{b}$ and B does not depend on $\vec{a}$. Without loss of generality we can say that the value $\lambda$ is achieved with probability density $\rho(\lambda)\geq0$ obeying
\begin{align}
\int{\rho(\lambda)d\lambda}=1 \ .
\end{align}
We define a correlation function for the joint spin measurement, that yields the value $1$ when the spins are parallel and $-1$ when the they are antiparallel
\begin{align}
C\left(A(\vec{a},\lambda),B(\vec{b},\lambda)\right)=A(\vec{a},\lambda) \cdot B(\vec{b},\lambda) \ .
\end{align}
The expectation value of this quantity therefore is
\begin{align}
\label{jointexp}
E(\vec{a},\vec{b})=\int{\rho(\lambda)A(\vec{a},\lambda)B(\vec{b},\lambda)d\lambda} \ .
\end{align}
If we combine different expectation values of different measurement directions in the following way
\begin{align}
E(\vec{a},\vec{b})-E(\vec{a},\vec{b'})&=\int{\rho(\lambda)\left[A(\vec{a},\lambda)B(\vec{b},\lambda)-A(\vec{a},\lambda)B(\vec{b'},\lambda)\right]d\lambda} \nonumber \\
&=\int{\rho(\lambda)A(\vec{a},\lambda)B(\vec{b},\lambda)\left[1 \pm A(\vec{a'},\lambda)B(\vec{b'},\lambda)\right]d\lambda} \nonumber \\
&-\int{\rho(\lambda)A(\vec{a},\lambda)B(\vec{b'},\lambda)\left[1 \pm A(\vec{a'},\lambda)B(\vec{b},\lambda)\right]d\lambda} \ ,
\end{align}
then the absolute value yields
\begin{align}
|E(\vec{a},\vec{b})-E(\vec{a},\vec{b'})| & \leq \left| \int{\rho(\lambda)\left[1 \pm A(\vec{a'},\lambda)B(\vec{b'},\lambda)\right]d\lambda} \right| \nonumber \\
&+ \left| \int{\rho(\lambda)\left[1 \pm A(\vec{a'},\lambda)B(\vec{b},\lambda)\right]d\lambda} \right| \nonumber \\
&=2 \pm |E(\vec{a'},\vec{b'})+E(\vec{a'},\vec{b})| \ .
\end{align}
We obtain the \textbf{CHSH inequality} by rewriting this in the form
\begin{align}
|E(\vec{a},\vec{b})-E(\vec{a},\vec{b'})+E(\vec{a'},\vec{b'})+E(\vec{a'},\vec{b})|\leq2 \ .
\end{align}
The derivation reveals that every local realistic theory has to fulfill the CHSH inequality. Quantum mechanics however predicts a violation under particular circumstances. Consider the quantum mechanical expectation value of $E(\vec{a},\vec{b})$ for the Bell state $\ket{\Psi^-}=\frac{1}{\sqrt{2}}\left(\ket{\uparrow\downarrow} - \ket{\downarrow\uparrow}\right)$
\begin{align}
E(\vec{a},\vec{b})=\bra{\Psi^-} \vec{a}\cdot\vec{\sigma}\otimes\vec{b}\cdot\vec{\sigma}\ket{\Psi^-} \ ,
\end{align}
with unit vectors $\vec{a}$ and $\vec{b}$. A short evaluation yields
\begin{align}
E(\vec{a},\vec{b})=-\vec{a}\cdot\vec{b}=-cos(\alpha-\beta) \ ,
\end{align}
wherein the angles $\alpha$ and $\beta$ substitute the vectors $\vec{a}$ and $\vec{b}$. Hence, we can write the CHSH inequality in the form
\begin{align}
|-cos(\alpha-\beta)+cos(\alpha-\beta')-cos(\alpha'-\beta)-cos(\alpha'-\beta')| \ ,
\end{align}
wherein $\alpha,\alpha',\beta$ and $\beta'$ are the angles of the four vectors $\vec{a},\vec{a'},\vec{b}$ and $\vec{b'}$ in a plane. For angles obeying $|\alpha-\beta|=|\alpha'-\beta|=|\alpha'-\beta'|=\frac{\pi}{4}$ and $|\alpha-\beta'|=\frac{3\pi}{4}$ we find
\begin{align}
|-cos(\alpha-\beta)+cos(\alpha-\beta')-cos(\alpha'-\beta)-cos(\alpha'-\beta')| \nonumber \\
=|-\frac{\sqrt{2}}{2}-\frac{\sqrt{2}}{2}-\frac{\sqrt{2}}{2}-\frac{\sqrt{2}}{2}|=2\sqrt{2}>2 \ .
\end{align}
As a result we have found that any local realistic theory cannot reproduce the statistics of quantum mechanics. Experiments with entangled photons confirm the violation of the CHSH inequality (see \cite{AspectEPR}, \cite{Weihs}).
\subsubsection{Horodecki violation criterion}\label{HorodeckiVIO}
The example with the singlet state $\ket{\Psi^-}$ demonstrates the violation of the CHSH inequality. Now we want to work out the whole set of states causing violations. For an arbitrary mixed state $\rho$ the expectation value $E(\vec{a},\vec{b})$ is
\begin{align}
E(\vec{a},\vec{b})=Tr(\rho \mbox{  } \vec{a}\cdot\vec{\sigma}\otimes\vec{b}\cdot\vec{\sigma} ) \ .
\end{align}
Thus, we can write the Bell inequality in the form
\begin{align}
\left|Tr\left(\rho \left[ \vec{a}\cdot\vec{\sigma}\otimes(\vec{b}-\vec{b'})\cdot\vec{\sigma}+\vec{a'}\cdot\vec{\sigma}\otimes (\vec{b}+\vec{b'})\cdot\vec{\sigma}\right]\right)\right|\leq2 \ .
\end{align}
We call the operator in square brackets the \textbf{Bell operator $\mathcal{B}$}
\begin{align}
\mathcal{B}(\vec{a},\vec{a'},\vec{b},\vec{b'})=\vec{a}\cdot\vec{\sigma}\otimes(\vec{b}-\vec{b'})\cdot{\sigma}+\vec{a'}\cdot\vec{\sigma}\otimes (\vec{b}+\vec{b'})\cdot\vec{\sigma} \ .
\end{align}
It is obvious that for a given state $\rho$ one has to find the right vectors $\vec{a},\vec{a'},\vec{b}$ and $\vec{b'}$ to show possible violation. Only if the global maximum with respect to all Bell operators $\mathcal{B}(\vec{a},\vec{a'},\vec{b},\vec{b'})$ is less or equal to 2, then the inequality is preserved
\begin{align}
\max_{\vec{a},\vec{a'},\vec{b},\vec{b'}} \left|Tr\left(\rho \left[\vec{a}\cdot\vec{\sigma}\otimes(\vec{b}-\vec{b'})\cdot\vec{\sigma}+\vec{a'}\cdot\vec{\sigma}\otimes (\vec{b}+\vec{b'})\cdot\vec{\sigma}\right]\right)\right|\leq2 \ .
\end{align}
Since the Bell operator $\mathcal{B}(\vec{a},\vec{a'},\vec{b},\vec{b'})$ is expressed in terms of Pauli matrices, it makes sense to express $\rho$ in the way we introduced it in ($\ref{generaldensity}$)
\begin{align}
\rho=\frac{1}{4}\left( \mathbbm{1}\otimes\mathbbm{1} + \vec{r}\cdot\vec{\sigma}\otimes\mathbbm{1} + \mathbbm{1}\otimes \vec{s} \cdot \vec{\sigma} + \sum_{n,m=1}^{3}t_{nm}\sigma_n\otimes \sigma_m \right) \ .
\end{align}
The computation of $\left|Tr(\rho \mathcal{B})\right|$ then yields
\begin{align}
\left|\vec{a}\cdot T(\rho)(\vec{b}-\vec{b'})+\vec{a'}\cdot T(\rho)(\vec{b}+\vec{b'})\right| \ ,
\end{align}
wherein $T(\rho)$ denotes the $3\times3$ correlation matrix with the coefficients $t_{nm}$. We replace $\vec{b}-\vec{b'}$ and $\vec{b}+\vec{b'}$ by mutually orthogonal unit vectors $\vec{c}$ and $\vec{c'}$
\begin{align}
\vec{b}-\vec{b'}=2\mbox{sin}\theta \vec{c'} \mbox{ \ \ \ \ } \vec{b}+\vec{b'}=2\mbox{cos}\theta \vec{c} \ .
\end{align}
Now the maximum of $\left|Tr(\rho \mathcal{B})\right|$ has to be determined with respect to $\theta, \vec{a},\vec{a'},\vec{c}$ and $\vec{c'}$
\begin{align}
\max_{\theta, \vec{a},\vec{a'},\vec{c},\vec{c'}}2\left|\mbox{sin}\theta \vec{a}\cdot T(\rho)\vec{c'}+\mbox{cos}\theta \vec{a'}\cdot T(\rho)\vec{c}\right| \ .
\end{align}
The scalar products are maximal for parallel vectors. Consequently, we choose the unit vectors $\vec{a}_{max}= \frac{T(\rho)\vec{c'}}{\|T(\rho)\vec{c'}\|}$ and $\vec{a'}_{max}= \frac{T(\rho)\vec{c}}{\|T(\rho)\vec{c}\|}$
\begin{align}
\max_{\theta,\vec{c},\vec{c'}}2\left|\mbox{sin}\theta \|T(\rho)\vec{c'}\|+\mbox{cos}\theta \|T(\rho)\vec{c}\|\right| \ .
\end{align}
Maximisation with respect to $\theta$ yields
\begin{align}
\max_{\vec{c},\vec{c'}}2\sqrt{\|T(\rho)\vec{c'}\|^2+\|T(\rho)\vec{c}\|^2} \ .
\end{align}
Since $\vec{c}$ and $\vec{c'}$ are mutually orthogonal unit vectors, the expressions $\|T(\rho)\vec{c'}\|^2= \vec{c'} \cdot T^T(\rho) T(\rho) \vec{c'}$ and $\|T_{\rho}\vec{c}\|^2= \vec{c} \cdot T^T(\rho) T(\rho) \vec{c}$ are maximal when $\vec{c'}$ and $\vec{c}$ are eigenvectors of the two largest eigenvalues $\lambda_1(\rho)$ and $\lambda_2(\rho)$ of $K(\rho)=T^T(\rho) T(\rho)$. Thus the CHSH inequality reads
\begin{align}
2\sqrt{\lambda_1(\rho)+\lambda_2(\rho)}\leq2 \ .
\end{align}
Now we have proven that if and only if the two largest eigenvalues of the matrix $K(\rho)=T^T(\rho) T(\rho)$ comply
\begin{align}
\label{CHSHVIO}
\lambda_1(\rho)+\lambda_2(\rho)>1 \ ,
\end{align}
then the state $\rho$ violates the CHSH inequality for particular vectors $\vec{a},\vec{a'},\vec{b}$ and $\vec{b'}$. This necessary and sufficient condition was found by the Horodeckis in 1995 \cite{HorodeckiCHSHNSC}.
\subsection{A Bell inequality for bipartite qudit systems: CGLMP}
The CHSH inequality is a Bell inequality for bipartite qubit systems. In this section we present Bell inequalities for bipartite qudit systems based on logical constraints established by Collins, Gisin, Linden, Massar and Popescu  \textbf{(CGLMP)}\cite{CGLMPineq} .
\subsubsection{Derivation of the CGLMP inequality}
\label{CGLMPder}
Once again, we consider the standard situation with the two parties Alice and Bob spatially separated, only having access onto their local Hilbert spaces $\mathcal{H}^A=\mathcal{H}^B=\mathbb{C}^d$ of dimension d, while the composite Hilbert space is once again $\mathcal{H}^{AB}=\mathbb{C}^d\otimes\mathbb{C}^d$. Consequently, measurements on each side have d possible outcomes. Let us assume again that we could reproduce the statistical predictions with a local realistic theory. In analogy to the CHSH Bell experiment, each party has two apparatuses with different settings. Due to the fact that in a local realistic theory all observables have definite values simultaneously, the state of a system induces a probability distribution of form
\begin{align}
\label{pdist}
P(A_1=j,A_2=k,B_1=l,B_2=m)
\end{align}
that apparatus $A_1$ gives measurement result $j \in \left\{0,..,d-1 \right\}$, apparatus $A_2$ gives $k \in \left\{0,..,d-1 \right\}$ and so on. In sum, we have $d^4$ values determining the statistics of the system.  As usual we normalise these probabilities
\begin{align}
\sum_{jklm}{P(A_1=j,A_2=k,B_1=l,B_2=m)}=1 \ .
\end{align}
If we are only interested in measurement results of certain apparatuses then we have to sum over all ignored observables. For example the probability of $A_1$ giving j and $B_1$ giving $l$ is $ P(A_1=j,B_1=l)=\sum_{km}{P(A_1=j,A_2=k,B_1=l,B_2=m)}$. We introduce some variables $r',s',t'$ and $u'$ defined by relations of measurement outcomes 
\begin{align}
r'&=B_1-A_1=l-j \ , \\
s'&=A_2-B_1=k-l \ , \\
t'&=B_2-A_2=m-k \ , \\
u'&=A_1-B_2=j-m \ .
\end{align}
The definition induces the constraint
\begin{align}
(r'+s'+t'+u') \mbox{ mod } d=0 \ .
\end{align}
We introduce the quantity $I$ defined by
\begin{align}
I=P(A_1=B_1)+P(B_1=A_2+1)+P(A_2=B_2)+P(B_2=A_1) \ ,
\end{align}
with the probabilities $P(A_a=B_b+k)$ that the outcome of $A_a$ differs from outcome $B_b$ by k
\begin{align}
P(A_a=B_b+k)=\sum_{j=0}^{d-1}{P(A_a=(j+k)\mbox{ mod }d, B_b=j)} \ .
\end{align}
Then $I$ is the sum of the probabilities that $r'=0,s'=-1,t'=0$ and $u'=0$. Nevertheless, because of the logical constraint $r'+s'+t'+u'=0$ only three of those four relations can be valid. Thus we have
\begin{align}
I\leq3
\end{align}
for all local realistic theories. We introduce another quantity $I_3$
\begin{align}
I_3=&+\left[P(A_1=B_1)+P(B_1=A_2+1)+P(A_2=B_2)+P(B_2=A_1)\right]\\
&-\left[P(A_1=B_1-1)+P(B_1=A_2)+P(A_2=B_2-1)+P(B_2=A_1-1)\right]\nonumber
\end{align}
Consider the following table which shows how the logical constraint affects the maximum of $I_3$.
\begin{figure*}[htp]
\centering
  \setlength{\unitlength}{1bp}%
  \begin{picture}(220.37, 114.44)(0,0)
  \put(0,0){\includegraphics{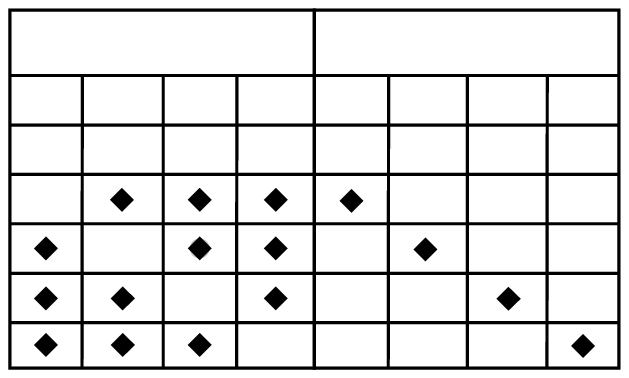}}
  \put(49.24,98.80){\fontsize{8.83}{10.60}\selectfont \makebox(0,0)[]{+1\strut}}
  \put(138.39,98.23){\fontsize{8.83}{10.60}\selectfont \makebox(0,0)[]{-1\strut}}
  \put(16.08,80.86){\fontsize{7.07}{8.48}\selectfont \makebox[0pt]{r'}}
  \put(38.25,80.86){\fontsize{7.07}{8.48}\selectfont \makebox[0pt]{s'}}
  \put(60.37,80.86){\fontsize{7.07}{8.48}\selectfont \makebox[0pt]{t'}}
  \put(82.26,80.86){\fontsize{7.07}{8.48}\selectfont \makebox[0pt]{u'}}
  \put(104.37,80.86){\fontsize{7.07}{8.48}\selectfont \makebox[0pt]{r'}}
  \put(125.93,80.86){\fontsize{7.07}{8.48}\selectfont \makebox[0pt]{s'}}
  \put(149.06,80.86){\fontsize{7.07}{8.48}\selectfont \makebox[0pt]{t'}}
  \put(171.35,80.86){\fontsize{7.07}{8.48}\selectfont \makebox[0pt]{u'}}
  \put(16.08,66.60){\fontsize{7.07}{8.48}\selectfont \makebox[0pt]{0}}
  \put(38.25,66.60){\fontsize{7.07}{8.48}\selectfont \makebox[0pt]{-1}}
  \put(60.37,66.60){\fontsize{7.07}{8.48}\selectfont \makebox[0pt]{0}}
  \put(82.26,66.60){\fontsize{7.07}{8.48}\selectfont \makebox[0pt]{0}}
  \put(104.37,66.60){\fontsize{7.07}{8.48}\selectfont \makebox[0pt]{1}}
  \put(125.93,66.60){\fontsize{7.07}{8.48}\selectfont \makebox[0pt]{0}}
  \put(149.06,66.60){\fontsize{7.07}{8.48}\selectfont \makebox[0pt]{1}}
  \put(171.35,66.60){\fontsize{7.07}{8.48}\selectfont \makebox[0pt]{1}}
  \put(201.42,54.49){\fontsize{7.07}{8.48}\selectfont \makebox(0,0)[]{$\Rightarrow I_3=2$\strut}}
  \put(201.42,39.68){\fontsize{7.07}{8.48}\selectfont \makebox(0,0)[]{$\Rightarrow I_3=2$\strut}}
  \put(201.42,25.44){\fontsize{7.07}{8.48}\selectfont \makebox(0,0)[]{$\Rightarrow I_3=2$\strut}}
  \put(201.42,11.49){\fontsize{7.07}{8.48}\selectfont \makebox(0,0)[]{$\Rightarrow I_3=2$\strut}}
  \end{picture}%
\end{figure*}\\
In consequence
\begin{align}
I_3\leq2.
\end{align}
Additional quantities $I_d$ can be introduced\footnote{the bracket [ ] stands for the floor function}
\begin{align}
I_d = \sum_{k=0}^{[d/2]-1}\bigl(1-\frac{2k}{d-1}\bigr)\bigl\{+&\bigl[P(A_1=B_1+k)+P(B_1=A_2+k+1)\\
&+P(A_2=B_2+k)+P(B_2=A_1+k)\bigr]\nonumber\\
-&\bigl[P(A_1=B_1-k-1)+P(B_1=A_2-k)\nonumber\\
&+P(A_2=B_2-k-1)+P(B_2=A_1-k-1)\bigr]\bigr\}\nonumber
\end{align}
For local realistic theories the upper bound is 2
\begin{align}
I_d\leq2 \ .
\end{align}
Proof: We introduce the variables
\begin{align}
r&=A_1-B_1 \ , \\
s&=B_1-A_2-1 \ , \\
t&=A_2-B_2 \ , \\
u&=B_2-A_1 \ , 
\end{align}
that obey
\begin{align}
\label{rstu}
(r+s+t+u+1) \mbox{ mod } d=0 \ .
\end{align}
Since all computations are done modulo d, without loss of generality we can restrict $r,s,t$ and $u$ to lie in the interval
\begin{align}
\label{interval}
-[\frac{d}{2}]\leq r,s,t,u \leq [\frac{(d-1)}{2}] \ . 
\end{align}
$I_d$ can be written as a function depending on $r,s,t$ and $u$ in the following way
\begin{align}
I_d=f(r)+f(s)+f(t)+f(u) \ ,
\end{align}
where $f(x)$ is given by
\begin{align}
f(x)\, = \,\left\{\begin{array}{l} \displaystyle{\frac{-2x}{d-1}+1 \hspace{3cm} x \geq 0} \\
		\displaystyle{\frac{-2x}{d-1} - \frac{d+1}{d-1} \hspace{2.2cm} x <0 \ .} \end{array}\right.
\end{align}
We now have to check all combinations of algebraic signs of $r,s,t$ and $u$. With help of (\ref{rstu}) and (\ref{interval}) we find
\begin{itemize}
	\item $r,s,t,u \geq 0$ :\\
	$\Rightarrow r+s+t+u+1=d \Rightarrow I_d=2$
	\item three of $r,s,t,u$ are $\geq 0$ and one is $<0$ :\\
	either $r+s+t+u+1=d \Rightarrow I_d=2$\\
	or $r+s+t+u+1=0 \Rightarrow I_d=\frac{-2}{d-1}$
	\item two of $r,s,t,u$ are $\geq 0$ and two are $<0$ :\\
	$\Rightarrow r+s+t+u+1=0 \Rightarrow I_d=\frac{-2}{d-1}$
	\item one of $r,s,t,u$ is $\geq 0$ and three are $<0$ :\\
	either $r+s+t+u+1=0 \Rightarrow I_d=\frac{-2(d+1)}{d-1}$\\
	or $r+s+t+u+1=-d \Rightarrow I_d=\frac{-2}{d-1}$
	\item $r,s,t,u<0$ :\\
	$\Rightarrow r+s+t+u+1=-d \Rightarrow I_d=\frac{-2(d+1)}{d-1}$	
\end{itemize}
Hence, $I_d\leq2$. \hspace{1cm} $\Box$\\
\\
\newpage
\noindent
In the following, we investigate the predictions of quantum mechanics for this family of inequalities. First, we have to insert the quantum mechanical probabilities
\begin{align}
\label{PAB}
P(A_a=k,B_b=l)=Tr\left(\rho \ket{k}_{Aa}\bra{k}_{Aa}\otimes \ket{l}_{Bb}\bra{l}_{Bb} \right) \ ,
\end{align}
wherein $\left\{\ket{k}_{Aa}\right\}$ and $\left\{\ket{l}_{Bb}\right\}$ denote orthonormal eigenvectors of the observables with $k,l\in\left\{0,..,d-1\right\}$. To confirm the violation of these inequalities consider the maximally entangled state
\begin{align}
\ket{\Omega_{0,0}}=\frac{1}{\sqrt{d}}\sum_{s=0}^{d-1}\ket{s}_A \otimes \ket{s}_B 
\end{align}
and orthonormal eigenvectors
\begin{align}
\label{ONBBell}
\ket{k}_{Aa}&=\frac{1}{\sqrt{d}}\sum_{s=0}^{d-1}\exp\left(\frac{2\pi i}{d}s(k+\alpha_a)\right)\ket{s}_A \ ,\\
\ket{l}_{Bb}&=\frac{1}{\sqrt{d}}\sum_{s=0}^{d-1}\exp\left(\frac{2\pi i}{d}s(-l+\beta_b)\right)\ket{s}_B \ ,
\end{align}
with $\alpha_1=0$ , $ \alpha_2=\frac{1}{2}$ , $ \beta_1=\frac{1}{4}$ and $\beta_2=-\frac{1}{4}$. Then (\ref{PAB}) becomes
\begin{align}
&P(A_a=k,B_b=l) \nonumber \\
=&Tr\left(\ket{\Omega_{0,0}}\bra{\Omega_{0,0}} \ket{k}_{Aa}\bra{k}_{Aa}\otimes \ket{l}_{Bb}\bra{l}_{Bb} \right) \nonumber \\
=&\left| \bra{ \Omega_{0,0}} \ket{k}_{Aa}\otimes \ket{l}_{Bb} \right|^2 \nonumber \\
=&\frac{1}{d^2}\left| \bra{ \Omega_{0,0}} \left[ \sum_{r=0}^{d-1}\exp\left(\frac{2\pi i}{d}r(k+\alpha_a)\right)\ket{r}_A    \right] \otimes \left[\sum_{m=0}^{d-1}\exp\left(\frac{2\pi i}{d}m(-l+\beta_b)\right)\ket{m}_A \right] \right|^2 \nonumber \\
=&\frac{1}{d^3}\left| \sum_{s,r,m=0}^{d-1} \underbrace{\bra{s}_A \ket{r}_A}_{\delta_{sr}} \underbrace{\bra{s}_B \ket{m}_B}_{\delta_{sm}} \exp\left(\frac{2\pi i}{d}r(k+\alpha_a)\right)\exp\left(\frac{2\pi i}{d}m(-l+\beta_b)\right)\right|^2 \nonumber\\
=&\frac{1}{d^3}\left| \sum_{s=0}^{d-1} \exp \left( \frac{2\pi i}{d} s(k-l+\alpha_a+\beta_b)\right) \right|^2 \nonumber\\
=&\frac{\sin^2\left(\pi (k-l+\alpha_a+\beta_b)\right)}{d^3\sin^2\left(\pi (k-l+\alpha_a+\beta_b/d)\right)} \ .
\end{align}
With the values of $\alpha_1$, $ \alpha_2$, $ \beta_1$ and $\beta_2$ given above we find
\begin{align}
P(A_a=k,B_b=l)=\frac{1}{2d^3\sin^2\left(\pi (k-l+\alpha_a+\beta_b)/d\right)}  \ . 
\end{align}
Inserting this into $I$ and $I_d$ leads to 
\begin{align}
I&=\frac{2}{d^2\sin^2\left(\frac{\pi}{4d}\right)} \ ,\\
I_d&=\frac{2}{d^2}\sum_{k=0}^{[\frac{d}{2}]-1}\left(1-\frac{2k}{d-1}\right)\left(\frac{1}{\sin^2\left(\frac{\pi}{d}(k+\frac{1}{4})\right)}-\frac{1}{\sin^2\left(\frac{-\pi}{d}(k+\frac{3}{4})\right)}\right) \ .
\end{align}
We obtain the following values for $I$ and $I_d$
\\
\\
$\begin{array}{|c|c|c|c|c|c|c|c|c|}
  \hline
  d & 2 & 3&4&5&6& 7& & \rightarrow\infty \\
  \hline
  I &3.41421& 3.31738& 3.28427& 3.26908& 3.26086& 3.25592 & & 3.24228\\
  \mbox{violation} \left[ \% \right]& 13.8071 & 10.5793 & 9.47559 & 8.96922 & 8.69533 &  8.53059 & & 8.07593\\
  \hline
  I_d & 2.82843& 2.87293& 2.89624& 2.91054& 2.92020&2.92716& & 2.96981\\
  \mbox{violation} \left[ \% \right]&41.4214& 43.6467& 44.8122& 45.5272& 46.0102& 46.358 & & 48.4906\\
  \hline
\end{array}$\\
\\
\\
As we can see, the \textbf{CGLMP inequalities} $I\leq3$ and $I_d\leq2$ are violated. The table also reveals that by using $I_d$ instead of $I$ we achieve much stronger violations. This is in close relation to the resistance against noise.
Before we analyse this, we present a concise notation for $I$ and $I_d$. Due to (\ref{PAB}) the probabilities $P(A_a=B_b+k)$ become
\begin{align}
P(A_a=B_b+k)=\sum_{j=0}^{d-1}{Tr\left( \rho \ket{(j+k)\mbox{ mod } d}_{Aa} \bra{(j+k) \mbox{ mod } d}_{Aa} \otimes \ket{j}_{Bb} \bra{j}_{Bb} \right)} \ . \nonumber
\end{align}
We rewrite $I$ by exploiting linearity of the trace in the following way
\begin{align}
I&=P(A_1=B_1)+P(A_2=B_2)+P(B_1=A_2+1)+P(B_2=A_1 \nonumber )\\
&=Tr\Biggl( \rho \sum_{j=0}^{d-1} \Bigl\{ \ket{j}_{A1} \bra{j}_{A1} \otimes \ket{j}_{B1} \bra{j}_{B1} +\ket{j}_{A2} \bra{j}_{A2} \otimes \ket{j}_{B2} \bra{j}_{B2} \nonumber \\
&+ \ket{(j-1)\mbox{ mod } d}_{A2} \bra{(j-1)\mbox{ mod } d}_{A2} \otimes \ket{j}_{B1} \bra{j}_{B1}
+\ket{j}_{A1} \bra{j}_{A1} \otimes \ket{j}_{B2} \bra{j}_{B2} \Bigl\} \Biggl)\nonumber\\
&=Tr\left(\rho \mathcal{B}_I \right) \ .
\end{align}
We identify the sum of operators as our new \textbf{Bell operator} $\mathcal{B}_I$ for $I$. In the same way we can define Bell operators $\mathcal{B}_{I_d}$ for all quantities $I_d$ so that
\begin{align}
I_d=Tr\left(\rho \mathcal{B}_{I_d}\right) \ .
\end{align}
Now, let us consider the state $\ket{\Omega_{0,0}}$ in presence of uncolored noise
\begin{align}
\rho=(1-r)\ket{\Omega_{0,0}}\bra{\Omega_{0,0}}+r\frac{\mathbbm{1}}{d^2} \ ,
\end{align}
wherein $r\in[0,1]$ is the amount of noise. We obtain
\begin{align}
I=&Tr\left(\rho \mathcal{B}_{I}\right)=(1-r)Tr\left(\ket{\Omega_{0,0}}\bra{\Omega_{0,0}} \mathcal{B}_{I}\right)+\frac{r}{d^2} Tr\mathcal{B}_{I} \ ,\\
I_d=&Tr\left(\rho \mathcal{B}_{I_d}\right)=(1-r)Tr\left(\ket{\Omega_{0,0}}\bra{\Omega_{0,0}} \mathcal{B}_{I_d}\right)+\frac{r}{d^2}Tr\mathcal{B}_{I_d} \ .
\end{align}
$\mathcal{B}_I$ contains $4 d$ projectors with prefactor $1$, hence $Tr\mathcal{B}_I=4d$. Whereas the addends in $\mathcal{B}_{I_d}$ contain $4d$ projectors with prefactor $\bigl(1-\frac{2k}{d-1}\bigr)$ and $4d$ projectors with prefactor $-\bigl(1-\frac{2k}{d-1}\bigr)$, hence $Tr\mathcal{B}_{I_d}=0$. In consequence, we obtain
\begin{align}
I=&(1-r)Tr\left(\ket{\Omega_{0,0}}\bra{\Omega_{0,0}} \mathcal{B}_{I}\right)+r\frac{4}{d} \ ,\\
I_d=&(1-r)Tr\left(\ket{\Omega_{0,0}}\bra{\Omega_{0,0}} \mathcal{B}_{I_d}\right) \ .
\end{align}
Above a certain amount of noise $r=r_{max}$, the state $\rho$ ceases to violate the Bell inequality. The values are given by $I(r_{max})=3$ and $I_{d}(r_{max})=2$, and are summarised in the following table.\\
\\
$\begin{array}{|c|c|c|c|c|c|c|c|c|}
  \hline
  d & 2 & 3&4&5&6& 7& & \rightarrow\infty \\
  \hline
  I:r_{max} \left[ \%\right]  &29.2893 & 15.9965 & 12.4446 & 10.8979 & 10.0555 & 9.5332& & 7.47246\\
    \hline
  I_d:r_{max} \left[ \% \right] &29.2893& 30.3848& 30.9450& 31.2843& 31.5116& 31.6744& & 32.6557\\
  \hline
\end{array}$
\\
\\
We conclude that one should give preference to the inequalities $I_d\leq2$ due to their higher resistance to noise.
\subsubsection{Optimisation of the Bell operator}
\label{numopt}
The orthonormal bases given in (\ref{ONBBell}) are optimal for the state $\ket{\Omega_{0,0}}$. In other words, there exists no better Bell operator for $\ket{\Omega_{0,0}}$ causing higher violation. This has not been proven so far, though numerical optimisation indicates that this is true (see \cite{CGLMPineq},\cite{DKZvio}). The lack of a proof lies in the fact that the finding of an optimal Bell operator for an arbitrary $\rho$ is a nonlinear optimisation problem: The finding of solutions for grad$(Tr(\rho\mathcal{B}))=0$ with respect to all vector components is a nonlinear task and is further complicated by nonlinear constraints since the vectors $\left\{\ket{k}_{Aa}\right\}$ and $\left\{\ket{l}_{Bb}\right\}$ have to form an orthonormal basis (observables are hermitian operators and their eigenvectors are orthogonal)
\begin{align}
\label{orthonorm}
\bra{k}_{Aa}\ket{j}_{Aa}&=\delta_{kj} \ , \\
\bra{l}_{Bb}\ket{m}_{Bb}&=\delta_{lm} \ .
\label{orthonorm2}
\end{align}
Using the Weyl operators or the Gell-Mann matrices to express $\rho$ and $\mathcal{B}$ does not lead to a simplification like in \S \ref{HorodeckiVIO} for the CHSH inequality. For this reason, we have developed a numerical optimisation algorithm which reliably finds optimal Bell operators for any $\rho$. Let us first investigate the number of variables of the Bell operator $\mathcal{B}$. We have 4 orthonormal bases $\left\{\ket{k}_{A1}\right\},\left\{\ket{k}_{A2}\right\}, \left\{\ket{l}_{B1}\right\}$ and $\left\{\ket{l}_{B2}\right\}$ with each d vectors. Each vector is an element of $\mathbb{C}^d$ and can be described by 2d real numbers (d vector components, each described by 2 real numbers, one for the amplitude and one for the phase). Altogether we have $4\times d \times2d=8d^2$ real variables. Orthonormality restricts the values of the variables, thus we have to optimise $\mathcal{B}$ under the constraints (\ref{orthonorm}) and (\ref{orthonorm2}). For all practical purposes however this way is quite impractical. We now show how to satisfy the constraints by choosing the right set of variables. We begin with some mathematical considerations. Any basis transformation can be realised via a unitary transformation, i.e. any basis $\left\{\ket{k}\right\}$ of $\mathbb{C}^d$ can be transformed into any other basis $\left\{\ket{k'}\right\}$ of $\mathbb{C}^d$ by means of a certain unitary transformation $\left\{\ket{k'}\right\}=\left\{U\ket{k}\right\}$. Since it is our goal to find the right orthonormal bases $\left\{\ket{k}_{A1}\right\},\left\{\ket{k}_{A2}\right\}, \left\{\ket{l}_{B1}\right\}$ and $\left\{\ket{l}_{B2}\right\}$, we can select arbitrary orthonormal bases and seek the unitary transformations $U_{A1},U_{A2},U_{B1}$ and $U_{B1}$ that maximise $I_d$ and $I$. These transformations are elements of the unitary group $U(d)$. Due to the fact that global phases do not affect probabilities and that $U(d)$ can be written as a semidirect product $U(d)\cong SU(d) \times U(1)$, it suffices to regard the special unitary group $SU(d)$. We use the \textbf{generalised Euler angle parametrisation} of this group, wherein the Euler angles embody all degrees of freedom (see \cite{ToddSUN},\cite{ToddUN}). For this kind of parametrisation the following antisymmetric and diagonal generalised Gell-Mann matrices (\ref{GGM1},\ref{GGM2}) are needed
\begin{align}
		\lambda_{k^2-1} & \equiv \Lambda^{k-2}=\sqrt{\frac{2}{(k-1)k}}\left(\sum_{j=0}^{k-2}\ket{j}\bra{j}-\left(k-1\right)\ket{k-1}\bra{k-1}\right) \hspace{0.2cm}  & 2\leq k \leq d  \ , \nonumber \\
	\lambda_{k^2+1} & \equiv \Lambda_a^{0k}=-i\ket{0}\bra{k}+i\ket{k}\bra{0}  & 1\leq k \leq d-1  \ . \nonumber
\end{align}
In this notation the explicit form of the parametrisation of $U \in SU(d)$ by generalised Euler angles $\left\{\alpha_i\right\}$ reads (see \cite{ToddSUN}, \cite{ToddUN}\footnote{Note that our notation differs slightly. We chose an indexing where all indices are increasing and the sequence of the product is $\prod_{i=1}^{N}A_i=A_1 \cdot A_{2} \cdots A_N$})
\begin{align}
U=\left[ \prod_{x=0}^{d-2} \left(\prod_{k=2}^{d-x} A(k,j(x)) \right)\right] \left[\prod_{n=2}^{d}\exp({i\lambda_{n^2-1}\alpha_{d^2-(d+1-n)}})\right] \ ,
\end{align}
with $A(k,j(x))$ and $j(x)$ given by
\begin{align}
A(k,j(x))=\exp(i\lambda_3\alpha_{(2k-3)+j(x)}) \cdot \exp(i\lambda_{(k-1)^2+1}\alpha_{2(k-1)+j(m)}) \ ,
\end{align}
\begin{align}
j(x)\, = \,\left\{\begin{array}{l} \displaystyle{0 \hspace{4.1cm}  x=0} \\
		\displaystyle{2\sum_{l=0}^{x-1}(d-x+l) \hspace{1.5cm} x>0 \ . } \end{array}\right.
\end{align}
For instance, for $d=2,3,4$ we find
\begin{align}
U\in SU(2) \Longleftrightarrow U=& \exp(i\lambda_3\alpha_1)\cdot \exp(i\lambda_2\alpha_2)\cdot \exp(i\lambda_3\alpha_3)\ , \\
U\in SU(3) \Longleftrightarrow U=& \exp(i\lambda_3\alpha_1)\cdot \exp(i\lambda_2\alpha_2)\cdot \exp(i\lambda_3\alpha_3)\cdot \exp(i\lambda_5\alpha_4) \nonumber \\ 
\cdot & \exp(i\lambda_3\alpha_5)\cdot \exp(i\lambda_2\alpha_6)\cdot \exp(i\lambda_3\alpha_7)\cdot \exp(i\lambda_8\alpha_8) \ , \\
U\in SU(4) \Longleftrightarrow U=& \exp(i\lambda_3\alpha_1)\cdot \exp(i\lambda_2\alpha_2)\cdot \exp(i\lambda_3\alpha_3)\cdot \exp(i\lambda_5\alpha_4) \nonumber \\
\cdot & \exp(i\lambda_3\alpha_5)\cdot \exp(i\lambda_{10}\alpha_6)\cdot \exp(i\lambda_3\alpha_7)\cdot \exp(i\lambda_2\alpha_8) \nonumber \\
\cdot & \exp(i\lambda_3\alpha_9)\cdot \exp(i\lambda_5\alpha_{10})\cdot \exp(i\lambda_3\alpha_{11})\cdot \exp(i\lambda_2\alpha_{12}) \nonumber \\  \cdot & \exp(i\lambda_3\alpha_{13})\cdot \exp(i\lambda_8\alpha_{14})\cdot \exp(i\lambda_{15}\alpha_{15}) \ .
\end{align}
The parametrisation guarantees compliance with the constraints (\ref{orthonorm}) and (\ref{orthonorm2}) and reduces the number of variables to $4(d^2-1)$ without discarding any solution. However, the optimisation of $\mathcal{B}$ with respect to the Euler angles still requires a numerical optimisation algorithm. While popular algorithms like \textbf{Differential Evolution}, \textbf{Simulated Annealing} or \textbf{gradient based methods} have been very time-consuming, the \textbf{Nelder-Mead method} \cite{NelderMead} has performed this task relatively fast and reliable. In general this method can be advantageous when the number of variables is very large, because it manages to find a maximum without computing derivatives, which can possibly be computationally intensive. For maximisation the algorithm proceeds as follows: Assume a function $f(x_1,..,x_n)$ which has to be maximised with respect to n variables $y=(x_1,..,x_n)$. At the beginning a simplex with $n+1$ vertices $y_1,..,y_{n+1}$ is created at random. In the first step (1) the vertices are being arranged by their values $f(y)$ and labeled according to
\begin{align}
f(y_1)\geq f(y_2) \geq \ldots \geq f(y_{n+1}) \ .
\end{align}
According to this, $f(y_1)$ is the best and $f(y_{n+1})$ the worst vertex. Then a reflection of the worst point through the centroid of the remaining n points $y_0=\frac{1}{n}\sum_{i=1}^n{y_i}$ is performed
\begin{align}
y_r=y_0+\alpha(y_0-y_{n+1}) \ ,
\end{align}
where $\alpha >0$ is called the reflection parameter. The next step of the procedure depends on the value $f(y_r)$: 
\begin{itemize}
	\item{If $f(y_r)$ is better than $f(y_n)$ but not better than $f(y_1)$, i.e. $f(y_1)\geq f(y_r) > f(y_n)$ then a new simplex with $y_{n+1}$ substituted by $y_r$ is built and step (1) is repeated.}
	\item{If $f(y_r)$ is the best point, i.e. $f(y_r)>f(y_1)$ then an expansion $y_e=y_0+\gamma(y_0-y_{n+1})$ with expansion parameter $\gamma >0$ is performed. If $f(y_e )>f(y_r)$ then a new simplex with $y_{n+1}$ substituted by $y_e$ is built and step (1) is repeated. Otherwise $y_r$ is used instead of $y_e$.}
	\item{If $f(y_r)$ does not result in an improvement, i.e. $f(y_r) \leq f(y_{n+1})$ then a contraction $y_c=y_{n+1}+ \rho (y_0-y_{n+1})$ with contraction parameter $\rho>0$ is performed. If $f(y_c) \geq f(y_{n+1})$, then a new simplex with $y_{n+1}$ substituted by $y_c$ is built and step (1) is repeated. Else the simplex is being shrinked: The vertices are being replaced by $y_1'=y_1$ and $y_i'=y_1+\sigma(y_i-y_1)$ with shrink parameter $\sigma > 0$ for all $i\in\left\{2,..,n+1\right\}$, afterwards step (1) is repeated.}	 
\end{itemize}
The rules are repeated until the convergence criteria $|f(y_1')-f(y_1)|< c_1$ and $\|y_1'-y_1\|< c_2$ are satisfied, where $y_1'$ is the new and $y_1$ the old best point, and $c_1,c_2>0$ are constants depending on the desired precision. As we can infer from the rules, the Nelder-Mead method is a hill climbing algorithm, which has the disadvantage that it could converge at a local maximum. We cannot completely avoid this problem; however, by varying the starting points we will find the global maximum in all likelihood.\\
The optimisation procedures for the Bell operators $\mathcal{B}_{I_3}$ and $\mathcal{B}_{I_4}$ via the Euler angle parametrisation and the Nelder-Mead method have been realised in MATHEMATICA 6 and can be found in the appendix \ref{OPBI3} and \ref{OPBI4}. To achieve good results it was necessary to find proper values for the parameters $\alpha,\gamma,\rho$ and $\sigma$. With the values $\alpha = 1.6, \gamma = 1.6, \rho = 0.8 $ and $\sigma = 0.8$ an agreeable compromise between robustness against local maxima and time exposure was attained (with the standard values the algorithm converges at local maxima more often). We have chosen to execute the algorithm ten times with different starting simplices to guarantee that the global maximum is obtained. The accuracy/precision goal has been set to MATHEMATICA 6 standard values $c_1=c_2=10^{-8}$.
\subsubsection{Properties}
Before we discuss some properties of the CGLMP inequalities we show that the inequality $I_d\leq2$ is equivalent to the CHSH inequality when the regarded Hilbert space is two dimensional $\mathcal{H}^A=\mathcal{H}^B=\mathbb{C}^2$. Correspondingly, all properties of the CGLMP inequalities $I_d\leq2$ likewise hold for the CHSH inequality.\\
The proof of equivalence is straight forward. We write all terms of $I_2$ explicitly
\begin{align*}
I_2=&+\left[P(A_1=B_1)+P(B_1=A_2+1)+P(A_2=B_2)+P(B_2=A_1)\right]\\ 
&-\left[P(A_1=B_1-1)+P(B_1=A_2)+P(A_2=B_2-1)+P(B_2=A_1-1)\right]\\
=&+[P(A_1=0,B_1=0)+P(A_1=1,B_1=1)+P(A_2=0,B_1=1)\\
&\mbox{ \ }+P(A_2=1,B_1=0)+P(A_2=0,B_2=0)+P(A_2=1,B_2=1)\\
&\mbox{ \ }+P(A_1=0,B_2=0)+P(A_1=1,B_2=1)]\\
&-[P(A_1=0,B_1=1)+P(A_1=1,B_1=0)+P(A_2=0,B_1=0)\\
&\mbox{ \ }+P(A_2=1,B_1=1)+P(A_2=0,B_2=1)+P(A_2=1,B_2=0)\\
&\mbox{ \ }+P(A_1=0,B_2=1)+P(A_1=1,B_2=0)] \ .
\end{align*}
Since the expectation value (\ref{jointexp}) can be written as $E(A,B)=\int{\rho(\lambda)A(\lambda)B(\lambda)d\lambda}=P(A=0,B=0)+P(A=1,B=1)-P(A=0,B=1)-P(A=1,B=0)$ we can summarise the terms of $I_2$ and find
\begin{align}
I_2=E(A_1,B_1)-E(A_2,B_1)+E(A_2,B_2)+E(A_1,B_2)\leq2 \ .
\end{align}
Due to the symmetry of positive and negative terms of $I_2$ for dimension two, the inequality $|I_2|\leq2$ holds too. Hence, this is exactly the CHSH inequality with $A$ and $B$ interchanged (which makes no difference).\\
Now we quote some important properties: The CGLMP inequalities have been proven to be \textbf{tight Bell inequalities} (see \cite{Masanes}). For the understanding of this, we have to return to the probability distribution (\ref{pdist}) on which the CGLMP inequalities are based on. As noted before, there are $d^4$ values determining the statistics of a local realistic theory. In contrast, in quantum mechanics we have probability distributions $P(A_a=k,B_b=l)$, with $k,l\in\left\{0,..,d-1\right\}$ for each of the four settings $(A_1B_1),(A_1B_2),(A_2B_1)$ and $(A_2B_2)$. As we have seen, those cannot be reproduced by summing over all unregarded observables of $P(A_1=j,A_2=k,B_1=l,B_2=m)$. To conclude, we only have $d^2$ probabilities for each setting and $4d^2$ in total. The quantum mechanical probability distributions are restricted by normalisation
\begin{align}
\sum_{k,l=0}^{d-1}{P(A_a=k,B_b=l)}=1 \hspace{2cm} a,b=1,2 \ ,
\end{align}
and the non-signaling condition (quantum mechanics cannot be used for superluminal communication)
\begin{align}
\sum_{l=0}^{d-1}{P(A_a=k,B_1=l)}=\sum_{l=0}^{d-1}{P(A_a=k,B_2=l)} \hspace{1cm} a=1,2
\end{align}
(it is clear, that the same has to hold for A and B interchanged)\ . \\
What have we gained? The quantum mechanical probabilities can be specified by $4d^2$ values, which can be regarded as entries of a vector $\textbf{P} \in \mathcal{R}^{4d^2}$. Due to the constraints, all physical relevant vectors lie in an affine space of dimension $4d^2-4d$ (see \cite{Masanes}). In this context, the normalisation coming from the local realistic description of the system restricts a vector $\textbf{P}$ to belong to the convex hull of $d^4$ vectors $\textbf{G}_i \in \mathcal{R}^{4d^2}$
\begin{align}
\sum_{j,k,k,m=0}^{d-1}P(A_1=j,A_2=k,B_1=l,B_2=m)=1 \ , \\
\label{polytopeLR}
\Rightarrow \textbf{P}_{LR}=\sum_{i=1}^{d^4}{c_i\textbf{G}_i} \hspace{1.5cm} c_i\geq0 \mbox{ and } \sum_{i=1}^{d^4}c_i=1 \ .
\end{align}
The extremal points $\textbf{G}_i \in \mathcal{R}^{4d^2}$ then are the generators of a convex polytope. Such a polytope can also be described by facets and their half-spaces. Facets induce inequalities of form
\begin{align}
\label{facets}
\textbf{P}\in \mbox{conv}\left( \left\{\textbf{G}_i\right\} \right) \Longleftrightarrow \textbf{X}_i \cdot \textbf{P} \leq x_i \hspace{1cm} \forall i \in \left\{1,2,..,n \right\} \ ,
\end{align} 
where $n$ is the necessary number of facets to define the polytope. Hence, with this set of inequalities we can determine all states that are in contradiction to local realism in an experiment with two apparatuses on each side. The problem involved is, that computing the facets of a high-dimensional polytope is a very difficult task that has only been completely solved for d=2 (see \cite{Fine}). Let us now compare this approach with our well-known Bell inequalities. We know that all states that violate a Bell inequality do not belong to the convex hull of $\left\{\textbf{G}_i\right\}$. However, states that fulfill a Bell inequality do not necessarily belong to it. This means that Bell inequalities define a region (half-space, sphere, polytope, etc.) that contains the convex polytope. The term \textbf{tight Bell inequality} is used if the boundary of the generated region at least partially coincides with at least one of the facets of the polytope. Hence, the inequalities (\ref{facets}) are themselves tight Bell inequalities as well as the CGLMP inequality given by $I_{\mathcal{B}_d}$. For the CGLMP it has been shown in \cite{Masanes} that it coincides with a family of equivalent facets. However, we do not know if the CGLMP inequality coincides with \textbf{all} facets of the polytope. Even though this might be the case, we cannot infer from this that all non-local states can be found with help of the CGLMP inequality because increasing the number of observables per party could define an improved polytope that could enable us to find even more non-local states. Please refer to \cite{Davis} for a more detailed discussion.\\
Now let us investigate another property of the CGLMP inequality concerning the \textbf{maximal violation}. As we stated before, for the maximally entangled state $\ket{\Omega_{0,0}}$ the violation is maximal when the measurement setup is configured according to ($\ref{ONBBell}$). However, in this case the largest eigenvalue of the corresponding Bell operator $\mathcal{B}_{I_d}$ is larger than the value $I_d\left(\ket{\Omega_{0,0}}\right)$ for all $d>2$. The largest eigenvalues are summarised in this table (see \cite{AcinDurt}):
\\
\\
$\begin{array}{|c|c|c|c|c|c|c|}
  \hline
  d & 2 & 3&4&5&6& 7 \\
    \hline
    I_d \left(\ket{\Psi_{mv}}\right)&2.8284& 2.9149& 2.9727& 3.0157& 3.0497& 3.0776 \\
   I_d\left(\ket{\Omega_{0,0}}\right) & 2.8284& 2.8729& 2.8962& 2.9105& 2.9202&2.9272\\ 
  \mbox{difference} \left[ \% \right]& 0 & 1.4591 & 2.6398 & 3.6133 & 4.4345 &  5.1411\\
  \hline
\end{array}$\\
\\
\\
wherein $\ket{\Psi_{mv}}$ \footnote{$mv$ stands for "maximal violation"} denotes the eigenvector of $\mathcal{B}_{I_d}$ with the largest eigenvalue. Since all maximally entangled states are equivalent in terms of local unitaries (as discussed in \S \ref{Biqudits}) the state $\ket{\Psi_{mv}}$ must be non-maximally entangled\footnote{If $\ket{\Omega_{0,0}}=U_A\otimes U_B \ket{\Psi_{mv}}$ then we could use the Bell operator $U_A \otimes U_B\mathcal{B}_{I_d}U_A^{\dagger}\otimes U_B^{\dagger}$ to obtain $I_d\left(\ket{\Omega_{0,0}}\right)=I_d \left(\ket{\Psi_{mv}}\right)$ in contradiction to obtained values.}. For instance, for qutrits the eigenvector is the non-maximally entangled state $\ket{\Psi_{mv}}=\frac{1}{\sqrt{n}}\left(\ket{00}+\gamma\ket{11}+\ket{22}\right)$ with $n=2+\gamma^2$ and $\gamma=\left(\sqrt{11}-\sqrt{3}\right)/2$. Knowing that the states $\ket{\Psi_{mv}}$ yield stronger violations, we could ask if the above given values are the highest obtainable for $I_d$ because the measurement configuration (\ref{ONBBell}) has only been intended to be the best for $\ket{\Omega_{0,0}}$. Numerical investigations with varying states and measurement configurations show that these values are indeed the highest (see \cite{AcinDurt}). How can we interpret this result? Does the higher violation and the consequential higher resistance against noise signify that the states $\ket{\Psi_{mv}}$ are more non-local than $\ket{\Omega_{0,0}}$? This and further questions on the comparison of nonlocality and entanglement will be discussed in the next section.
\subsection{Remarks on nonlocality and entanglement}
A state $\rho$ is said to be \textbf{non-local} if it violates a Bell inequality. Such a state is necessarily entangled, which follows from the fact that separable states cannot violate any Bell inequality. This can easily be seen by generalising the probability distribution (\ref{pdist}). Recall that ideal Bell inequalities correspond to the facets of the polytope of local realistic correlations (\ref{facets}). The number of observables on each site depends on the regarded Bell inequality but we do not want to restrict ourself to the case of two observables. Thus, we generalise (\ref{pdist}) to
\begin{align}
P(A_1=j,..,A_r=k,B_1=l,..,B_s=m) \ ,
\end{align}
with $r,s\in \mathbb{N}$. According to (\ref{polytopeLR}), the normalisation defines a polytope and all local states belong to it. It is not difficult to recognise that any separable pure state belongs to the polytope: Since all probabilities are of product form $P(A_n=a,B_m=b)=P(A_n=a) \cdot P(B_m=b)$ (which is not the case when a state is entangled) we can easily construct the above stated probability distribution
\begin{align}
&P(A_1=j,..,A_r=k,B_1=l,..,B_s=m) \nonumber \\
=&P(A_1=j)\cdots P(A_r=k)P(B_1=l)\cdots P(B_s=m) \ ,
\end{align}
which of course obeys the normalisation relation due to normalised state vectors. Since separable mixed states are only convex combinations of pure ones we conclude that they belong to the polytope as well. This completes the proof that  separable states cannot violate any Bell inequality. As a consequence, the operator $\left[2 \cdot \mathbbm{1}-\mathcal{B}_{I_d}\right]\in\mathcal{H}^{AB}_{HS}$ is a entanglement witness because $Tr(\rho \mathcal{B}_{I_d})\leq2$ holds for all separable states. However, in general this is not an optimal witness, even if $\mathcal{B}_{I_d}$ is optimised for $\rho$.\\
Let us go back to the main subject of this section. Entanglement is a necessary component for nonlocality, however not all entangled states violate a Bell inequality. This will become more obvious when we discuss the geometry of entanglement and nonlocality in \S \ref{Geometry}. At this point, we mention one famous example of an entangled state which does not violate the CHSH inequality: The Werner state $\rho_W=p\ket{\Psi^-}\bra{\Psi^-}+\frac{1-p}{4}\mathbbm{1}$ (for d=2 equivalent to the isotropic state) is entangled for $\frac{1}{3}<p\leq1$, however violation of the CHSH inequality is obtained only for $\frac{1}{\sqrt{2}}< p\leq1$\footnote{The values of the parameter p can easily be verified with the PPT criterion and the Horodecki violation criterion}. This means that for $\frac{1}{3}< p \leq \frac{1}{\sqrt{2}} $ the statistical predictions can be reproduced by a local realistic theory, even though the state is entangled. This brings us to the notion of \textbf{hidden nonlocality}. We can claim that any distillable state (\S \ref{distbound}) contains some amount of nonlocality since we could perform local operations and classical communication (\S \ref{LOCC}) until nonlocality is revealed. Despite this argument, the distinction between nonlocality and entanglement is not entirely clarified. Besides the fact that we do not know if any entangled state behaves non-locally in a particular measurement situation, there are some well-known examples that indicate that they might not be exactly the same. Eberhard has shown that non-maximally entangled states require lower detection efficiencies than maximally entangled ones, in order to close the detection loophole (see \cite{Eberhard}). We have also seen that some non-maximally entangled states seem to be more non-local than the Bell states, due to their higher violation of the CGLMP inequality. With regard to this issue, one might argue that the resistance against noise is not a good measure of nonlocality. Some remarks on this can be found in a publication by Acin, Durt, Gisin and Latorre \cite{AcinDurt}. Nevertheless, they also suggest that a Bell inequality with more than two measurement instruments on each site could avoid this peculiarity. Another hint that entanglement and nonlocality are different resources has been found by Brunner, Gisin and Scarani (see \cite{BrunnerGisin}). They have shown that the simulation of non-maximally entangled states via \textbf{non-local machines} requires more resources than the simulation of a Bell state. The hypothetical non-local machine\footnote{other common names are PR box (named after Popescu and Rohrlich) and non-local box} was constructed to obtain the algebraic bound of the CHSH inequality without violating the non-signaling condition. Alice and Bob each have an input ($x$ and $y$) and an output ($a$ and $b$). Alice can choose the value of $x\in\left\{0,1\right\}$ and gets $a\in\left\{0,1\right\}$ in return, while Bob can choose the value of $y\in\left\{0,1\right\}$ and gets $b\in\left\{0,1\right\}$ in return.
\begin{figure}[htp]
\centering
\setlength{\unitlength}{1bp}%
  \begin{picture}(401.29, 67.98)(0,0)
  \put(0,0){\includegraphics{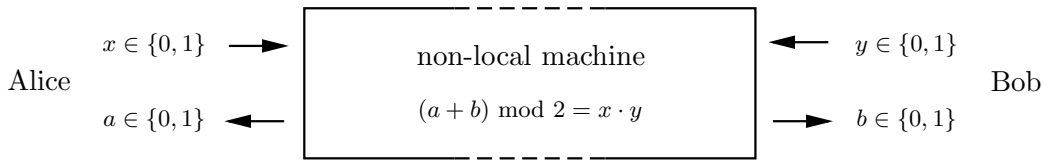}}
  \put(17.95,34.10){\fontsize{11.38}{13.66}\selectfont \makebox(0,0)[]{Alice\strut}}
  \put(386.17,33.75){\fontsize{11.38}{13.66}\selectfont \makebox(0,0)[]{Bob\strut}}
  \put(203.25,43.66){\fontsize{11.38}{13.66}\selectfont \makebox(0,0)[]{non-local machine\strut}}
  \put(61.27,48.09){\fontsize{8.54}{10.24}\selectfont \makebox(0,0)[]{$x\in \left\{0,1\right\}$\strut}}
  \put(61.27,19.88){\fontsize{8.54}{10.24}\selectfont \makebox(0,0)[]{$a\in \left\{0,1\right\}$\strut}}
  \put(344.76,48.26){\fontsize{8.54}{10.24}\selectfont \makebox(0,0)[]{$y\in \left\{0,1\right\}$\strut}}
  \put(344.76,20.05){\fontsize{8.54}{10.24}\selectfont \makebox(0,0)[]{$b\in \left\{0,1\right\}$\strut}}
  \put(202.85,23.20){\fontsize{8.54}{10.24}\selectfont \makebox(0,0)[]{$(a+b)\mbox{ mod } 2 = x \cdot y$\strut}}
  \end{picture}%
  \caption{Schematic illustration of the non-local machine}
\end{figure}\\
The machine creates a random output $P(a=0)=P(a=1)=P(b=0)=P(b=1)=\frac{1}{2}$, though with $a$ and $b$ correlated according to the rule
\begin{align}
(a+b)\mbox{ mod } 2 = x \cdot y \ .
\end{align}
For instance, if $x = y = 0$ then the output of the machine is $a = b = 0$ in half the cases and $a = b = 1$ in the other half. Due to the local randomness this box cannot be used for signaling. Let us investigate the violation of the CHSH inequality. We assume that for any observable $A(\vec{a})$ there is a corresponding value for $x(\vec{a})\in \left\{0,1\right\}$, and the same applies to $B(\vec{b})$ and $y(\vec{b}) \in \left\{0,1\right\}$. As before, we define a correlation function that yields the value $1$ when the outputs $a$ and $b$ are identical, and $-1$ when they are different
\begin{align}
C\left(A(\vec{a}),B(\vec{b})\right) \equiv \,\left\{\begin{array}{l} \displaystyle{\mbox{ \ } 1 \hspace{1cm} \mbox{ \ for \ }  a=b} \\
		\displaystyle{-1 \hspace{1cm} \mbox{ \ for \ } a\neq b\ .} \end{array}\right.
\end{align}
For the case $A(\vec{a}) \rightarrow x(\vec{a})=1, A(\vec{a'}) \rightarrow x(\vec{a'})=0, B(\vec{b}) \rightarrow y(\vec{b})=0$ and $B(\vec{b'}) \rightarrow y(\vec{b'})=1$ the expectation values of the correlation function are $E(\vec{a},\vec{b})=E(\vec{a'},\vec{b})=E(\vec{a'},\vec{b'})=1$ and $E(\vec{a},\vec{b'})=-1$. Under such circumstances, the CHSH inequality yields the algebraic bound
\begin{align}
|E(\vec{a},\vec{b})-E(\vec{a},\vec{b'})+E(\vec{a'},\vec{b'})+E(\vec{a'},\vec{b})|=4 \ .
\end{align}
This is the most non-local behaviour of a system possible. Thus, from a information theoretical point of view, this machine can be seen as a resource of nonlocality. Inspired by this idea, Cerf, Gisin, Massar and Popescu have shown that quantum entanglement can be simulated without communication by use of the non-local machine (see \cite{CerfGisin}). Subsequent work demonstrated that for the simulation of a Bell state only one non-local machine is required, while at least two are required to simulate non-maximally entangled states (see \cite{BrunnerGisin}). This is another fact that strengthens the conjecture that there is a difference between nonlocality and entanglement. Further open questions concerning Bell inequalities, including also the experimental point of view, can be found in a publication by N. Gisin \cite{BIquestions}. 
\newpage
\hfill
\vfill
\newpage
\section{Geometrical aspects of bipartite systems}
As the state space in quantum physics is a complex Hilbert-Schmidt space $\mathcal{H}_{HS}$ it is in general difficult to get a feeling for the properties of density matrices. However, in some cases they can be represented by vectors in a real vector space. For example, in \S \ref{basics} we have shown that density matrices can be described by real Bloch vectors. Due to the fact that the dimension of this vector is $d^2-1$, this representation is not very helpful if we intend to investigate bipartite systems. This is because even for the simplest case of two qubits the vector space is 15 dimensional. In this section we show that it is still possible to find attractive illustrations when it comes to studying entanglement and nonlocality of bipartite systems.
\label{Geometry}
\subsection{Geometry of the two-qubit system - the tetrahedron}
\subsubsection{Introduction}
\begin{figure}[H]
\centering
  \setlength{\unitlength}{1bp}
  \begin{picture}(320, 380)(0,0)
  \put(20,30){\includegraphics[scale=0.45]{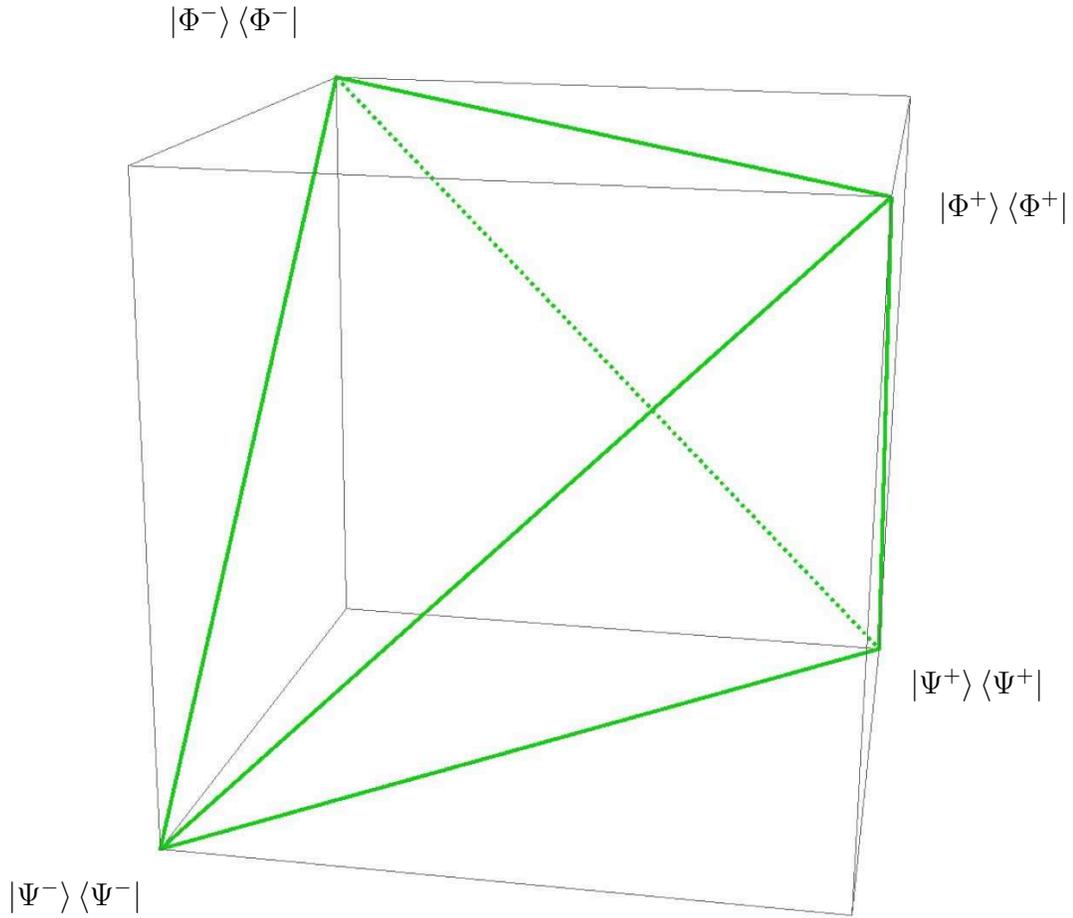}}

  \put(0,40){\fontsize{12}{20}\selectfont \makebox(0,0)[]{$\ket{\Psi^-}\bra{\Psi^-}$\strut}}
  \put(340,120){\fontsize{12}{20}\selectfont \makebox(0,0)[]{$\ket{\Psi^+}\bra{\Psi^+}$\strut}}
  \put(350,300){\fontsize{12}{20}\selectfont \makebox(0,0)[]{$\ket{\Phi^+}\bra{\Phi^+}$\strut}}
  \put(60,370){\fontsize{12}{20}\selectfont \makebox(0,0)[]{$\ket{\Phi^-}\bra{\Phi^-}$\strut}}
   \end{picture}
  \caption{Illustration of the regular tetrahedron (green)}
  \label{tetrafig}
\end{figure}

As usual, we begin with the bipartite qubit system (the following approach can be found in \cite{HorodeckiGEOMETRY} and \cite{BertlmannGEOMETRY}). According to (\ref{generaldensity}), any density matrix acting on $\mathcal{H}^{AB}=\mathbb{C}^2\otimes\mathbb{C}^2$ can be written in the form
\begin{align}
\rho=\frac{1}{4}( \mathbbm{1}\otimes\mathbbm{1} + \vec{r}\cdot\vec{\sigma}\otimes\mathbbm{1} + \mathbbm{1}\otimes \vec{s} \cdot \vec{\sigma} + \sum_{n,m=1}^{3}t_{nm}\sigma_n\otimes \sigma_m ) \ .
\end{align}
Separability and nonlocality are invariant under local unitary transformations $U_A\otimes U_B$, therefore we can define equivalence classes of states 
\begin{align}
\left[ \rho \right]=\{ \rho' | \rho'=U_A\otimes U_B \rho U_A^{\dagger} \otimes U_B^{\dagger}\}
\end{align} 
that have the same properties concerning separability and nonlocality. We exploit the group isomorphism between $SU(2)$ and $SO(3)$ which induces that for any $O \in SO(3)$ there exists a $U \in SU(2)$ that obeys
\begin{align}
U \vec{n} \cdot \vec{ \sigma} U^{\dagger}= ( O \vec{n}) \cdot \vec{\sigma} \ .
\end{align}
Under a transformation $U_A\otimes U_B \rho U_A^{\dagger} \otimes U_B^{\dagger}$ the vectors $\vec{r},\vec{s}$ and the matrix $T=(t_{nm})$ then become
\begin{align}
\vec{r'}=&O_A \vec{r} \ , \\
\vec{s'}=&O_A \vec{s} \ , \\
T'=&O_A T O_B^T \ .
\end{align}
According to the \textbf{singular value decomposition}, there exist orthogonal matrices $O_A$ and $O_B$ so that $T'$ becomes diagonal with real entries. We choose density matrices $\rho$ with diagonal $T=diag(t_{11},t_{22},t_{33})$ as the representatives of the equivalence classes $[\rho]$. Hence, for determining separability and nonlocality it suffices to investigate the set of states with diagonal $T$ which we denote by $\mathcal{D}$. We write the diagonal entries in a vector $\vec{t}=(t_{11},t_{22},t_{33})$, in this way a density matrix $\rho \in \mathcal{D}$ is described by three vectors $\vec{r},\vec{s},\vec{t}\in \mathbb{R}^3$. Now, let us look through the elements of $\mathcal{D}$. Non-negativity is a necessary condition for a density matrix $\rho$, i.e. $Tr\left(\ket{\Psi}\bra{\Psi}\rho\right)\geq 0$ for all $\ket{\Psi} \in \mathcal{H}^{AB}$. Consider the four projectors $P_1=\ket{\Psi^+}\bra{\Psi^+},P_2=\ket{\Psi^-}\bra{\Psi^-},P_3=\ket{\Phi^+}\bra{\Phi^+}$ and $P_4=\ket{\Phi^-}\bra{\Phi^-}$ given by the Bell states $\ket{\Psi ^\pm}=\frac{1}{\sqrt{2}}\left(\ket{01} \pm \ket{10}\right)$ and $\ket{\Phi ^\pm}=\frac{1}{\sqrt{2}}\left(\ket{00} \pm \ket{11}\right)$. The Pauli matrix decomposition (\ref{algebra}) yields $r_i=0$ and $s_i=0$ for $i\in\{1,2,3\}$ and
\begin{align}
P_1=\ket{\Psi^+}\bra{\Psi^+} \hspace{1cm} &\Leftrightarrow \hspace{1cm} \vec{t_1}=(+1,+1,-1) \ ,\\
P_2=\ket{\Psi^-}\bra{\Psi^-} \hspace{1cm} &\Leftrightarrow \hspace{1cm} \vec{t_2}=(-1,-1,-1) \ ,\\
P_3=\ket{\Phi^+}\bra{\Phi^+} \hspace{1cm} &\Leftrightarrow \hspace{1cm} \vec{t_3}=(+1,-1,+1) \ ,\\
P_4=\ket{\Phi^-}\bra{\Phi^-} \hspace{1cm} &\Leftrightarrow \hspace{1cm} \vec{t_4}=(-1,+1,+1) \ .
\end{align}
For two states $\rho$ and $\rho ' $ given in the Pauli matrix decomposition we can compute $Tr ( \rho \rho ' )$ easily, because it simplifies to $Tr ( \rho \rho' ) =\frac{1}{4}(1+\vec{r} \cdot \vec{r}' + \vec{s} \cdot \vec{s}'+Tr ( T T'^{\dagger}))$. Thus, the four inequalities $Tr\left(P_n\rho\right)\geq 0$ with $n\in\{1,..,4\}$ give
\begin{align}
\label{tet1}
1+t_{11}+t_{22}-t_{33}\geq0 \ ,\\
1-t_{11}-t_{22}-t_{33}\geq0 \ ,\\
1+t_{11}-t_{22}+t_{33}\geq0 \ ,\\
\label{tet4}
1-t_{11}+t_{22}+t_{33}\geq0 \ .
\end{align}
This restricts any $\vec{t}$ of $\rho\in\mathcal{D}$ to belong to a \textbf{regular tetrahedron} spanned by the Bell states at the vertices (see Fig.\ref{tetrafig}).
Note that only in the case where $\vec{r}=0$ and $\vec{s}=0$ the density matrices $\rho \in \mathcal{D}$ are diagonal in the Bell basis $\left\{ \ket{\Psi^+},\ket{\Psi^-},\ket{\Phi^+},\ket{\Phi^-} \right\}$ and only then the constraints (\ref{tet1})-(\ref{tet4}) are also sufficient for non-negativity. States with this property define a subset $\mbox{LMM}\subset \mathcal{D}$ and are called \textbf{locally maximally mixed} because the respective reduced density matrices are maximally disordered $Tr_A(\rho)=Tr_B(\rho)=\frac{1}{2}\mathbbm{1}$. The subset $\mbox{LMM} \subset \mathcal{D}$ contains only four pure states, namely the four Bell states, which can easily be seen by testing the purity condition
\begin{align}
Tr(\rho^2)=\frac{1}{4}(1+t_{11}^2+t_{22}^2+t_{33}^2)\stackrel{!}{=}1\\
\Rightarrow t_{11}^2+t_{22}^2+t_{33}^2=3 \ .
\end{align}
Geometrically spoken this defines a sphere with radius $\sqrt{3}$ that intersects the tetrahedron at the vertices, which therefore are the only pure sates of $\mbox{LMM}$.
\newpage
\subsubsection{Geometry of separable and entangled states}
\begin{figure}[htp]
\centering
\setlength{\unitlength}{1bp}
  \begin{picture}(340, 380)(0,0)
  \put(20,30){\includegraphics[scale=0.45]{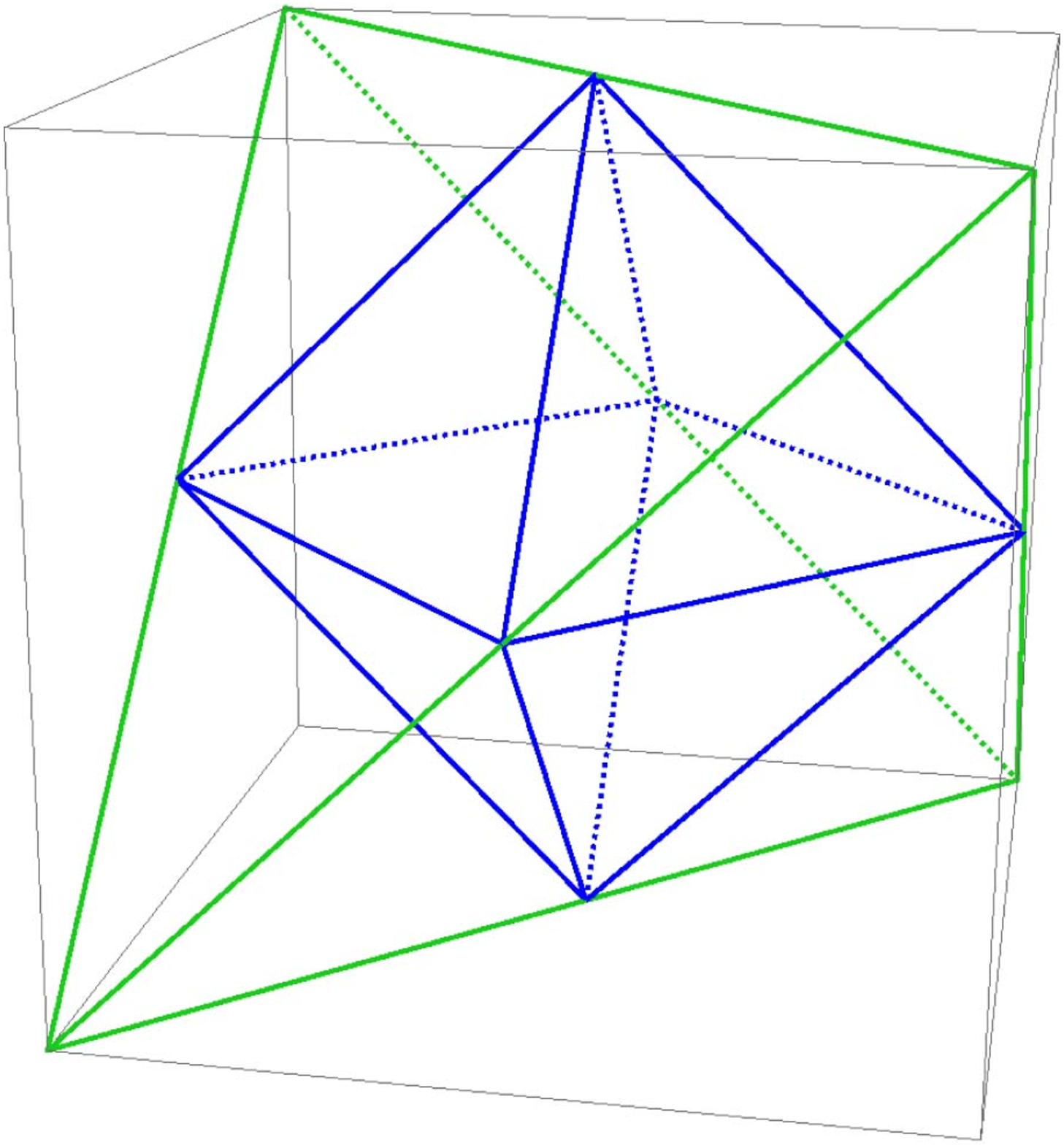}}

  \put(0,40){\fontsize{12}{20}\selectfont \makebox(0,0)[]{$\ket{\Psi^-}\bra{\Psi^-}$\strut}}
  \put(340,120){\fontsize{12}{20}\selectfont \makebox(0,0)[]{$\ket{\Psi^+}\bra{\Psi^+}$\strut}}
  \put(350,300){\fontsize{12}{20}\selectfont \makebox(0,0)[]{$\ket{\Phi^+}\bra{\Phi^+}$\strut}}
  \put(60,370){\fontsize{12}{20}\selectfont \makebox(0,0)[]{$\ket{\Phi^-}\bra{\Phi^-}$\strut}}
   \end{picture}
  \caption{Illustration of the tetrahedron (green) and the octahedron of separable states (blue)}
  \label{tetrasep}
\end{figure}

Now, let us study the separability of $\rho \in \mbox{LMM}$. In \S \ref{CPmaps} we stated that the PPT criterion is a necessary and sufficient criterion for separability for density matrices acting on $\mathcal{H}^{AB}=\mathbb{C}^2\otimes \mathbb{C}^2$. In the Pauli matrix decomposition it is trivial to compute the partial transposition. $\sigma_1$ is a symmetric matrix and therefore invariant under transposition, the same applies to $\sigma_3$ which is diagonal. Only $\sigma_2$ changes its algebraic sign $\sigma_2^T=-\sigma_2$, since it is anti-symmetric. Consequently, partial transposition results in a reflection of the tetrahedron where the coordinates are transformed according to $(t_{11},t_{22},t_{33}) \rightarrow (t_{11},-t_{22},t_{33})$. All operators $\rho \in \mbox{LMM}$ belonging to the original tetrahedron are non-negative, while the reflected tetrahedron contains operators $\rho$ with negative eigenvalues. This means that all states $\rho$ belonging to the cross section of the two tetrahedron are separable, while the other states are entangled. This intersection is an \textbf{octahedron} (see Fig.\ref{tetrasep}).
The extremal separable points have the coordinates $\vec{t}_{1/2}=(0,0,\pm1),\vec{t_{3/4}}=(0,\pm1,0)$ and $\vec{t_{5/6}}=(0,0,\pm1)$ and are 1:1 mixtures of two Bell states. A possible decomposition into a convex combination of separable density matrices can easily be found. Consider the mixture $\rho=\frac{1}{2}(\rho_1+\rho_2)$ of the separable states $\rho_1=\frac{1}{2}(\mathbbm{1}+\vec{a} \cdot \vec{\sigma})\otimes\frac{1}{2}(\mathbbm{1}+\vec{b}\cdot \vec{\sigma})$ and $\rho_2=\frac{1}{2}(\mathbbm{1}-\vec{a} \cdot \vec{\sigma})\otimes\frac{1}{2}(\mathbbm{1}-\vec{b}\cdot \vec{\sigma})$. A short computation gives
\begin{align}
\rho=\frac{1}{2}(\rho_1+\rho_2)=\frac{1}{4}(\mathbbm{1}+\vec{a}\cdot \vec{\sigma}\otimes \vec{b} \cdot \vec{\sigma}) \ .
\end{align}
With the vectors $\vec{a}=\vec{e}_n$ and $\vec{b}=\pm \vec{e}_n$, where $\vec{e}_n$ stands for the unit vectors $\vec{e}_1=(1,0,0)$, $\vec{e}_2=(0,1,0)$ and $\vec{e}_3=(0,0,1)$, we can explicitly demonstrate the separability of the extremal points of the octahedron, thus all states within the octahedron because they are just convex combinations of these points.
\subsubsection{Geometry of non-local states}
\begin{figure}[H]
\centering
\setlength{\unitlength}{1bp}
  \begin{picture}(300, 380)(0,0)
  \put(10,20){\includegraphics[scale=0.60]{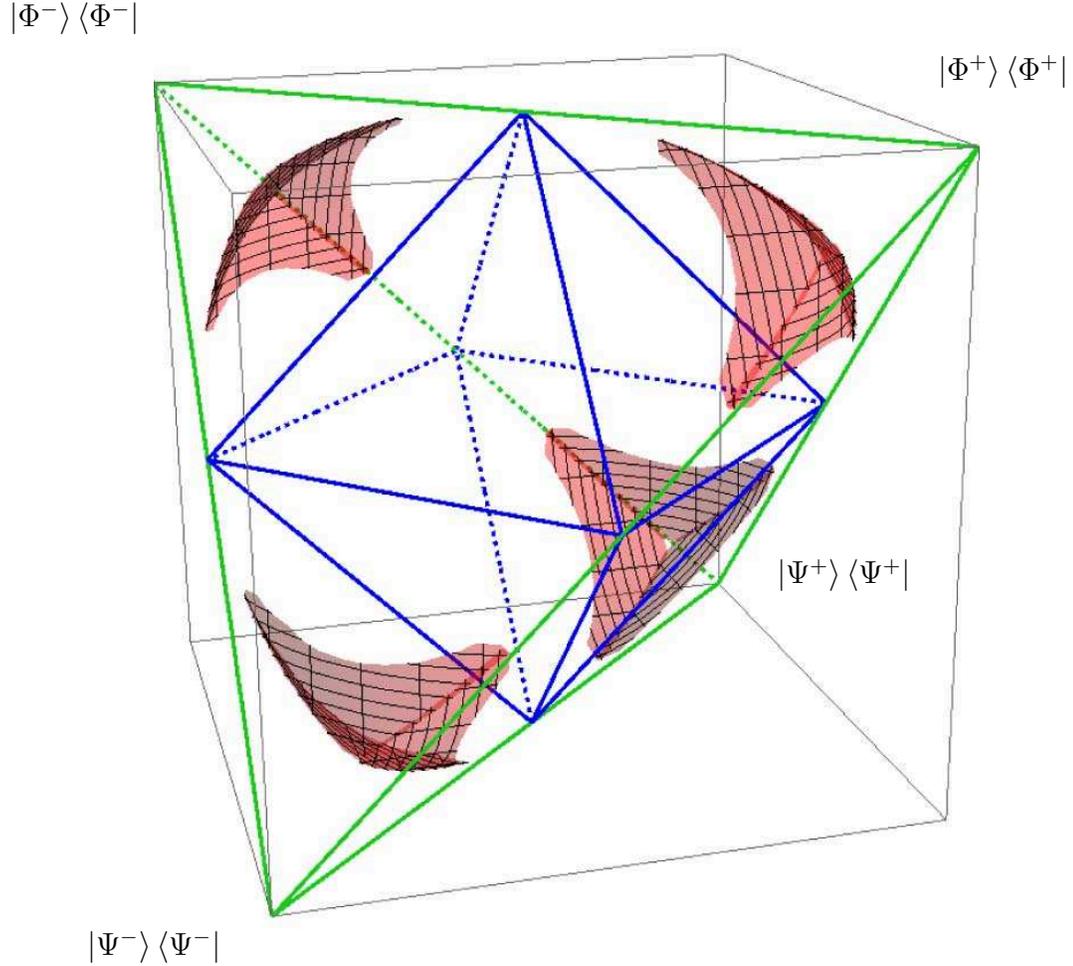}}

  \put(10,20){\fontsize{12}{20}\selectfont \makebox(0,0)[]{$\ket{\Psi^-}\bra{\Psi^-}$\strut}}
  \put(270,160){\fontsize{12}{20}\selectfont \makebox(0,0)[]{$\ket{\Psi^+}\bra{\Psi^+}$\strut}}
  \put(330,350){\fontsize{12}{20}\selectfont \makebox(0,0)[]{$\ket{\Phi^+}\bra{\Phi^+}$\strut}}
  \put(-20,370){\fontsize{12}{20}\selectfont \makebox(0,0)[]{$\ket{\Phi^-}\bra{\Phi^-}$\strut}}
   \end{picture}
  \caption{Illustration of the tetrahedron and the borders of nonlocality. States $\rho$ beyond the red/meshed surfaces violate the CHSH inequality.}
  \label{tetranl}
\end{figure}

We intend to find the non-local states within the tetrahedron, i.e. states that violate the CHSH inequality\footnote{CHSH is equivalent to CGLMP for d=2}. For the bipartite qubit system this is not demanding because we can exploit the Horodecki violation criterion (\ref{CHSHVIO}). First, we have to compute the eigenvalues of the matrix $K=T^T T$. Since $T$ is already diagonal we get $\lambda_1=t_{11}^2, \lambda_2=t_{22}^2$ and $\lambda_3=t_{33}^2$. A sufficient condition for nonlocality is that the sum of two eigenvalues is larger than 1 which implies the validity of at least one of the three inequalities
\begin{align}
t_{11}^2+t_{22}^2>&1 \ ,\\
t_{11}^2+t_{33}^2>&1 \ ,\\
t_{22}^2+t_{33}^2>&1 \ .
\end{align}
Each of the inequalities defines a cylinder with radius $1$ in the geometric picture and a density matrix $\rho$ that lies outside of one of them is non-local. The union of the exterior regions of these three cylinders is the set of non-local states (see Fig.\ref{tetranl}).
\noindent
This descriptively illustrates that not all entangled states violate the CHSH inequality and that Bell operators $\mathcal{B}_{CHSH}$ are non-optimal entanglement witnesses for $\mbox{LMM}$ states.
\newpage
\subsection{Geometry of bipartite qudit systems - the magic simplex}\label{MagicSimplex}
\subsubsection{Introduction}
The derivation of the octahedron was direct and straight forward. In seeking a similar geometric object for qudits we definitely try to proceed analogously. The first step is to express $\rho \in \mathcal{H}_{HS}^{AB}$ in the generalised form of (\ref{generaldensity}) which is (\ref{biqudit})
\begin{align}
\rho=\frac{1}{d^2} \left(\mathbbm{1} \otimes \mathbbm{1} +\sum_{i=1}^{d^2-1}r_i\Gamma_i\otimes\mathbbm{1} + \sum_{i=1}^{d^2-1} s_i \mathbbm{1}\otimes \Gamma_i + \sum_{i,j=1}^{d^2-1} t_{ij}\Gamma_i\otimes\Gamma_j\right) \ .
\end{align}
As we require a real vector space for a geometric structure, we use the generalised Gell-Mann matrices for $\{\Gamma_i\}$. In order to use the same notation as for qubits, we express $\rho$ with generalised Bloch vectors $\vec{r}_{\Lambda}$, $\vec{s}_{\Lambda}$ and a correlation matrix $T=(t_{ij})$
\begin{align}
\rho=\frac{1}{d^2} \left(\mathbbm{1} \otimes \mathbbm{1} +\vec{r}_{\Lambda} \cdot \vec{\Lambda}\otimes\mathbbm{1} + \mathbbm{1}\otimes \vec{s}_{\Lambda} \cdot \vec{\Lambda} + \sum_{i,j=1}^{d^2-1} t_{ij}\Gamma_i\otimes\Gamma_j\right) \ .
\end{align}
The next step would be the diagonalisation of the matrix $T=(t_{ij})$ via two orthogonal matrices $O_A$ and $O_B$. However, here we are confronted with the problem that the relation $U \vec{r}_{\Lambda} \cdot \vec{\Lambda} U^{\dagger}= (O\vec{r}_{\Lambda})\cdot \vec{\Lambda}$ is invalid for Gell-Mann matrices. For example, consider the state $\rho=\frac{1}{3}(\mathbbm{1}+\vec{b}_{\Lambda}\cdot\vec{\Lambda})$ with $\vec{b}_{\Lambda}^T=(0,0,1,0,0,0,0,0)$
\begin{align}
\rho=\frac{1}{3}(\mathbbm{1}+\lambda_3)=\left(
\begin{array}{lll}
 \frac{2}{3} & 0 & 0 \\
 0 & 0 & 0 \\
 0 & 0 & \frac{1}{3}
\end{array}
\right) \ ,
\end{align}
which is obviously a valid density matrix. Let us assume that there exists a unitary transformation $\rho'=U \rho U^{\dagger} =\frac{1}{3}(\mathbbm{1}+U\vec{b}_{\Lambda}\cdot\vec{\Lambda}U^{\dagger})=\frac{1}{3}(\mathbbm{1}+(O\vec{r}_{\Lambda})\cdot \vec{\Lambda})$ that corresponds to the orthogonal matrix $O$
\begin{align}
O=
\begin{pmatrix}
 0 & 0 & 0 & 0 & 0 & 0 & 0 & 0 \\
 0 & 0 & 0 & 0 & 0 & 0 & 0 & 0 \\
 0 & 0 & 0 & 0 & 0 & 0 & 0 & 1 \\
 0 & 0 & 0 & 0 & 0 & 0 & 0 & 0 \\
 0 & 0 & 0 & 0 & 0 & 0 & 0 & 0 \\
 0 & 0 & 0 & 0 & 0 & 0 & 0 & 0 \\
 0 & 0 & 0 & 0 & 0 & 0 & 0 & 0 \\
 0 & 0 & 1 & 0 & 0 & 0 & 0 & 0 
\end{pmatrix} \hspace{1cm} \Rightarrow \hspace{1cm} \vec{b}_{\Lambda}'=O \vec{r}_{\Lambda} = \begin{pmatrix}
 0 \\
 0 \\
 0 \\
 0 \\
 0 \\
 0 \\
 0 \\
 1 
\end{pmatrix} \ .
\end{align}
This transformation turns the state $\rho$ into $\rho'=\frac{1}{3}(\mathbbm{1}+\vec{b}_{\Lambda}'\cdot\vec{\Lambda})$
\begin{align}
\rho'=\frac{1}{3}(\mathbbm{1}+\lambda_8)=
\begin{pmatrix}
\frac{1}{3} \left(1+\frac{1}{\sqrt{3}}\right) & 0 & 0 \\
 0 & \frac{1}{3} \left(1+\frac{1}{\sqrt{3}}\right) & 0 \\
 0 & 0 & \frac{1}{3} \left(1-\frac{2}{\sqrt{3}}\right)
\end{pmatrix} \ .
\end{align}
As we can see the eigenvalues of $\rho$ and $\rho'$ are not the same, which in other words means that the orthogonal transformation $O$ results in a non-unitary transformation\footnote{Eigenvalues are invariant under unitary transformations} of $\rho$. Hence, the relation $U \vec{r}_{\Lambda} \cdot \vec{\Lambda} U^{\dagger}= (O\vec{r}_{\Lambda})\cdot \vec{\Lambda}$ cannot be valid. Even if $T$ can be diagonalised by product unitary transformations $U_A \otimes U_B$ for a subset of LMM states there is another problem. Consider a locally maximally mixed state $\rho\in\mbox{LMM}$ \footnote{$LMM=\{\rho\in\mathcal{H}_{HS}^{AB}|\rho_A=Tr_B \rho=\rho_B= Tr_A \rho = \frac{1}{d}\mathbbm{1}\}$} with diagonal $T$
\begin{align}
\rho=\frac{1}{d^2} \left(\mathbbm{1} \otimes \mathbbm{1} + \sum_{i=1}^{d^2-1} t_{ii}\Gamma_i\otimes\Gamma_i\right) \ .
\end{align}
In the qubit case the eigenvectors of $\rho$ are the Bell states for an arbitrary choice of $\{t_{ii}\}$, thus the non-negativity condition leads to a simple geometric figure, namely the tetrahedron. For qudits the situation is different because the eigenvectors of $\rho$ depend on $\{t_{ii}\}$ implying that the geometric figure given by non-negativity is much more difficult to determine. Due to this and the fact that the set of states whose correlation matrices $T$ can be diagonalised via local unitaries are merely a subset of LMM, we choose another subset of LMM whose properties concerning entanglement are interesting and whose non-negativity is easier to handle.\\
At this point we adopt the way of B. Baumgartner, B.C. Hiesmayr and H. Narnhofer, who introduced a generalisation of the tetrahedron for qudits \cite{BHNsimplex}. As the octahedron is spanned by the four Bell states, the expansion for higher dimensional systems is the so called \textbf{magic simplex} spanned by generalised Bell states $P_{k,l}=\ket{\Omega_{k,l}}\bra{\Omega_{k,l}}$ with  $\ket{\Omega_{k,l}}=\left(W_{k,l}\otimes\mathbbm{1}\right)\ket{\Omega_{0,0}}$ given by the Weyl operators (\ref{GBS})
\begin{align}
\mathcal{W}=\{\sum_{k,l=0}^{d-1} c_{k,l}P_{k,l} \mbox{ \ }  |  \mbox{ \ }  c_{k,l}\geq 0,\sum_{k,l=0}^{d-1}c_{k,l}=1\} \ .
\end{align}
This simplex is a convex set located in a $d^2-1$ dimensional hyperplane in a $d^2$ dimensional real vector space of hermitian operators spanned by the operators $\{P_{k,l}\}$. When the element $\rho \in \mathcal{W}\subset \mathcal{H}_{HS}^{AB}$ is expressed in the basis $\{\ket{\Omega_{k,l}}\} \in \mathcal{H}^{AB}$ it becomes obvious that the only pure states in $\mathcal{W}$ lie at the vertices $\{P_{k,l}\}$ (diagonal $\rho$). Before we investigate the states within the magic simplex let us specify the subset of LMM states belonging to it. One requirement for the existence of a representative in $\mathcal{W}$ surely is that $\rho\in \mbox{LMM}$ must be decomposable into orthogonal Bell states. However, we now show why this is only a necessary and not a sufficient condition. Consider the unitary transformation $U=U_A \otimes U_B = \sum_s\ket{s'}\bra{s}\otimes\mathbbm{1}$ transforming a Bell state $\ket{\Omega_{0,0}}=\frac{1}{\sqrt{d}} \sum_s \ket{s} \otimes \ket{s}$ into another orthonormal Bell state $\ket{\Phi}=\frac{1}{\sqrt{d}} \sum_s \ket{s'} \otimes \ket{s}$. Orthogonality implies $TrU_A=0$ meaning the sum of all eigenvalues $U_A$ has to be zero. Without loss of generality, we can set one of the eigenvalues $1$ because we are free in choosing a global phase. This implies the rest of the eigenvalues for qubits and qutrits, while for $d\geq4$ there are various ways for complying $TrU_A=0$. The Weyl operators $W_{k,l}$ have the eigenvalues $e^{i2\pi b/d}$ with $b\in\{0,..,d-1\}$ and if a state is decomposable into certain Bell states their intertwiners must have the same eigenvalues in order to have a representative in $\mathcal{W}$.
\subsubsection{Symmetries and equivalences inside $\mathcal{W}$}
\label{equiv}
We focus on local (anti-)unitary transformations $U_A \otimes U_B$ mapping $\mathcal{W}$ onto itself. In particular we are interested in equivalence classes $\left[ \rho \right]=\{ \rho' \in \mathcal{W} | \rho'=U_A\otimes U_B \rho U_A^{\dagger} \otimes U_B^{\dagger}\}$ of states within the magic simplex having the same properties concerning separability and nonlocality. These equivalences can best be studied when the group structure of the Weyl operators and the concept of a finite discrete classical phase space are used. The Weyl operators originate from the quantization of classical kinematics where they are used for translations between discrete states in a phase space. Each point within this phase space corresponds to an index pair $(k,l)$, where $l\in\{0,..,d-1\}$ denotes the quantum number of the position and $k\in\{0,..,d-1\}$ the quantum number of the momentum.
\begin{figure}[htp]
\centering
 \setlength{\unitlength}{1.2bp}%
  \begin{picture}(133.94, 112.34)(0,0)
  \put(0,0){\includegraphics[scale=1.2]{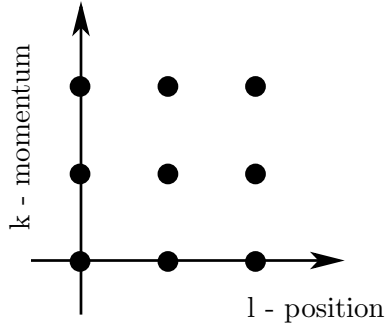}}
  \put(103.70,11.34){\fontsize{11.38}{13.66}\selectfont \makebox(0,0)[]{l - position\strut}}
  \put(11.34,65.76){\rotatebox{90.00}{\fontsize{11.38}{13.66}\selectfont \makebox(0,0)[]{k - momentum\strut}}}
  \end{picture}%
  \caption{Illustration of the finite discrete classical phase space for $d=3$.}
\end{figure}\\  
Due to the fact that the Weyl operators have been used as the intertwiners between the vertices $\{P_{k,l}\}$ of the simplex we can exploit this phase space structure for our investigations, i.e. to each point $(k,l)$ of the lattice in the phase space we assign a Bell state $P_{k,l}$. We know by construction that any Bell state $P_{k,l}$ can be mapped onto any $P_{k',l'}$ by use of a certain Weyl operator $U=W_{m,n}\otimes \mathbbm{1}$ with $k'=(k+m) \mbox{ mod } d$ and $l'=(l+n) \mbox{ mod } d$, meaning there are no restrictions on translations $\mathcal{T}$ in the phase space
\begin{align}
\mathcal{T}:\begin{pmatrix} k \\ l \end{pmatrix} \longrightarrow \begin{pmatrix} k' \\ l' \end{pmatrix} \ .
\end{align}
There are even more local unitary transformations mapping $\mathcal{W}$ onto itself such as $U_{\mathcal{R}}\otimes U_{\mathcal{R}}^*$ with
\begin{align}
U_{\mathcal{R}}=\frac{1}{d}\sum_{s,t=0}^{d-1} w^{-st}\ket{t}\bra{s} \ .
\end{align}
Under this transformation the Bell states $P_{k,l}$ become
\begin{align}
U_{\mathcal{R}}\otimes U_{\mathcal{R}}^* P_{k,l} U_{\mathcal{R}}^{\dagger}\otimes U_{\mathcal{R}}^T =P_{l,k} \ .
\end{align}
Thus, this is a \textbf{quarter rotation} around the origin in the phase space picture
\begin{align}
\mathcal{R}: \begin{pmatrix} k \\ l \end{pmatrix} \longrightarrow \begin{pmatrix} l \\ k \end{pmatrix} \ .
\end{align}
The next transformation we consider is $U_{\mathcal{V}}\otimes U_{\mathcal{V}}^*$ with
\begin{align}
U_{\mathcal{V}}=\sum_{s=0}^{d-1} w^{-s(s+d)/2} \ket{s} \bra{s} \ .
\end{align}
This affects the Bell states in the following way
\begin{align}
U_{\mathcal{V}}\otimes U_{\mathcal{V}}^* P_{k,l} U_{\mathcal{V}}^{\dagger}\otimes U_{\mathcal{V}}^T=P_{k+l,l}  \ .
\end{align}
In the phase space this is a \textbf{vertical shear}
\begin{align}
\mathcal{V}: \begin{pmatrix} k \\ l \end{pmatrix} \longrightarrow \begin{pmatrix} k+l \\ l \end{pmatrix} \ .
\end{align}
Another realisable mapping is the \textbf{vertical reflection} $\mathcal{S}$
\begin{align}
\mathcal{S}: \begin{pmatrix} k \\ l \end{pmatrix} \longrightarrow \begin{pmatrix} -k \\ l \end{pmatrix} \ .
\end{align}
The corresponding anti-unitary transformation on the Hilbert space $\mathcal{H}^{AB}$ is the tensorial product of local complex conjugation C
\begin{align}
\sum_{s=0}^{d-1} a_{s} \ket{s} \longrightarrow \sum_{s=0}^{d-1} a^*_{s} \ket{s} \ .
\end{align}
The composite application $C \otimes C$ of this anti-unitary transformation onto the $P_{k,l}$ yields the desired vertical reflection
\begin{align}
C\otimes C P_{k,l} C^{\dagger} \otimes C^{\dagger} =  P_{-k,l} \ .
\end{align}    
The three transformations $\mathcal{R}$,$\mathcal{V}$,$\mathcal{S}$ together with the translation $\mathcal{T}$ are the generating elements of an arbitrary phase space transformation of form (see \cite{BHNsimplex})
\begin{align}
\label{phasetrans}
\begin{pmatrix} k \\ l \end{pmatrix} \longrightarrow \begin{pmatrix} m & n \\ p&q \end{pmatrix} \begin{pmatrix} k \\ l \end{pmatrix} + \begin{pmatrix} j \\ r \end{pmatrix} \hspace{1.5cm} M=\begin{pmatrix} m & n \\ p&q \end{pmatrix} \ ,
\end{align}
with $det(M)=1$ or $det(M)=d-1$. For $det(M)=1$ the corresponding transformation acting on the Hilbert space $\mathcal{H}^{AB}$ is unitary, for $det(M)=d-1$ it is anti-unitary. Hence, further transformations as for instance horizontal shear, squeezing, horizontal or diagonal reflection with varying origins can be decomposed into $\mathcal{T}$,$\mathcal{R}$,$\mathcal{V}$ and $\mathcal{S}$ and a representation acting on $\mathcal{H}^{AB}$ can be constructed out of them. Transformations that cannot be written in the form (\ref{phasetrans}) or that do not obey $det(M)=1$ (or $d-1$) are excluded because they do not possess local (anti-)unitary representations. A detailed proof of this fact from a group theoretical point of view can be found in \cite{BHNsimplex}.\\
Let us briefly point out some major consequences of these equivalences. In most cases we study low dimensional sections of the simplex. These slices are mainly mixtures of particular $P_{k,l}$'s and the unity $\mathbbm{1}$ which is an equally weighted mixture of all Bell states $\mathbbm{1}=\sum_{k,l=0}^{d-1}P_{k,l}$ and can be regarded as uncolored noise. The transformation rules imply that all one parameter states $\rho=\frac{1-\alpha}{d^2} \mathbbm{1} + \alpha P_{k,l}$ with arbitrary $k,l\in\{0,..,d-1\}$ but same $\alpha$ have the properties in terms of separability and (non-)locality (Identity $\mathbbm{1}$ is mapped onto itself for all unitary transformations and the single Bell state $P_{k,l}$ can be translated freely). These are the so-called \textbf{isotropic states}. The same applies to two-parameter families of states of form $\rho=\frac{1-\alpha-\beta}{d^2} \mathbbm{1} + \alpha P_{k,l} + \beta P_{m,n}$ with arbitrary $k,l,m,n\in\{0,..,d-1\}$. We translate the first point $(k,l)$ to the origin $(0,0)$ and then bring the second point $(m-k, n-l)$ to $(0,1)$. This can be done by the matrix $M=\begin{pmatrix} n-l & k-m\\ 0 & q \end{pmatrix}$ with $q(n-l)=1=det(M)$. As a result we have found that such states with different $P_{k,l}$ and $P_{m,n}$ but same $\alpha$ and $\beta$ are equivalent and furthermore they are symmetric in $\alpha$ and $\beta$ (i.e. apply the same procedure but interchange the roles of the points). For three-parameter states $\rho=\frac{1-\alpha-\beta-\gamma}{d^2} \mathbbm{1} + \alpha P_{k,l} + \beta P_{m,n} + \gamma P_{p,q}$ the situation changes. We can proceed as before bringing two points to $(0,0)$ and $(0,1)$. If the third point was on a line $\{ (k,l),(m,n),(p,q)=(a(m-k)+k,a(n-l)+l)\}$ then it is now on the line with $k=0$ due to linearity of $M$, for instance for d=3 this point takes on (0,2). If it was not an element of this line it can be brought to $(1,0)$ with horizontal shear and vertical reflection without influencing the points on $k=0$. However, it cannot be brought to $k=0$ without transforming the other two points. This shows that not all three parameter states with same $\alpha, \beta$ and $\gamma$ necessarily possess the same attributes. We do not discuss cases with more than three points in detail but want to stress that all complete lines $\{ (k,l),(m, n), ..,((d-1)\cdot (m-k)+k,(d-1) \cdot (n-l)+l)\}$ are equivalent because they can all be mapped onto the line $\{(0,0),(0, 1),..,(0,d-1)\}$ by transforming two of the points onto $(0,0)$ and $(0,1)$ with the above stated method.
\begin{figure}[htp]
\centering
  \setlength{\unitlength}{1.2bp}%
  \begin{picture}(182.42, 154.54)(0,0)
  \put(0,0){\includegraphics[scale=1.2]{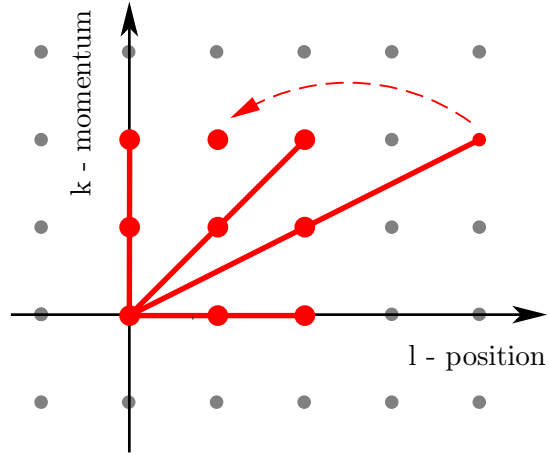}}
  \put(152.18,35.38){\fontsize{11.38}{13.66}\selectfont \makebox(0,0)[]{l - position\strut}}
  \put(28.42,115.49){\rotatebox{90.00}{\fontsize{11.38}{13.66}\selectfont \makebox(0,0)[]{k - momentum\strut}}}
  \end{picture}%
  \caption{Illustration of all possible complete lines through the point $(0,0)$ for $d=3$. Lines can be completed by points with $k,l \notin \{0,..,d-1\}$ because of the periodicity of the Weyl operators implying $P_{k,l}=P_{k+m \cdot d,l+n \cdot d}$ for all $n,m \in \mathbb{Z}$}
\end{figure}
\subsubsection{Geometry of separable and entangled states}
Theoretical strategies for determining the separable and entangled states within $\mathcal{W}$ are discussed and applied onto sections of the simplex for $d=3$. We begin with entangled states $\rho$ that can be revealed by the PPT criterion (see \S \ref{CPmaps}) and provide some simplifications for determining definiteness of operators $\rho^{T_A}$ that are inferred from the structure of the simplex $\mathcal{W}$. The explicit form of a projector $P_{k,l}$ is
\begin{align}
P_{k,l}=\frac{1}{d}\sum_{s,t=0}^{d-1}w^{k(s-t)}\ket{s-l,s}\bra{t-l,t} \ .
\end{align}
Therefore the elements of $\mathcal{W}$ are
\begin{align}
\rho=\frac{1}{d}\sum_{s,t,k,l=0}^{d-1}c_{k,l}w^{k(s-t)}\ket{s-l,s}\bra{t-l,t}
\end{align}
and partial transposition on subsystem A gives
\begin{align}
\rho^{T_A}=\frac{1}{d}\sum_{s,t,k,l=0}^{d-1}c_{k,l}w^{k(s-t)}\ket{t-l,s}\bra{s-l,t} \ .
\end{align}
We substitute the running index $l$ by $l=s+t-m$ which leads to the expression
\begin{align}
\rho^{T_A}&=\frac{1}{d}\sum_{s,t,k,m=0}^{d-1}c_{k,s+t-m}w^{k(s-t)}\ket{m-s,s}\bra{m-t,t} \nonumber \\
&=\sum_{m=0}^{d-1}\left[\frac{1}{d}\sum_{s,t,k=0}^{d-1}c_{k,s+t-m}w^{k(s-t)}\ket{m-s,s}\bra{m-t,t}\right] \nonumber \\
&=\sum_{m=0}^{d-1}B_m \ .
\end{align}
As we observe, each operator $B_m$ acts on a subspace $\mathcal{H}_{m}=\mathbb{C}^d$ spanned by the vectors $\{\ket{m-s} \otimes \ket{s} \}_{s=0,..,d-1}$. Thus, for determining definiteness of $\rho^{T_A}$ it suffices to consider the definiteness of the $d\times d$ matrices $B_m$ having the entries
\begin{align}
\left(B_m\right)_{s,t}=\frac{1}{d}\sum_{k=0}^{d-1}c_{k,s+t-m}w^{k(s-t)} \ .
\end{align}
Relations between the matrices $B_m$ lead to further simplifications of this problem. For example
\begin{align}
\left(B_{m-2}\right)_{s,t}=\frac{1}{d}\sum_{k=0}^{d-1}c_{k,s+t-m+2}w^{k(s-t)}=\left(B_{m}\right)_{s+1,t+1}
\end{align}
holds for even and odd $d$ and signifies that $B_{m-2}$ and $B_{m}$ are unitarily equivalent. Moreover, for odd $d$ all $B_m$ are unitarily equivalent which can be proven by using the periodicity (modulo $d$)
\begin{align}
\left(B_{m-1}\right)_{s,t}&=\frac{1}{d}\sum_{k=0}^{d-1}c_{k,s+t-m+1}w^{k(s-t)} \nonumber \\
&=\frac{1}{d}\sum_{k=0}^{d-1}c_{k,s+t-m+1+d}w^{k(s+(d+1)/2-t-(d+1)/2)} \nonumber \\
&=\left(B_m\right)_{s+(d+1)/2,t+(d+1)/2} \ .
\end{align}
Since unitarily equivalent matrices have the same eigenvalues it suffices to compute the eigenvalues of a single $B_m$ when the dimension $d$ is odd and two $B_m$ (one with odd $m$ and one with even $m$) when the dimension $d$ is even.\\
As we already know, we can exclude states $\rho$ with non-positive $\rho^{T_A}$ from being separable but positivity $\rho^{T_A}\geq0$ does not guarantee their separability. We now show which states $\rho \in \mathcal{W}$ are separable with certainty. Consider functions $k_n(x)$ and $l_n(x)$ with $x\in\{0,..,d-1\}$ defining complete lines $\{(k_n(0),l_n(0)),..,(k_n(d-1),l_n(d-1))\}$. We state that all states that are equally weighted mixtures of states forming a complete line are separable (in particular these are the outermost separable states of $\mathcal{W}$ as is shown in \cite{NarnhoferWigner}).
\begin{align}
\label{linestates}
\lambda_{n}=\frac{1}{d}\sum_{x=0}^{d-1} P_{k_n(x),l_n(x)} \hspace{0.5cm} \in \  \mbox{SEP} \ .
\end{align}
We have to prove this for one particular complete line only, for instance $k(x)=x$ and $l=0$ because of their equivalence in terms of local unitaries (\S \ref{equiv})
\begin{align}
\lambda&=\frac{1}{d}\sum_{x=0}^{d-1} P_{k=x,0} \nonumber \\
&=\frac{1}{d^2}\sum_{x,s,t=0}^{d-1} w^{x(s-t)}\ket{s,s}\bra{t,t} \ .
\end{align} 
Once again we use $\frac{1}{d} \sum_{x=0}^{d-1} w^{x(s-t)}=\delta_{s,t}$ and get
\begin{align}
\rho_{line}=\frac{1}{d}\sum_{s=0}^{d-1} \ket{s,s}\bra{s,s} \ ,
\end{align}
which is obviously separable. All possible convex combinations of all line states $\lambda_n$ (\ref{linestates}) form the \textbf{kernel polytope (KP)}
\begin{align}
\mbox{KP} = \left\{ \rho \in \mathcal{W} \ | \  \rho=\sum_n c_n \lambda_n  \ , \  c_n \geq 0 \  , \  \sum_n c_n=1  \right\} \ ,
\end{align}
a subset of separable states of PPT$\cap \mathcal{W}$\footnote{When we speak of PPT as a set we mean PPT=$\{\rho \in\mathcal{H}_{HS}^{AB} \ | \ Tr(\rho)=1, \rho \geq 0, \rho^{T_A} \geq 0$ \} }.\\
At this point the separability of states within the kernel polytope and the non-separability of NPPT states are ensured. For the remaining states that are PPT but do not lie within the kernel polytope we have to construct optimal entanglement witnesses $W_{opt.}$ if we want to completely clarify the question of separability. As we already know for all states $\rho$ that lie on the boundary of the convex set of separable states $\partial$SEP there exists an optimal entanglement witness with $Tr(W_{opt.}\rho)=0$ (\S \ref{witnesses}). To determine the boundary states of SEP$\cap \mathcal{W}$ and their witnesses the symmetry of the simplex is of great help. Consider a state $\rho$ which is invariant under the symmetry group $\mathcal{G}$ of unitary or anti-unitary operators $V_g \in \mathcal{G}$, meaning $V_g \rho V_g^{-1}=\rho$ for all $g$. 
Thus,
\begin{align}
Tr(W\rho)=Tr(W V_g^{-1} V_g \rho V_g^{-1} V_g)=Tr(V_g W V_g^{-1} V_g \rho V_g^{-1})=Tr(V_g W V_g^{-1} \rho) \nonumber
\end{align}
shows that the symmetries $\mathcal{G}$ of a state $\rho$ are reflected in its witnesses. In other words we can restrict the search for witnesses on $\mathcal{G}$-invariant operators $V_g W V_g^{-1}=W$ (for all $g$). This is a significant restriction on the form of the witness when the regarded state $\rho$ is of high symmetry. In our case where all states belong to the magic simplex $\mathcal{W}$ all states are invariant under the symmetry group $\mathcal{G}=\{2P_{k,l}-\mathbbm{1}\}_{k,l=0,..,d-1}$ and the most general form of an $\mathcal{G}$-invariant $W$ is
\begin{align}
W=\sum_{k,l} \kappa_{k,l} P_{k,l} \hspace{2cm} \kappa_{k,l} \in \mathbb{R} \ .
\end{align}
According to (\ref{poswitness}) W must have non-negative expectation values for all product states
\begin{align}
& \bra{\psi}\otimes \bra{\eta} W \ket{\psi}\otimes \ket{\eta} \nonumber \\
= & \frac{1}{d} \sum_{k,l,s,t=0}^{d-1} \kappa_{k,l} \bra{\psi} \otimes \bra{\eta} W_{k,l} \otimes \mathbbm{1} \ket{s} \otimes \ket{s} \bra{t} \otimes \bra{t} W^{\dagger}_{k,l} \otimes \mathbbm{1} \ket{\psi} \otimes \ket{\eta} \nonumber 
\\
= & \frac{1}{d} \sum_{k,l,s,t=0}^{d-1} \kappa_{k,l} \bra{\psi}W_{k,l} \ket{s} \left\langle \eta | s \right\rangle  \left\langle t | \eta \right\rangle \bra{t} W^{\dagger}_{k,l} \ket{\psi}  \nonumber\\
= & \frac{1}{d}  \bra{\psi}  \left[ \sum_{k,l=0}^{d-1} \kappa_{k,l}W_{k,l} \ket{\phi} \bra{\phi} W^{\dagger}_{k,l}  \right] \ket{\psi}  \nonumber \\
= & \frac{1}{d}  \bra{\psi}  M_{\phi} \ket{\psi} \geq 0 \hspace{2cm} \forall \ket{\psi} , \ket{\phi} \in \mathbb{C}^{d} \ .
\end{align}
Here we have introduced a matrix $M_{\phi}$ that depends on the vector $\ket{\phi}=\sum_{s=0}^{d-1}  \left\langle \eta | s \right\rangle \ket{s}$ which is merely an anti-unitary transformation of $\ket{\eta}$ (complex conjugation of the expansion coefficients of $\ket{\eta}$ in the basis $\{ \ket{s} \}$). Consequently, if non-negativity of $W$ holds for all $\ket{\phi}$ then it also holds for all $\ket{\psi}$ and vice versa. Iff $W$ is not only an entanglement witness but also optimal, then there exists a product state $\ket{\psi'} \otimes \ket{\eta'} \in \mathcal{H}^{AB}$ so that
\begin{align}
\bra{\psi'} \otimes \bra{\eta'} W \ket{\psi'} \otimes \ket{\eta'}=\frac{1}{d}  \bra{\psi'}  M_{\phi'} \ket{\psi'} = 0 \ .
\end{align}
Since $M_{\phi}$ is non-negative the eigenvalue for $\ket{\psi'}$ must be zero and if this is the case then $W$ is an optimal entanglement witness for $\rho=\ket{\psi',\eta'}\bra{\psi',\eta'}$ and all incoherent superpositions that are compatible with the symmetry of $W$
\begin{align}
\rho_{\mathcal{G}}=\sum_g c_g V_g  \ket{\psi',\eta'}\bra{\psi',\eta'} V_g^{-1}   \hspace{1.5cm} c_g \geq 0 \ , \ \sum_g c_g=1 \ .
\end{align}
It follows that the boundary of SEP$\cap \mathcal{W}$ is determined by the innermost states of the set $\mathcal{W}$ obeying $Tr(W\rho)=0$, where $W$ is an operator $W=\sum_{k,l} \kappa_{k,l} P_{k,l}$ whose associated non-negative matrices $M_{\phi}$ have at least one vanishing eigenvalue, i.e. $det(M_{\phi})=0$. Finding those innermost states is still a very difficult task even though symmetries have narrowed down the search for their optimal witnesses and in many cases one must perform a numerical variation of the parameters $\{\kappa_{k,l}\}$ and the vector $\ket{\phi}$. Analytical solutions can be obtained for states $\rho$ with further symmetries. For instance, if we intend to find the boundary of the one-parameter family of states 
\begin{align}
\rho=\frac{1-\alpha}{d^2}\mathbbm{1}+\alpha P_{0,0} \ ,
\end{align}
which is invariant under all phase space transformations except translations, we can restrict the search for an optimal witness on $W=a\mathbbm{1}+bP_{0,0}$ having the same symmetries. For this witness the associated matrix $M_{\phi}$ is 
\begin{align}
M_{\phi}=da\mathbbm{1}+b\ket{\phi}\bra{\phi} \ .
\end{align}
Here the eigenvalues do not depend on $\ket{\phi}$ and are $da+b$, $da$ and $da$ so one must choose $a>0$ and $b=-da$ in order to get an optimal $W$ ($det(M_{\phi})=0$). The state on the boundary of SEP is then given by
\begin{align}
Tr(W\rho)&=a Tr\left[ (\mathbbm{1}-dP_{0,0})(\frac{1-\alpha}{d^2}\mathbbm{1}+\alpha P_{0,0})\right] \nonumber \\
				&= a \left[ \frac{d-1-\alpha(d^2-1)}{d}  \right]\stackrel{!}{=}0 \ ,
\end{align} 
which is achieved for $\alpha=\frac{1}{d+1}$. It follows that all isotropic states
$
\rho=\frac{1-\alpha}{d^2}\mathbbm{1}+\alpha P_{p,q}
$
have this bound for $\alpha$ due to their equivalence by local unitaries (see \S \ref{equiv}). Their optimal entanglement witnesses $W=a\mathbbm{1}-daP_{p,q}$ with $a>0$ define the enclosure polytope. For an arbitrary state $\rho\in \mathcal{W}$ we compute
\begin{align}
Tr(W\rho)&=aTr\left[(\mathbbm{1}-dP_{p,q})(\sum_{k,l=0}^{d-1}c_{k,l}P_{k,l})\right] \nonumber \\
				&=a(-dc_{p,q}+\sum_{k,l}^{d-1}c_{k,l}) \nonumber \\
				&=a(-dc_{p,q}+1) \ .
\end{align} 
This implies that any $\rho$ that has at least one component $c_{k,l}> \frac{1}{d}$ is detected to be entangled by one of the isotropic witnesses $W=a\mathbbm{1}-daP_{p,q}$. Consequently, separable states lie within the so-called \textbf{enclosure polytope (EP)} defined by
\begin{align}
\mbox{EP} = \left\{ \rho \in \mathcal{W} \ | \  \frac{1}{d} \geq   c_{k,l} \geq 0 \ \ \forall c_{k,l}  \right\} \ .
\end{align}
We now study two and three dimensional sections of the simplex for $d=3$ (qutrits) in order to illustrate regions of separable and entangled states that result from these concepts. The first family of states we consider is a mixture of two Bell states which are all locally unitarily equivalent to $\rho=\frac{1-\alpha-\beta}{9}\mathbbm{1}+\alpha P_{0,0}+\beta P_{1,0}$. This means we have $c_{0,0}=\frac{1-\alpha-\beta}{9}+\alpha$ , $c_{1,0}=\frac{1-\alpha-\beta}{9}+\beta$ and all other components are $c_{k,l}=\frac{1-\alpha-\beta}{9}$ which implies $\alpha \geq \frac{\beta-1}{8}$, $\beta \geq \frac{\alpha-1}{8}$ and $\beta \leq 1 - \alpha$ for positivity of $\rho$. We obtain the boundary of PPT by setting $det(B_0)=0$ which yields $(2 \alpha + 2 \beta + 1) (8 \alpha^2 + 8 \beta^2 - 11 \beta \alpha + 2 \alpha + 2 \beta - 1)=0$ (details on this and other computations regarding $\rho$ can be found in the Appendix \ref{compdet1}). Due to invariance of $\rho$ under horizontal reflection we restrict on entanglement witnesses of form $W=\lambda \frac{1}{3} \mathbbm{1} + a P_{0,0} + b P_{1,0} + c P_{2,0}$. As it has been shown in this section, line states for example $\rho_{line}=\frac{1}{3}(P_{0,1} + P_{1,1} +  P_{2,1})$ are separable. Hence, $Tr(W\rho_{line})=\frac{\lambda}{3} \geq0$ must always be valid and therefore $\lambda$ cannot be negative. When we set $\lambda=1$ (which only fixes the scaling of the witness) the associated matrix $M_{\phi}$ becomes $M_{\phi}=\mathbbm{1}+a W_{0,0}\ket{\phi}\bra{\phi}W^{\dagger}_{0,0} + b W_{1,0}\ket{\phi}\bra{\phi}W^{\dagger}_{1,0} + c W_{2,0}\ket{\phi}\bra{\phi}W^{\dagger}_{2,0}$. A numerical search for solutions of $det(M_{\phi})=0$ by varying the parameters $a,b,c$ and the vector $\ket{\phi}$ was done in $\cite{BHNsimplexqutrits}$. It is shown in \cite{BeatrixUNPUB} that there exist optimal witnesses $W_{opt.}$ so that $Tr(W_{opt.}\rho)=0$ yields the boundaries $4 \alpha^2-5 \alpha+40 \beta^2+(17 \alpha-14) \beta+1=0$ and $4 \beta^2-5 \beta+40 \alpha^2+(17 \beta-14) \alpha+1=0$. Graphically this is illustrated in the following figures.

\begin{figure}[H]
\centering
\includegraphics[scale=0.45]{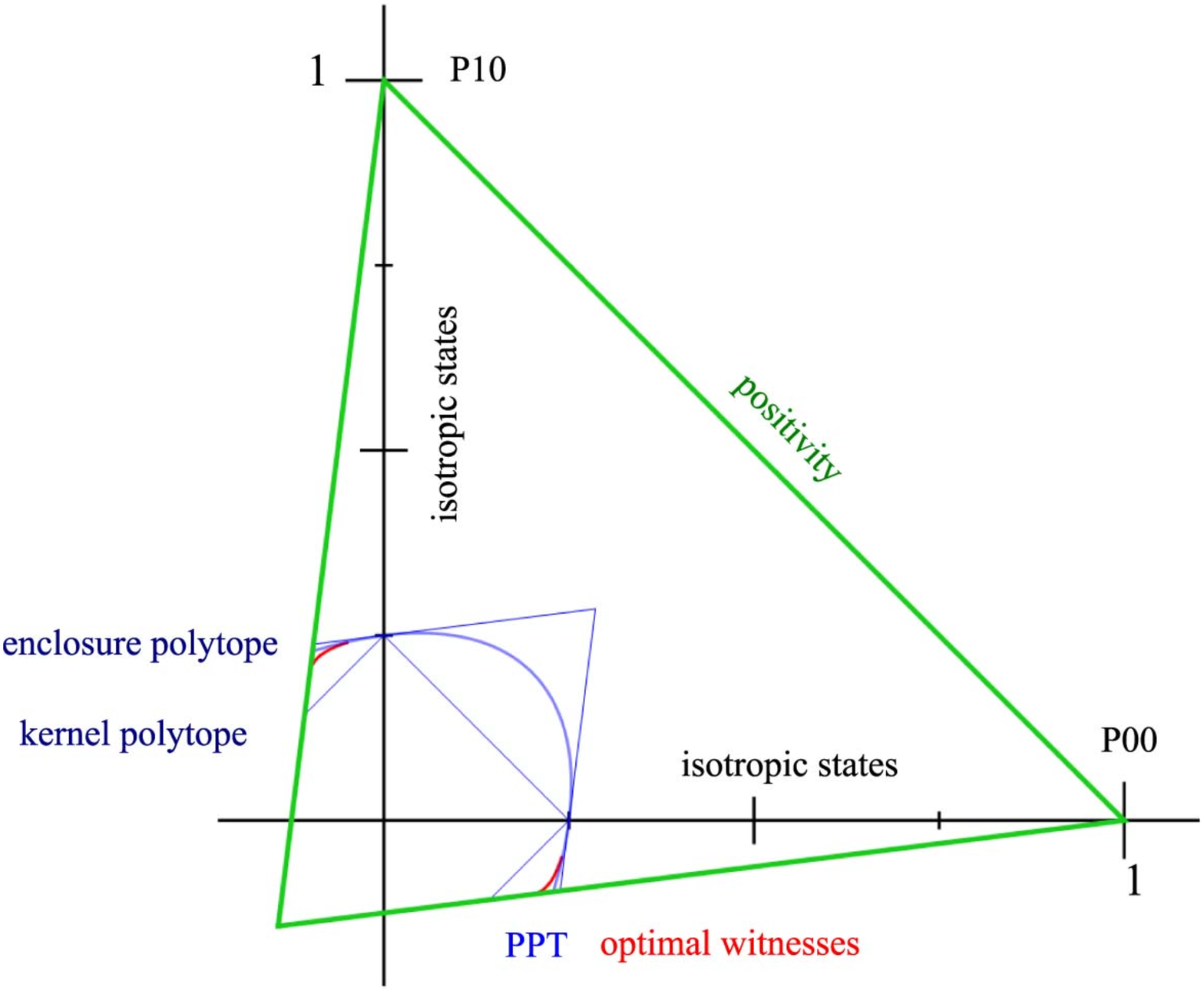}
\setlength{\unitlength}{1bp}%
  \begin{picture}(0, 0)(0,0)
\put(-10,40){\fontsize{11.38}{13.66}\selectfont \makebox(0,0)[]{$\alpha$ \strut}}
\put(-280,320){\fontsize{11.38}{13.66}\selectfont \makebox(0,0)[]{$\beta$ \strut}}
\end{picture}
  \caption{Illustration of the state $\rho=\frac{1-\alpha-\beta}{9}\mathbbm{1}+\alpha P_{0,0}+\beta P_{1,0}$. All physical states lie within the green triangle which represents the border of positivity. The blue lines correspond to the enclosure polytope (the outer one) and the kernel polytope (the inner one). PPT states lie within the blue ellipse and there is also a small region of bound entanglement (region between the red curve given by optimal witnesses and the PPT boundary) }
\end{figure}

\begin{figure}[H]
\centering
\includegraphics[scale=0.45]{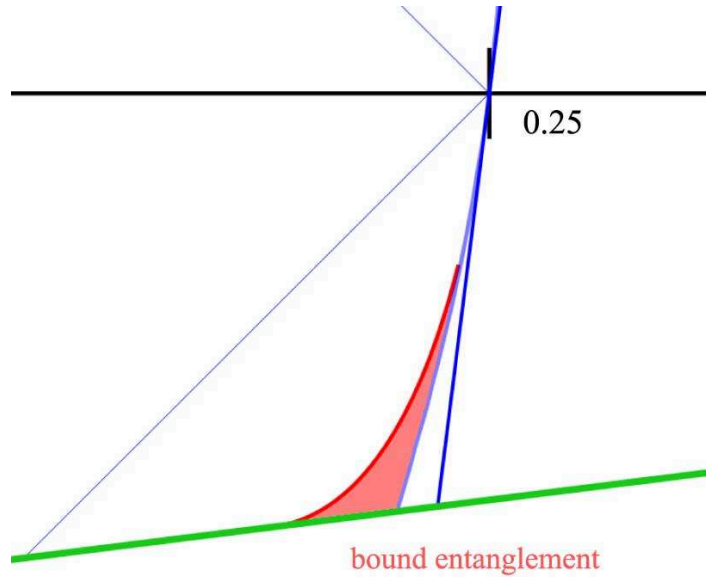}
  \caption{Enlarged illustration of the region of the state $\rho=\frac{1-\alpha-\beta}{9}\mathbbm{1}+\alpha P_{0,0}+\beta P_{1,0}$ where bound entanglement appears (filled red) }
\end{figure}

\newpage
\noindent
Next, we consider three-parameter families of states of form $\rho=\frac{1-\alpha-\beta-\gamma}{9} \mathbbm{1} + \alpha P_{k,l} + \beta P_{m,n} + \gamma P_{p,q}$ with $k,l,m,n,p,q \in \{0,..,2\}$. In section \S \ref{equiv} it has been shown that for fixed parameters $\alpha, \beta, \gamma$ any state of this family is either locally unitarily equivalent to the state $\rho=\frac{1-\alpha-\beta-\gamma}{9} \mathbbm{1} + \alpha P_{0,0} + \beta P_{1,0} + \gamma P_{2,0}$ or $\rho=\frac{1-\alpha-\beta-\gamma}{9} \mathbbm{1} + \alpha P_{0,0} + \beta P_{1,0} + \gamma P_{0,1}$ depending on whether the index pairs $\{(k,l),(m,n),(p,q)\}$ form a line or not. In both cases positivity restricts $\alpha, \beta, \gamma$ to lie within the region $\alpha \geq \frac{\beta+\gamma-1}{8}$ (and all parameter permutations of this term) and $\gamma \leq 1-\alpha-\beta$. The PPT boundary for states on a line reads (see Appendix \ref{compdet2}) 
\begin{align*}
(2 \alpha + 2 \beta + 2 \gamma + 1) (8 \alpha^2 + 8 \beta^2 + 8 \gamma^2 + 2 \alpha  + 2 \beta + 2 \gamma - 11 \beta \alpha  -    11 \alpha \gamma - 11 \beta \gamma  - 1)=0
\end{align*}
and for states off a line we get (see Appendix \ref{compdet3})
\begin{align*}
&-16 \alpha^3-16 \beta^3-16 \gamma^3
+6 \beta \alpha^2 + 6 \gamma \alpha^2+6 \gamma^2 \alpha+6 \beta^2 \alpha+6 \beta^2 \gamma+6 \beta \gamma^2\\
&-12 \alpha^2-12 \beta^2-12 \gamma^2
+3 \beta \alpha+3 \gamma \alpha+3 \beta \gamma
-15 \beta \gamma \alpha
+1=0 \ .
\end{align*}
Like in the previous case, the state $\rho=\frac{1-\alpha-\beta-\gamma}{9} \mathbbm{1} + \alpha P_{0,0} + \beta P_{1,0} + \gamma P_{2,0}$ is invariant under horizontal reflection and for this reason the search for optimal witnesses can once again be restricted on $W= \frac{1}{3} \mathbbm{1} + a P_{0,0} + b P_{1,0} + c P_{2,0}$ with $M_{\phi}=\mathbbm{1}+a W_{0,0}\ket{\phi}\bra{\phi}W^{\dagger}_{0,0} + b W_{1,0}\ket{\phi}\bra{\phi}W^{\dagger}_{1,0} + c W_{2,0}\ket{\phi}\bra{\phi}W^{\dagger}_{2,0}$. In \cite{BeatrixUNPUB} a choice for the parameters $a,b,c$ in compliance with the constraints for optimal witnesses was found that yields
\begin{align*}
40 \alpha^2+(17 \beta+17 \gamma-14) \alpha+4 \beta^2&+\gamma (4 \gamma-5)-\beta (19 \gamma+5)+1=0\\
\mbox{and permutations: \ } &(\alpha\leftrightarrow\beta), (\alpha\leftrightarrow\gamma),(\beta\leftrightarrow\gamma)
\end{align*}
for the boundary $Tr(W_{opt.}\rho)=0$. Unfortunately, for states off a line there exists no such solution because obtaining it is much more difficult due to the fact that all nine parameters $\kappa_{k,l}$ of $W=\sum_{k,l} \kappa_{k,l} P_{k,l}$ have to be taken into account because the state has fewer symmetries. Regardless of this, for both states the boundaries of positivity and PPT are graphically illustrated in the following figures.
\newpage
\begin{figure}[H]
\centering
\includegraphics[scale=0.47]{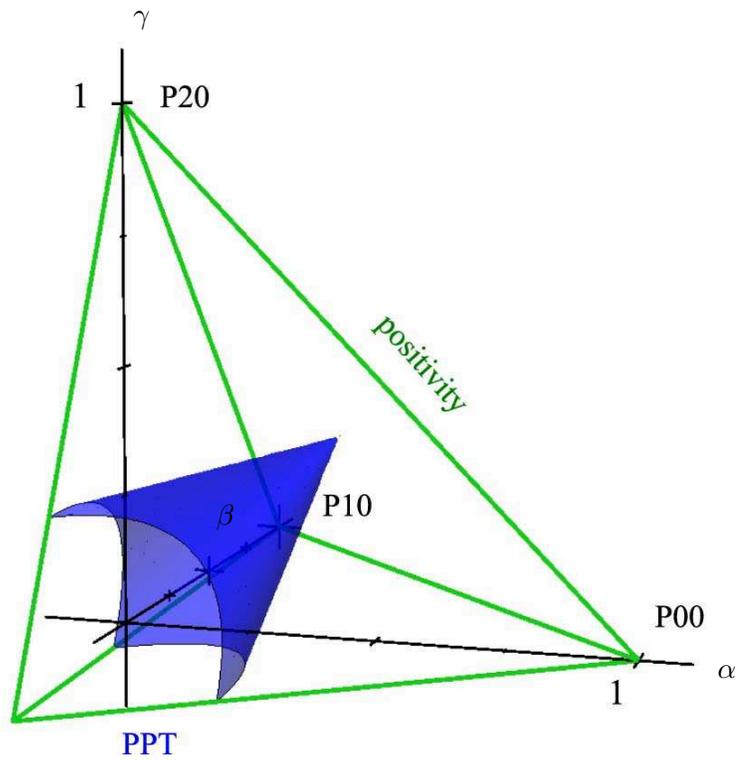}
\setlength{\unitlength}{1.05bp}%
  \begin{picture}(0, 0)(0,0)
\put(0,30){\fontsize{11.38}{13.66}\selectfont \makebox(0,0)[]{$\alpha$ \strut}}
\put(-210,265){\fontsize{11.38}{13.66}\selectfont \makebox(0,0)[]{$\gamma$ \strut}}
\put(-180,85){\fontsize{11.38}{13.66}\selectfont \makebox(0,0)[]{$\beta$ \strut}}
\end{picture}
  \caption{Illustration of the state $\rho=\frac{1-\alpha-\beta-\gamma}{9} \mathbbm{1} + \alpha P_{0,0} + \beta P_{1,0} + \gamma P_{2,0}$. Physical states lie within the green tetrahedron (positivity). The boundary of PPT states is a cone (blue) and thus all states beyond this surface are entangled. The tip of the cone touches the surface of positivity at $\alpha=\beta=\gamma=\frac{1}{3}$ illustrating the separability of the line state $\rho_{line}$}
\end{figure}
\begin{figure}[H]
\centering
\includegraphics[scale=0.47]{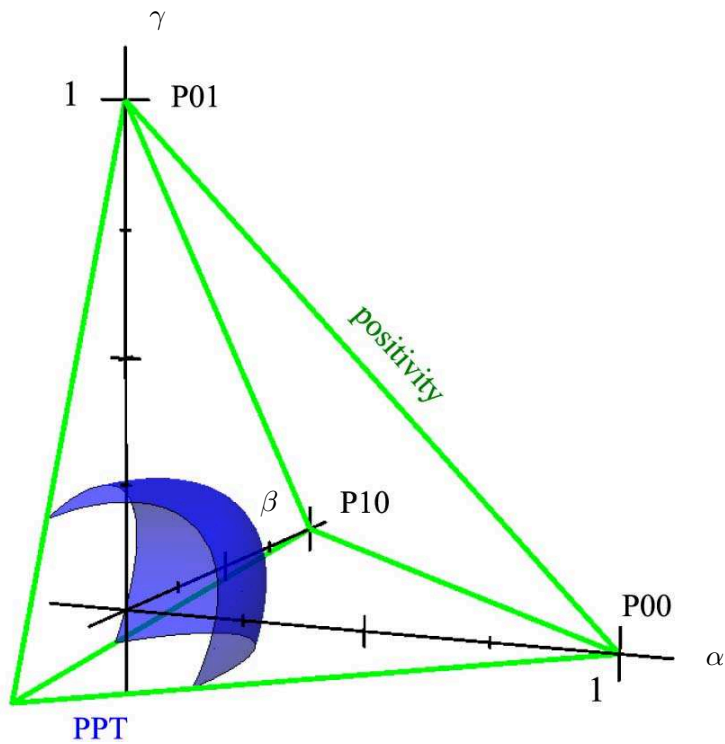}
\setlength{\unitlength}{1.05bp}%
  \begin{picture}(0, 0)(0,0)
\put(0,30){\fontsize{11.38}{13.66}\selectfont \makebox(0,0)[]{$\alpha$ \strut}}
\put(-200,260){\fontsize{11.38}{13.66}\selectfont \makebox(0,0)[]{$\gamma$ \strut}}
\put(-160,85){\fontsize{11.38}{13.66}\selectfont \makebox(0,0)[]{$\beta$ \strut}}
\end{picture}
  \caption{Illustration of the state $\rho=\frac{1-\alpha-\beta-\gamma}{9} \mathbbm{1} + \alpha P_{0,0} + \beta P_{1,0} + \gamma P_{0,1}$. Physical states lie within the green tetrahedron (positivity). The surface of the PPT state region is a complex geometric object (blue surface)}
\end{figure}

\subsubsection{Geometry of non-local states}
Our purpose is to determine the set of states $\rho$ of the magic simplex $\mathcal{W}$ that violate the CGLMP inequality. For any state of this kind there exists a Bell operator $\mathcal{B}_{I_d}$ such that $I_d=Tr(\rho \mathcal{B}_{I_d})>2$ and thus a state $\rho$ for which the inequality $\max_{\mathcal{B}_{I_d}}Tr(\mathcal{B}_{I_d}) \leq 2$ holds is an element of the complementary set. This means that the complement and thereof the set itself can be determined by use of the optimisation procedure that has been introduced in \S \ref{numopt}. In general, it is computationally intensive to determine the boundary with high precision because states of $\mathcal{W}$ have to be parameterised and varied until a required precision $\Delta I$ of $\max I_d=\max_{\mathcal{B}_{I_d}}Tr(\mathcal{B}_{I_d}\rho) = 2 \pm \Delta I$ is reached. However, in the case when for a given state $\mu \in \mathcal{W}$ the maximal value of $\max I_d(\mu)=\max_{\mathcal{B}_{I_d}}Tr(\mathcal{B}_{I_d}\mu)$ is known with high accuracy it is not necessary to perform further numerical investigations in order to obtain the state on the boundary for the family $\rho=\frac{1-a}{d^2}\mathbbm{1}+a\mu$ because of the relation $Tr(\mathcal{B}_{I_d})=0$ (see \S \ref{CGLMPder}),
\begin{align}
\max_{\mathcal{B}_{I_d}}Tr(\mathcal{B}_{I_d}\rho)&=\max_{\mathcal{B}_{I_d}}Tr(\mathcal{B}_{I_d}\left[\frac{1-a}{d^2}\mathbbm{1}+a\mu\right]) \nonumber \\
&= \max_{\mathcal{B}_{I_d}}Tr(\mathcal{B}_{I_d}a\mu) \nonumber \\
&=a \max I_d(\mu) \ .
\end{align}
Hence, $\rho=\frac{1-a}{d^2}\mathbbm{1}+a\mu$ lies on the boundary of (non-)locality for $a=2/\max I_d(\mu)$. For our numerical investigations of the three families of states $\rho=\frac{1-\alpha-\beta}{9} \mathbbm{1} + \alpha P_{0,0} + \beta P_{1,0}$, $\rho=\frac{1-\alpha-\beta-\gamma}{9} \mathbbm{1} + \alpha P_{0,0} + \beta P_{1,0} + \gamma P_{2,0}$ and $\rho=\frac{1-\alpha-\beta-\gamma}{9} \mathbbm{1} + \alpha P_{0,0} + \beta P_{1,0} + \gamma P_{0,1}$ this fact is of great help because out of any numerical obtained value $\max I_3(\rho)$ we can derive a state on the boundary. More precisely, for a certain choice of $\alpha,\beta$ and $\gamma$\footnote{$\gamma$=0 for the first family $\rho=\frac{1-\alpha-\beta}{9} \mathbbm{1} + \alpha P_{0,0} + \beta P_{1,0}$} with resulting value $\max I_3(\rho)$ it implies that the state with the parameter values $\alpha_b=[2/ \max I_3(\rho)] \cdot \alpha$, $\beta_b=[2/\max I_3(\rho)] \cdot \beta$ and $\gamma_b=[2/\max I_3(\rho)] \cdot \gamma$ lies on the boundary. For instance, the subsequent illustrations of the boundaries of \\ (non-)locality were deduced from values $\max I_3(\rho)$ of states on the boundaries of positivity ($\alpha = \frac{\beta+\gamma-1}{8}$ (and all parameter permutations of this term) and $\gamma = 1-\alpha-\beta$). For the two-parameter family we calculated $\max I_3(\rho)$ for $60$ such equally spaced points and for each of the two three-parameter families we calculated $\max I_3(\rho)$ for $920$ of them.
\newpage
\begin{figure}[H]
\centering
\includegraphics[scale=0.47]{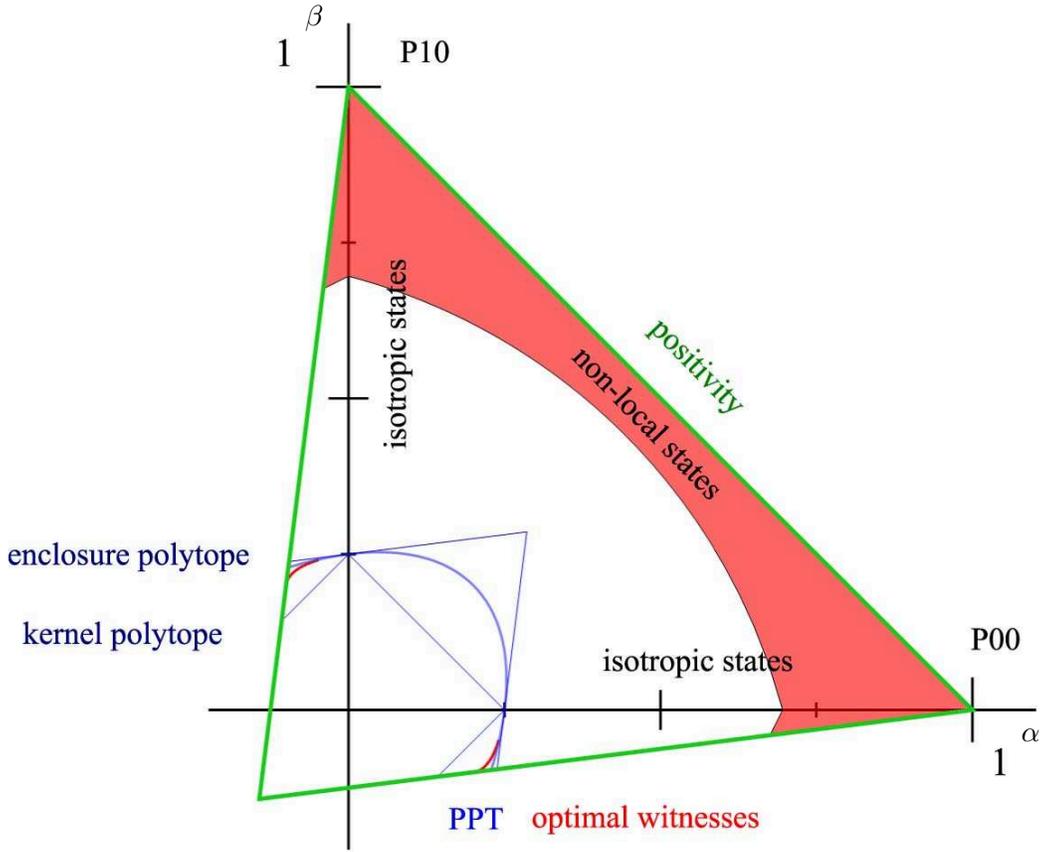}
\setlength{\unitlength}{1bp}%
  \begin{picture}(0, 0)(0,0)
\put(-10,50){\fontsize{11.38}{13.66}\selectfont \makebox(0,0)[]{$\alpha$ \strut}}
\put(-280,320){\fontsize{11.38}{13.66}\selectfont \makebox(0,0)[]{$\beta$ \strut}}
\end{picture}
  \caption{Illustration of the state $\rho=\frac{1-\alpha-\beta}{9}\mathbbm{1}+\alpha P_{0,0}+\beta P_{1,0}$. All physical states lie within the green triangle which represents the border of positivity. States $\rho$ in the red filled area violate the CGLMP inequality $(\max I_3(\rho)>2)$}
  \label{NL1}
\end{figure}
\noindent
The result of the calculations for the state $\rho=\frac{1-\alpha-\beta}{9}\mathbbm{1}+\alpha P_{0,0}+\beta P_{1,0}$ is illustrated in the above figure. The calculated points on the boundary are connected through lines and the area of non-local states is filled red. The boundary of (non-)locality seems to describe a circle for $\alpha,\beta > 0$ and a line if one of the parameters is negative, i.e. $\alpha<0$ or $\beta<0$. Suggestions on their specifications are given after the next figures illustrating the family $\rho=\frac{1-\alpha-\beta-\gamma}{9} \mathbbm{1} + \alpha P_{0,0} + \beta P_{1,0} + \gamma P_{2,0}$.
\newpage
\begin{figure}[H]
\centering
\includegraphics[scale=0.49]{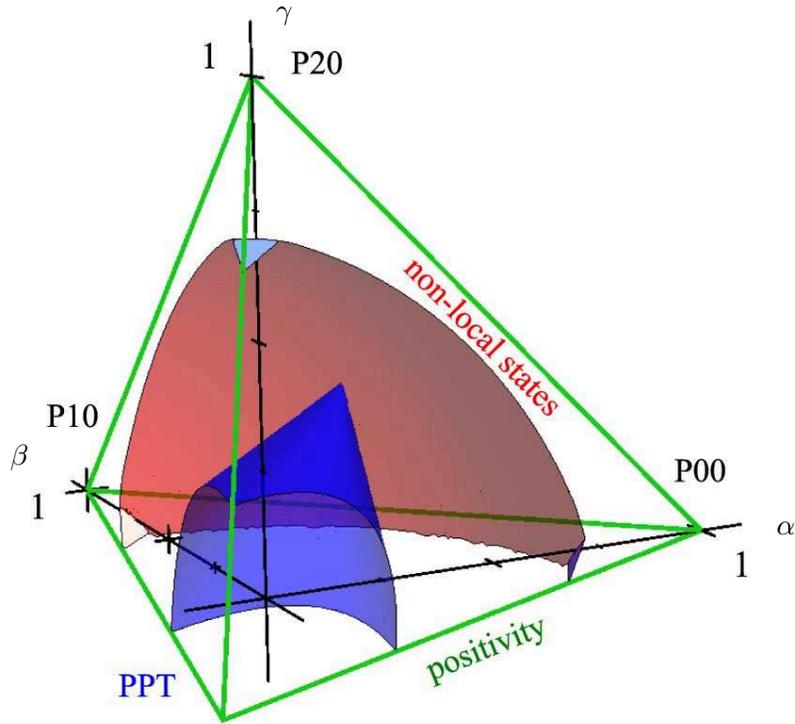}
\setlength{\unitlength}{1.05bp}%
  \begin{picture}(0, 0)(0,0)
\put(0,80){\fontsize{11.38}{13.66}\selectfont \makebox(0,0)[]{$\alpha$ \strut}}
\put(-180,265){\fontsize{11.38}{13.66}\selectfont \makebox(0,0)[]{$\gamma$ \strut}}
\put(-275,105){\fontsize{11.38}{13.66}\selectfont \makebox(0,0)[]{$\beta$ \strut}}
\end{picture}
  \caption{Illustration of the state $\rho=\frac{1-\alpha-\beta-\gamma}{9} \mathbbm{1} + \alpha P_{0,0} + \beta P_{1,0} + \gamma P_{2,0}$. States $\rho$ beyond the red/blue shaded surface violate the CGLMP inequality $(\max I_3(\rho)>2)$. }
\end{figure}

\begin{figure}[H]
\centering
\includegraphics[scale=0.49]{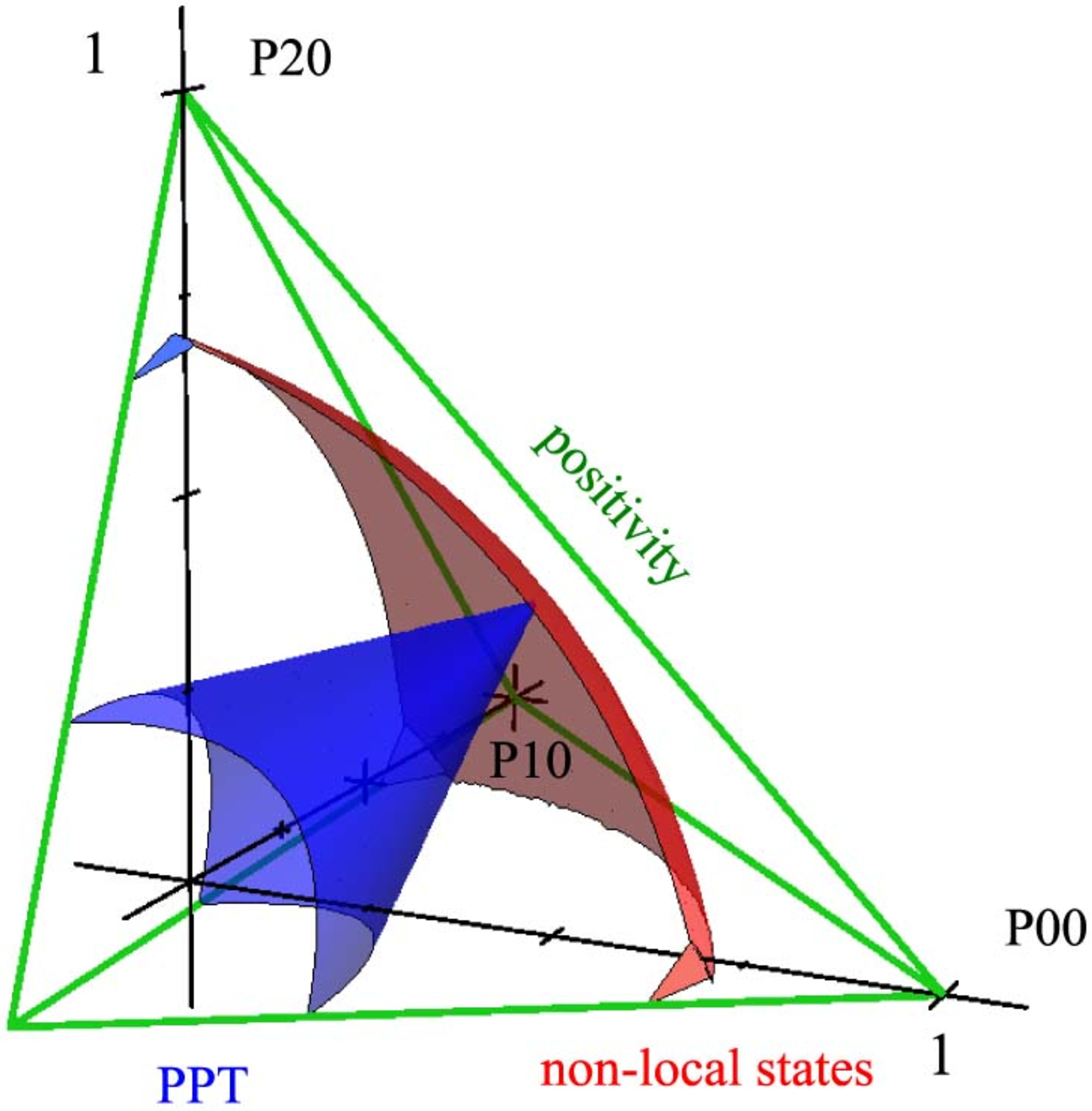}
\setlength{\unitlength}{1.05bp}%
  \begin{picture}(0, 0)(0,0)
\put(-30,30){\fontsize{11.38}{13.66}\selectfont \makebox(0,0)[]{$\alpha$ \strut}}
\put(-220,260){\fontsize{11.38}{13.66}\selectfont \makebox(0,0)[]{$\gamma$ \strut}}
\put(-170,105){\fontsize{11.38}{13.66}\selectfont \makebox(0,0)[]{$\beta$ \strut}}
\end{picture}
  \caption{Illustration of the state $\rho=\frac{1-\alpha-\beta-\gamma}{9} \mathbbm{1} + \alpha P_{0,0} + \beta P_{1,0} + \gamma P_{2,0}$ form a different perspective. States $\rho$ beyond the red/blue shaded surface violate the CGLMP inequality $(\max I_3(\rho)>2)$.}
\end{figure}

\newpage
\noindent
The result of the calculations for the state $\rho=\frac{1-\alpha-\beta-\gamma}{9} \mathbbm{1} + \alpha P_{0,0} + \beta P_{1,0} + \gamma P_{2,0}$ is illustrated in the figures on the previous page. The calculated points on the boundary tendentially describe a spherical surface that is intersected by planes in the region of negative parameters. Based on symmetries and the facts that the boundaries of positivity, PPT and (non-)locality meet at the point $\alpha=\beta=\gamma=\frac{1}{3}$ and that the suggested analytical ideal measurements (\ref{ONBBell}) yield the boundary parameter $\frac{1}{2} (6 \sqrt{3}-9)$ for the isotropic states, we were able to derive the radius $r=\frac{1}{156} (413 \sqrt{3}-558)$ and the center of the sphere $\alpha=\beta=\gamma=\frac{1}{156} (-361+186 \sqrt{3})$. These specifications coincide with the numerical data up to the order $10^{-5}$ and we believe that these discrepancies should decrease for better accuracy and precision goals of the numerical optimisation (see \S \ref{numopt}). We suppose that the intersecting planes are given by the functions $\gamma=\frac{1}{2}(\alpha+\beta+6 \sqrt{3}-9)$ (and permutations $\alpha\leftrightarrow\gamma$, $\beta\leftrightarrow\gamma$) because of compliance with the boundaries of the isotropic states and the numerical data (also up the order $10^{-5}$). It should be noted that the circle and line boundaries in figure \ref{NL1} can easily be obtained on the basis of these specifications because it solely illustrates the special case $\gamma=0$.\\
For the remaining family $\rho=\frac{1-\alpha-\beta-\gamma}{9} \mathbbm{1} + \alpha P_{0,0} + \beta P_{1,0} + \gamma P_{0,1}$ the geometric form of the boundary seems to be more complex and therefore we do not want to make any uncovered suggestions on the exact form. To get an impression of this, we have illustrated the raw data points in the following figures. Besides the complex shape of the boundary, we recognised an interesting peculiarity, namely, in contrast to the mixtures of states on a line where there is only one state $\alpha=\beta=\gamma=\frac{1}{3}$ on the boundary of positivity $1-\alpha-\beta-\gamma=0$ that does not violate the CGLMP inequality, here we have a whole region of local states for $1-\alpha-\beta-\gamma=0$. Moreover, in comparison the entire region of non-local states is smaller while the region of entangled states is larger. This can be seen as a further example for the diverging behaviour of these properties.
\newpage
\begin{figure}[H]
\centering
\includegraphics[scale=0.49]{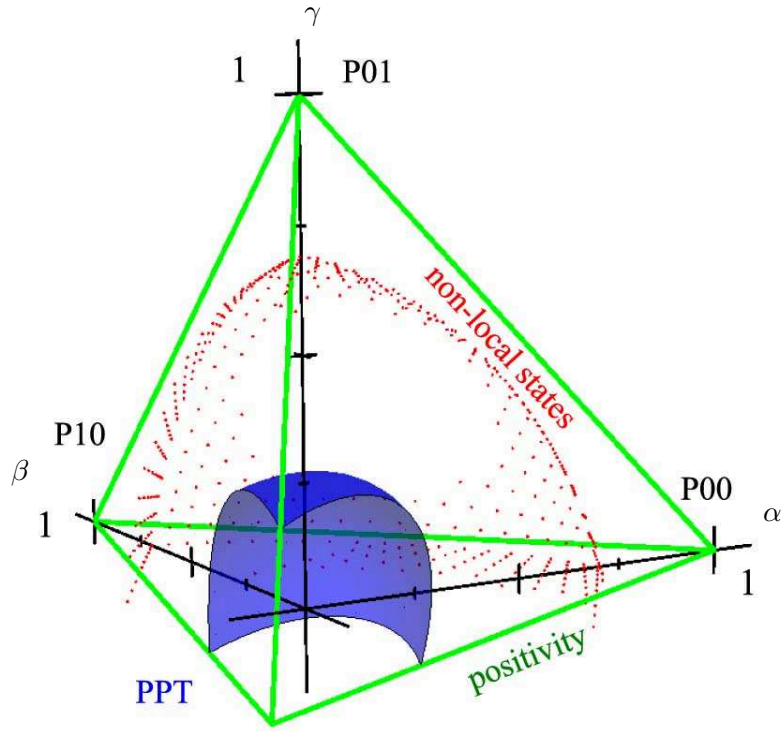}
\setlength{\unitlength}{1.05bp}%
  \begin{picture}(0, 0)(0,0)
\put(-15,85){\fontsize{11.38}{13.66}\selectfont \makebox(0,0)[]{$\alpha$ \strut}}
\put(-180,265){\fontsize{11.38}{13.66}\selectfont \makebox(0,0)[]{$\gamma$ \strut}}
\put(-285,100){\fontsize{11.38}{13.66}\selectfont \makebox(0,0)[]{$\beta$ \strut}}
\end{picture}
  \caption{Illustration of the state $\rho=\frac{1-\alpha-\beta-\gamma}{9} \mathbbm{1} + \alpha P_{0,0} + \beta P_{1,0} + \gamma P_{0,1}$. Red points denote states on the boundary of nonlocality $(\max I_3(\rho)=2)$.}
\end{figure}

\begin{figure}[H]
\centering
\includegraphics[scale=0.49]{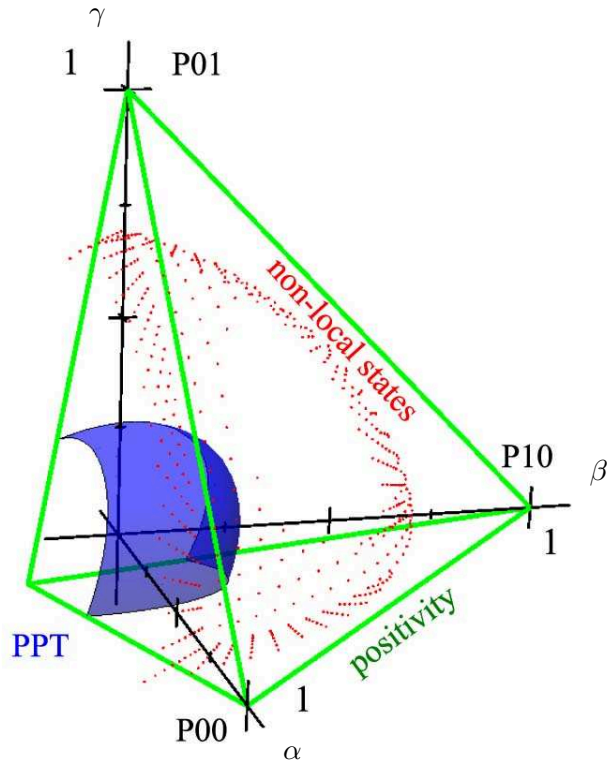}
\setlength{\unitlength}{1.05bp}%
  \begin{picture}(0, 0)(0,0)
\put(-130,10){\fontsize{11.38}{13.66}\selectfont \makebox(0,0)[]{$\alpha$ \strut}}
\put(-200,275){\fontsize{11.38}{13.66}\selectfont \makebox(0,0)[]{$\gamma$ \strut}}
\put(-20,110){\fontsize{11.38}{13.66}\selectfont \makebox(0,0)[]{$\beta$ \strut}}
\end{picture}
  \caption{Illustration of the state $\rho=\frac{1-\alpha-\beta-\gamma}{9} \mathbbm{1} + \alpha P_{0,0} + \beta P_{1,0} + \gamma P_{0,1}$. Red points denote states on the boundary of nonlocality $(\max I_3(\rho)=2)$.}
\end{figure}

\newpage
\subsection{Geometry of multipartite qubit systems (publication)}
In this section we present a paper that was published on \textit{Physical Review A 78} in collaboration with Beatrix Hiesmayr, Marcus Huber, Florian Hipp and Philipp Krammer. In this, we discuss a generalisation of the tetrahedron for multipartite qubit systems. We investigate separability, nonlocality and distillability with regard to a multipartite entanglement measure. In the context of this diploma thesis, this should be regarded as a further example of a geometric investigation of the state space. We emphasise that, because it contains several concepts for the study of multipartite systems that have not been introduced. Most of them should be self-explanatory or easily comprehensible with the knowledge gained in this thesis. Nonetheless, we recommend to read the cited publications.

\newpage
\hfill
\vfill

\newpage
\vspace*{1cm}
\includegraphics[scale=0.8]{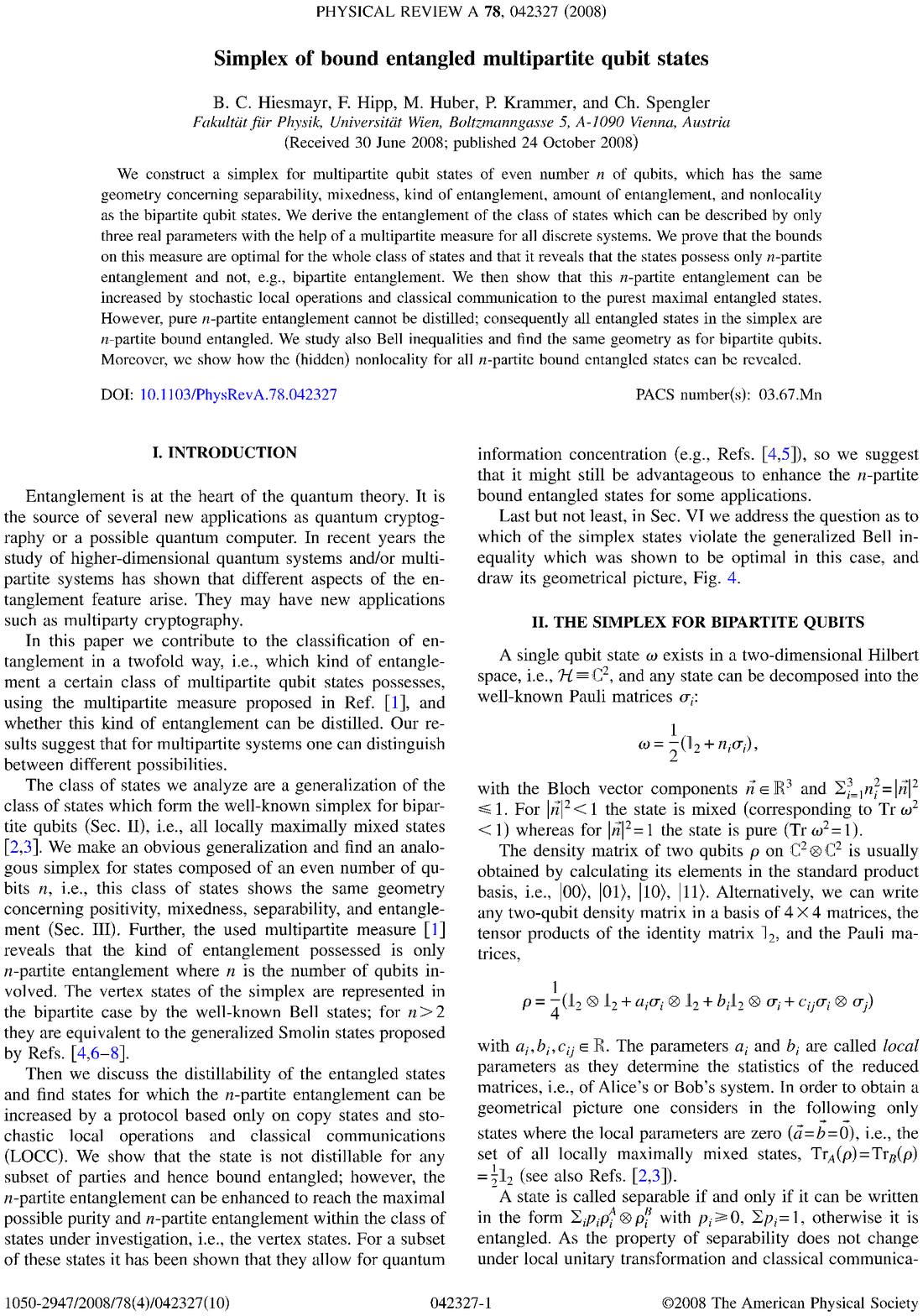}

\newpage
\vspace*{1cm}
\includegraphics[scale=0.8]{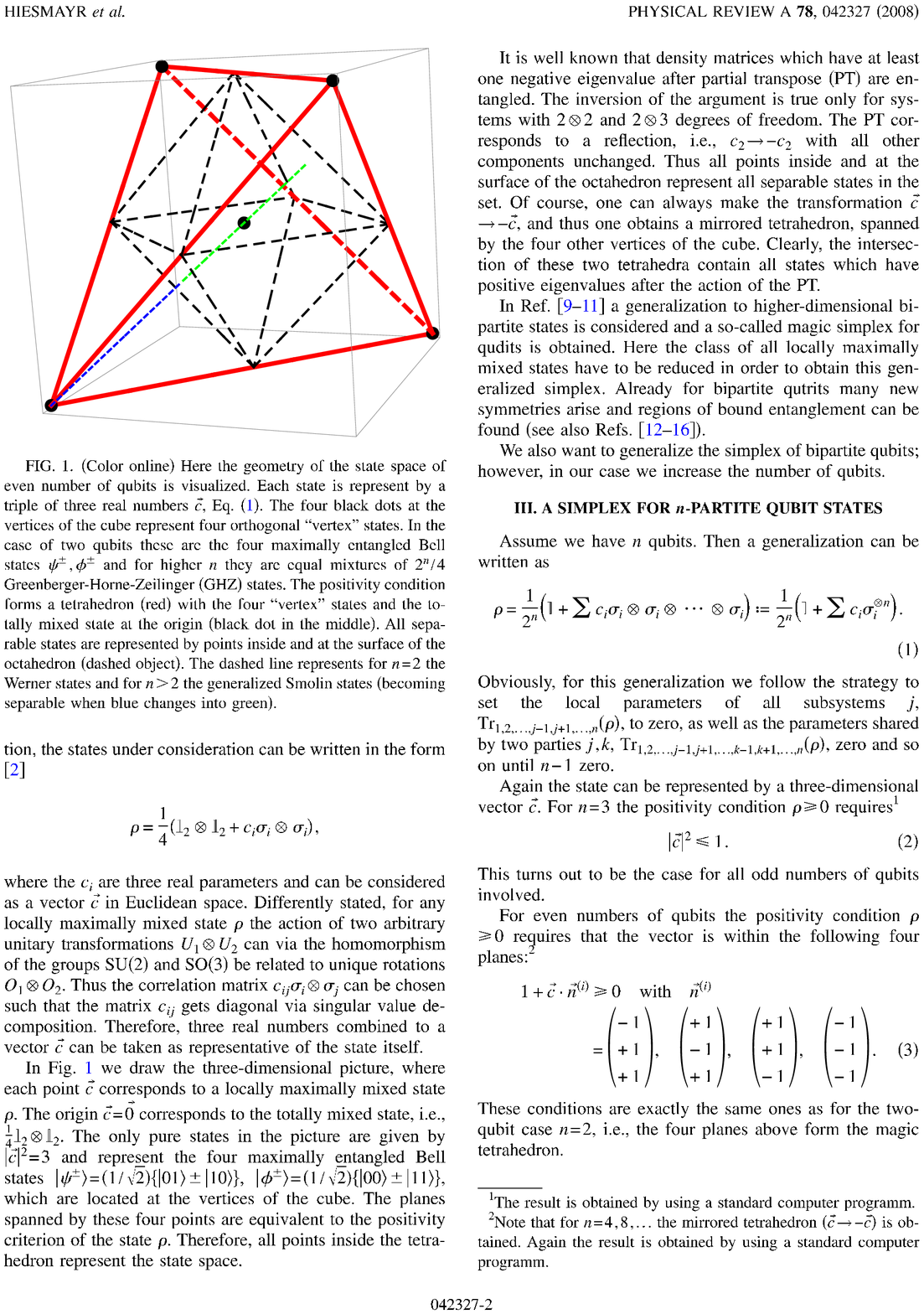}
\newpage
\vspace*{1cm}
\includegraphics[scale=0.8]{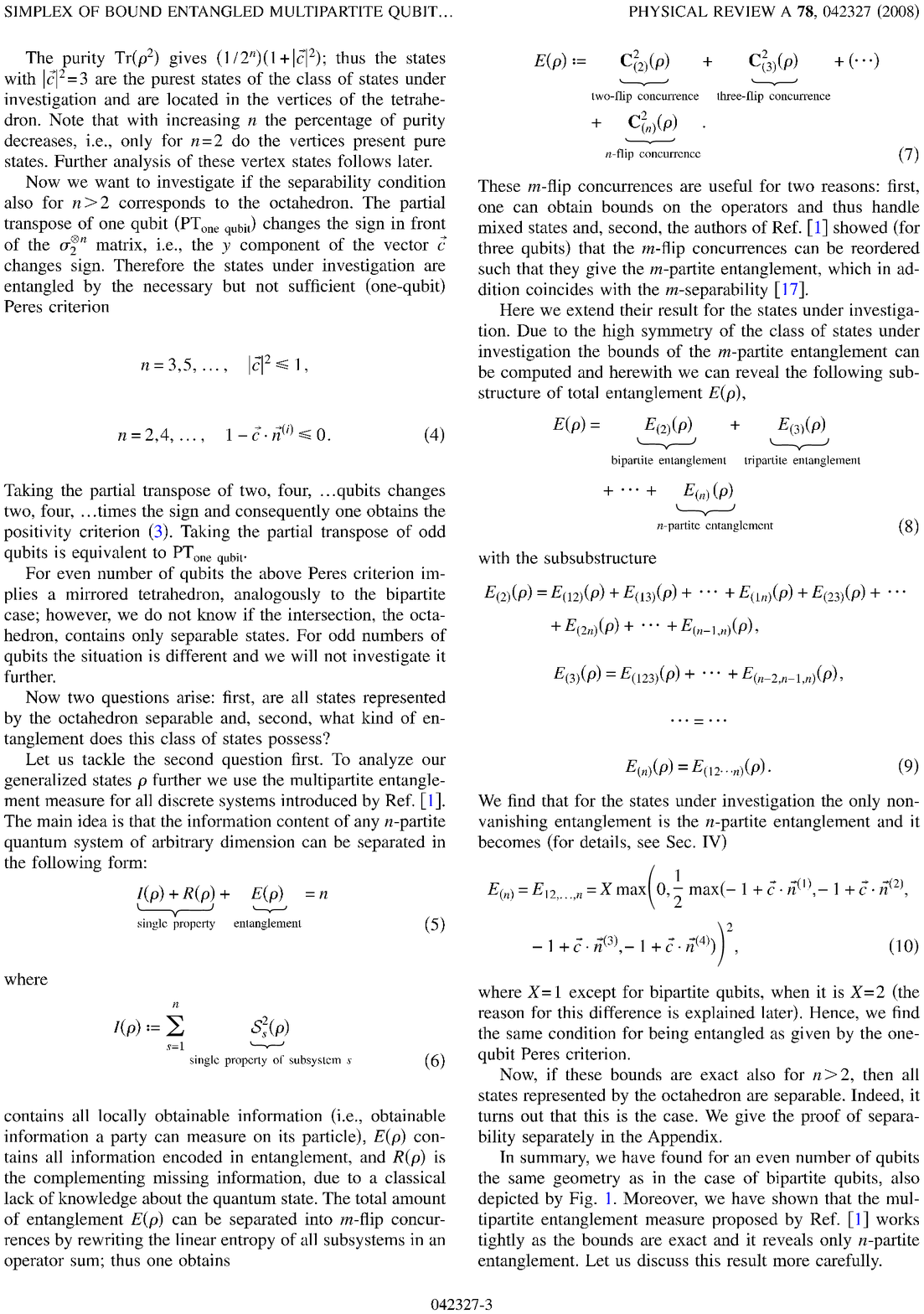}
\newpage
\vspace*{1cm}
\includegraphics[scale=0.8]{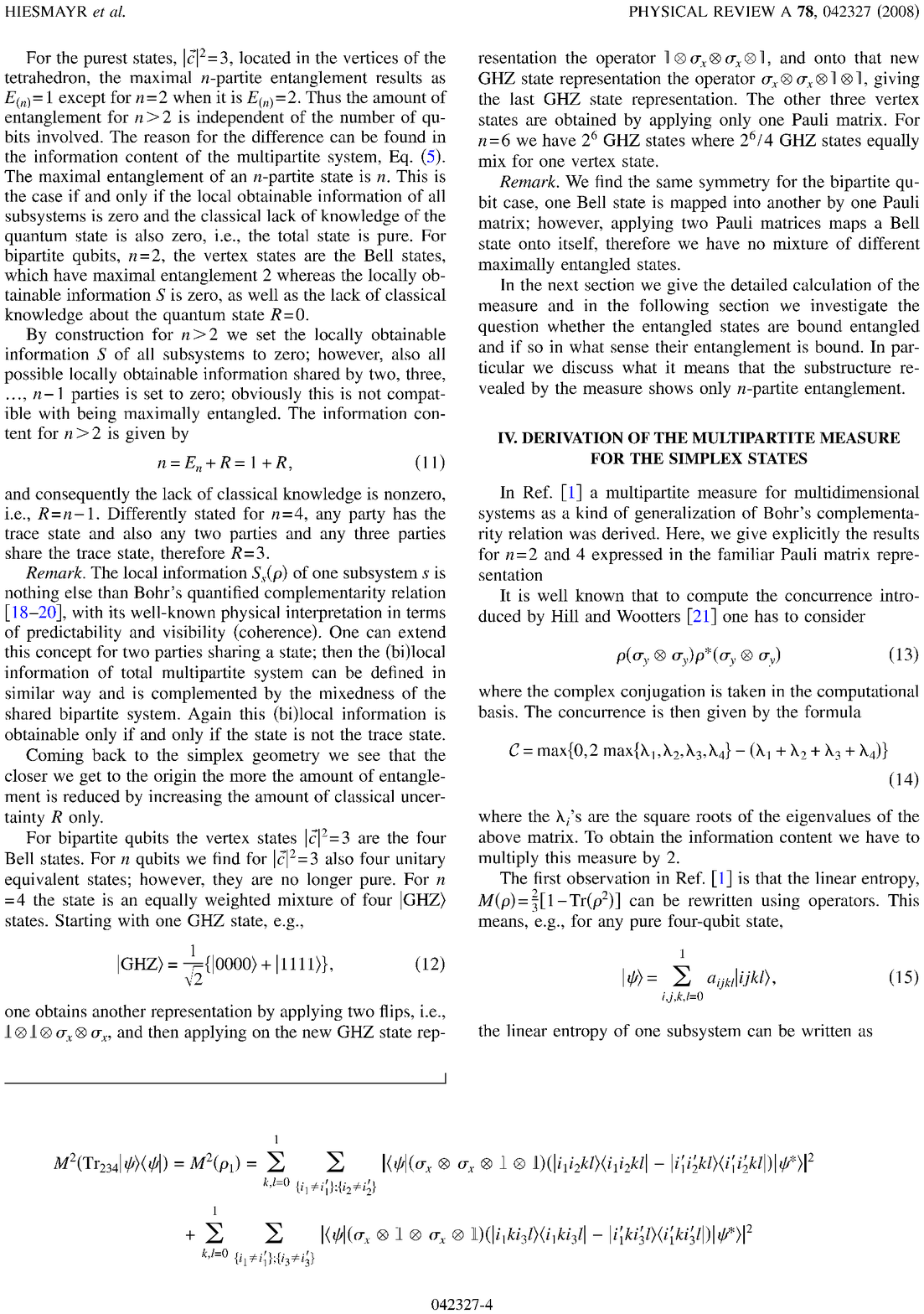}
\newpage
\vspace*{1cm}
\includegraphics[scale=0.8]{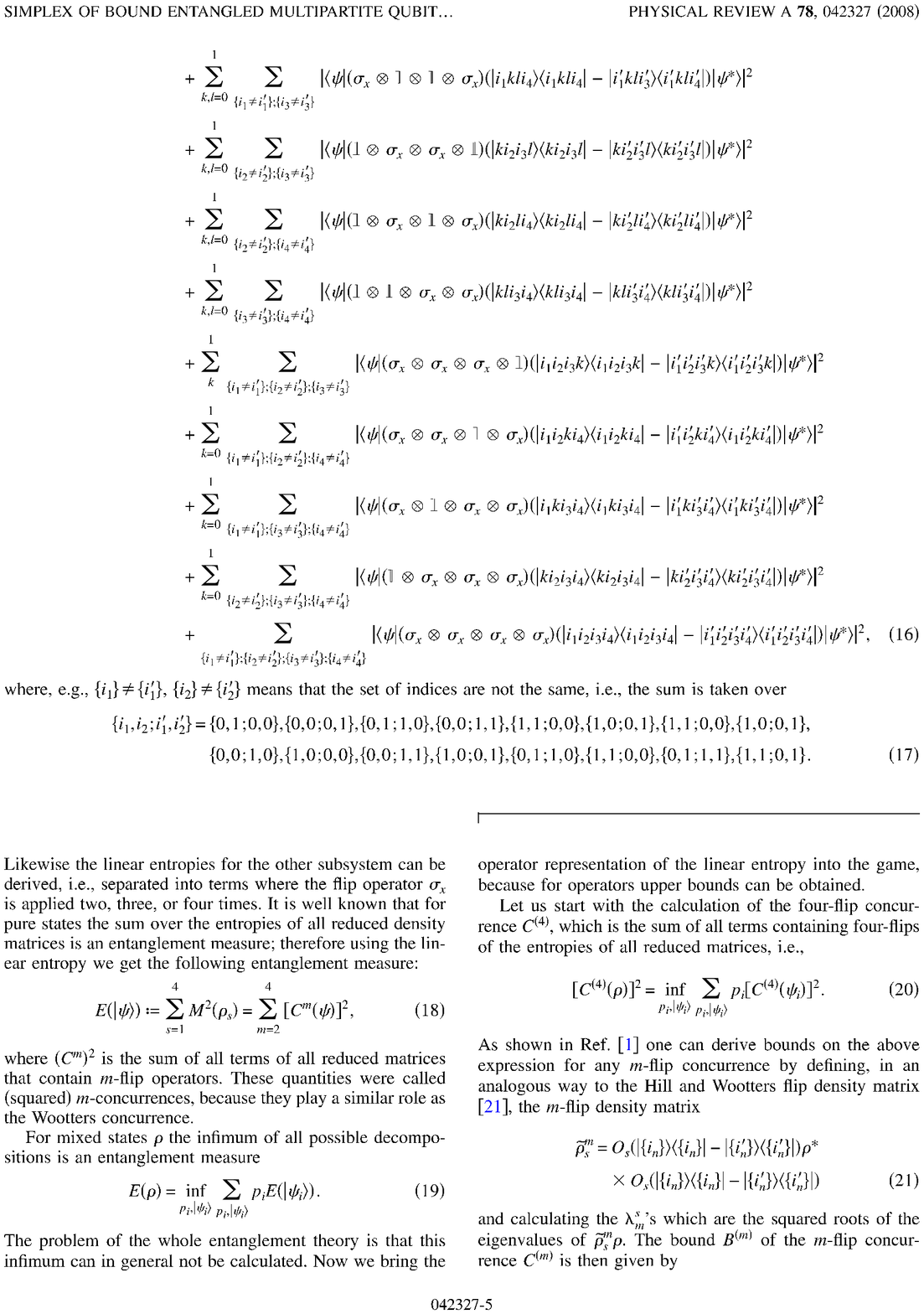}
\newpage
\vspace*{1cm}
\includegraphics[scale=0.8]{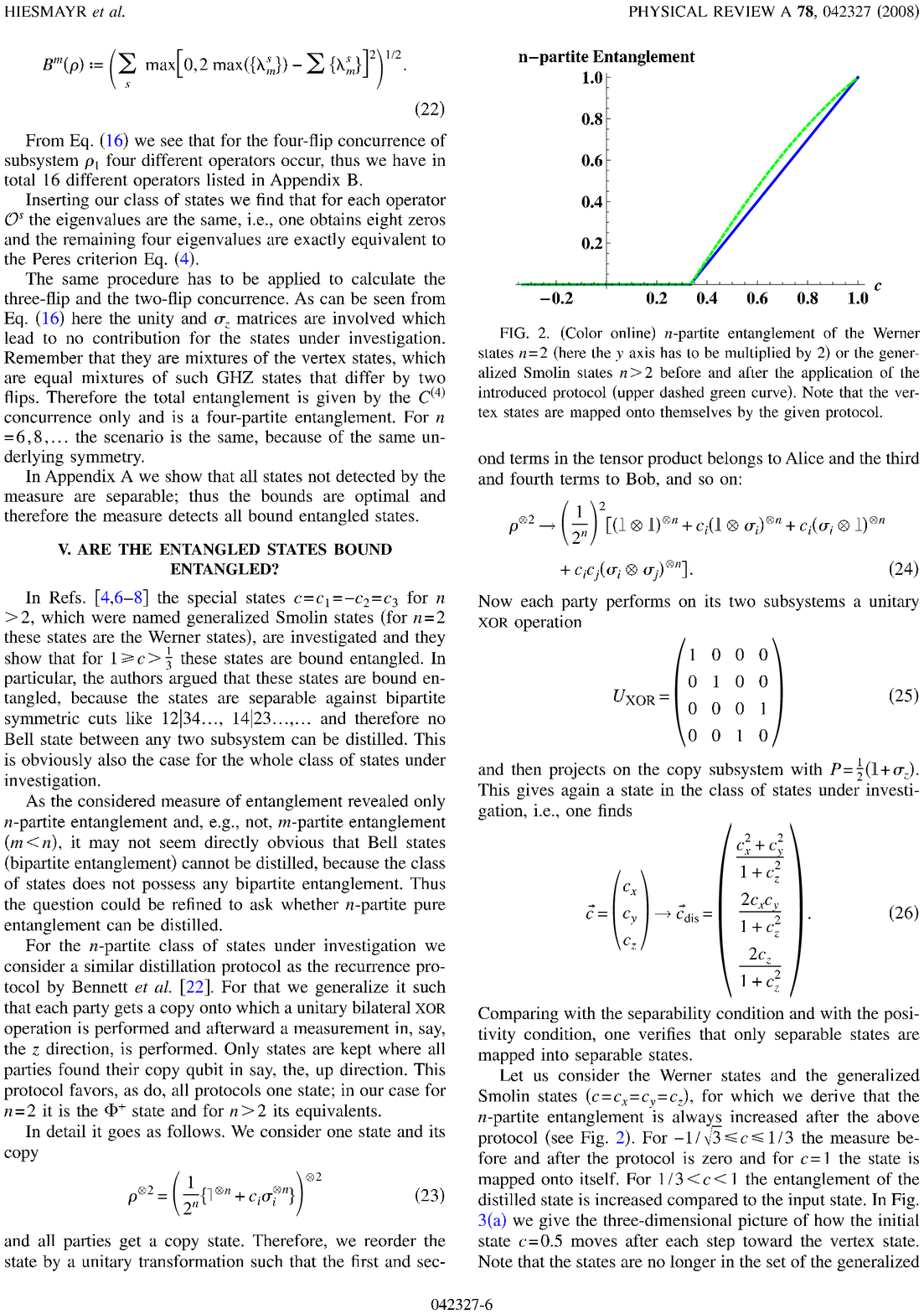}
\newpage
\vspace*{1cm}
\includegraphics[scale=0.8]{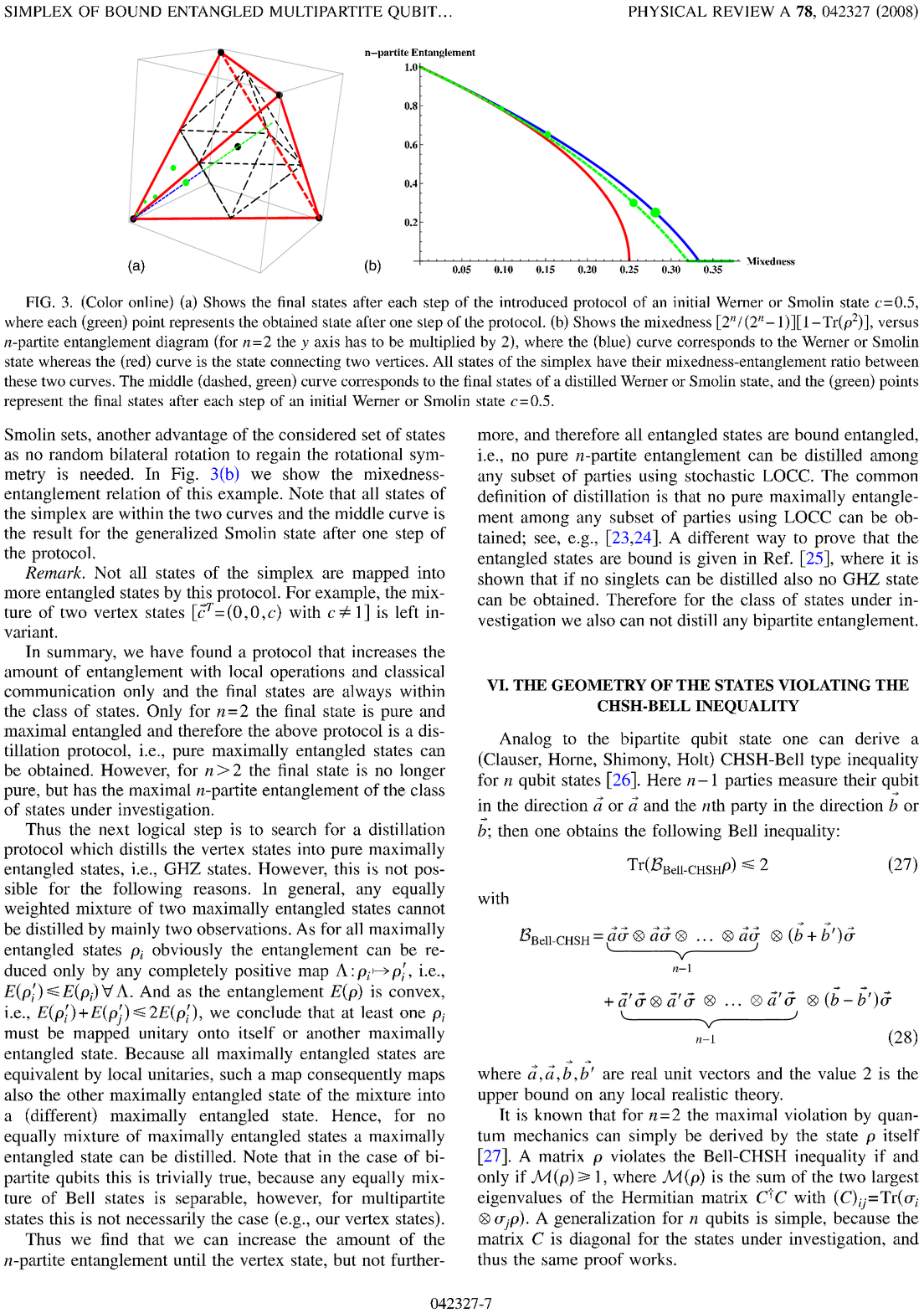}
\newpage
\vspace*{1cm}
\includegraphics[scale=0.8]{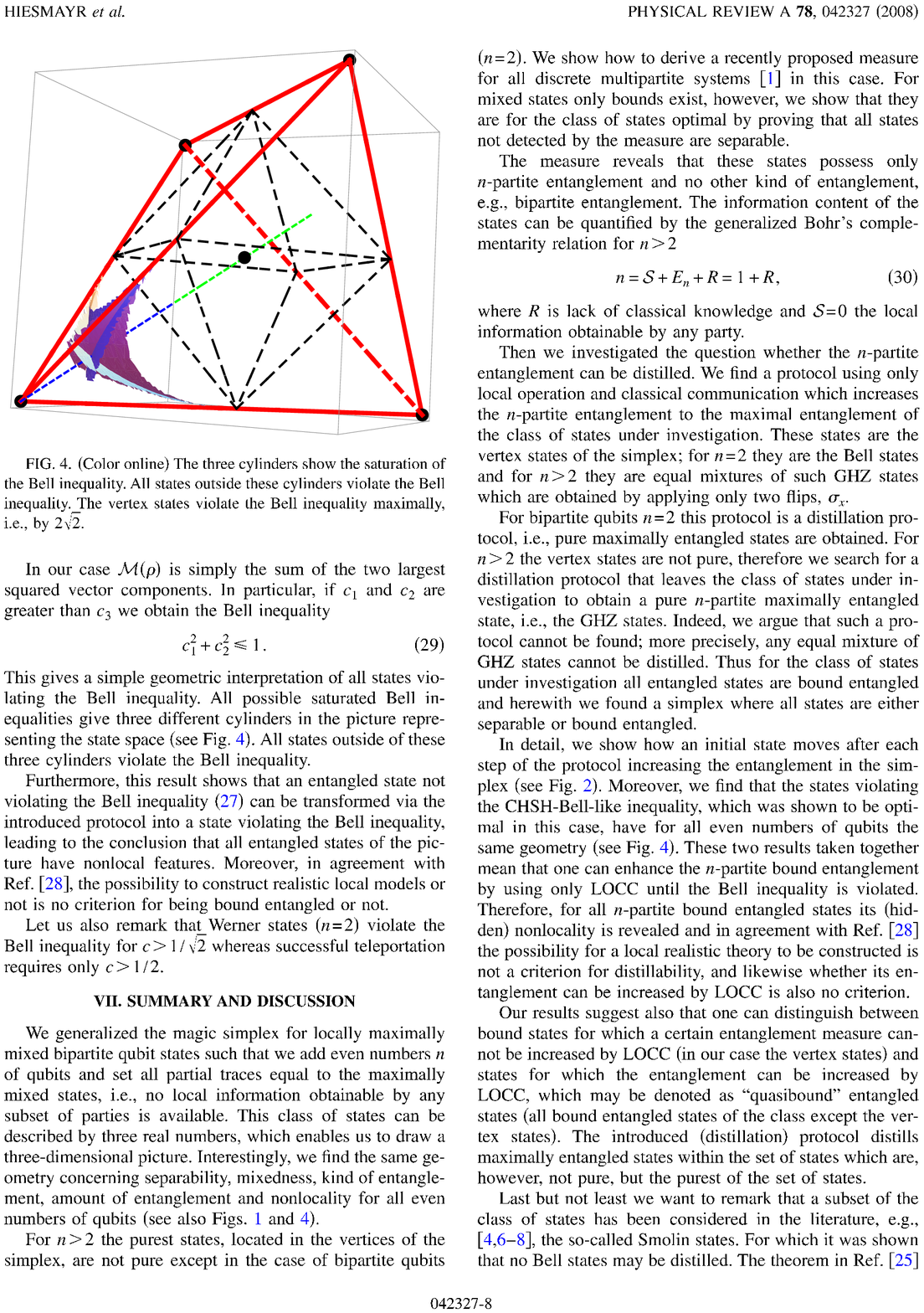}
\newpage
\vspace*{1cm}
\includegraphics[scale=0.8]{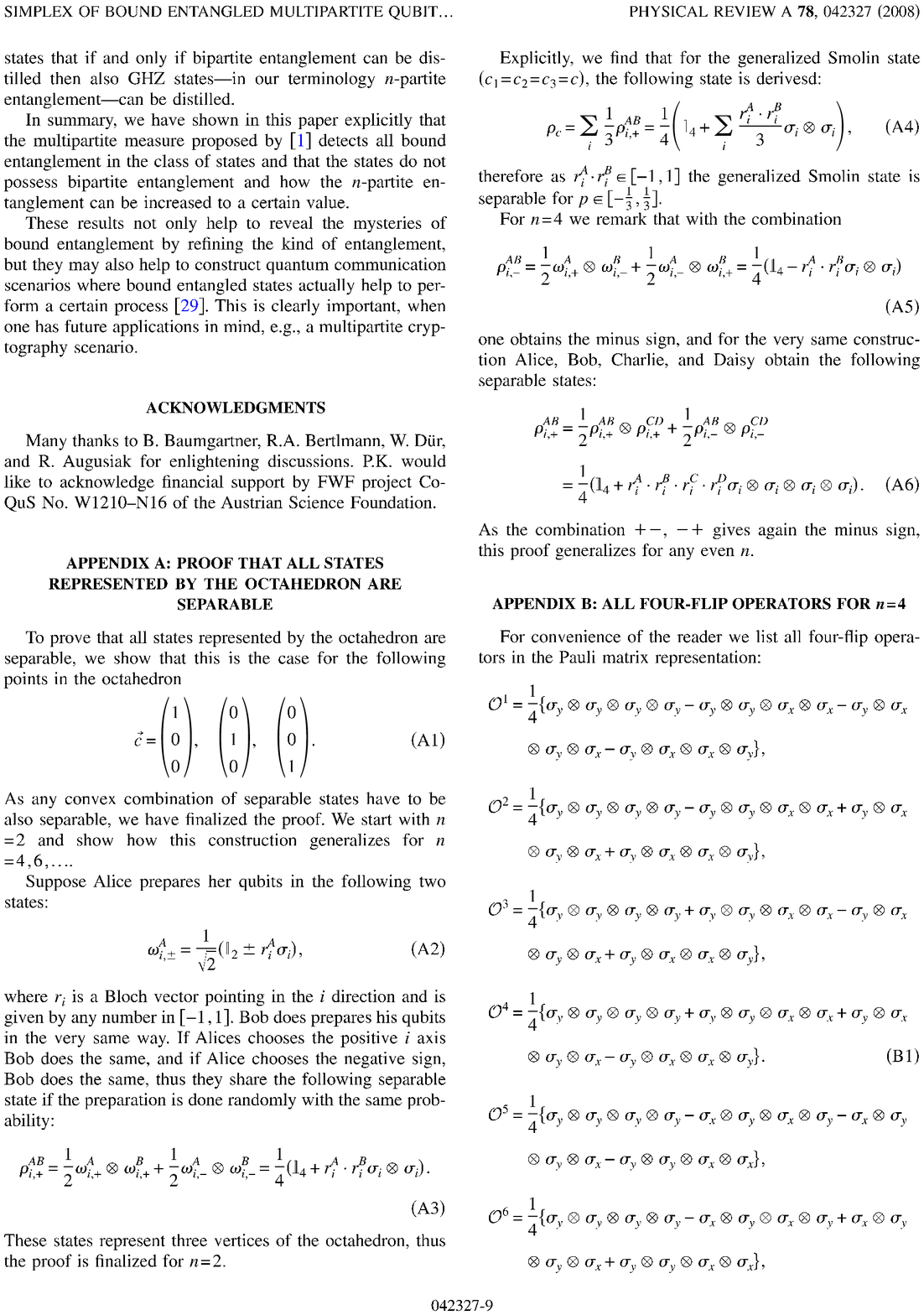}
\newpage
\vspace*{1cm}
\includegraphics[scale=0.8]{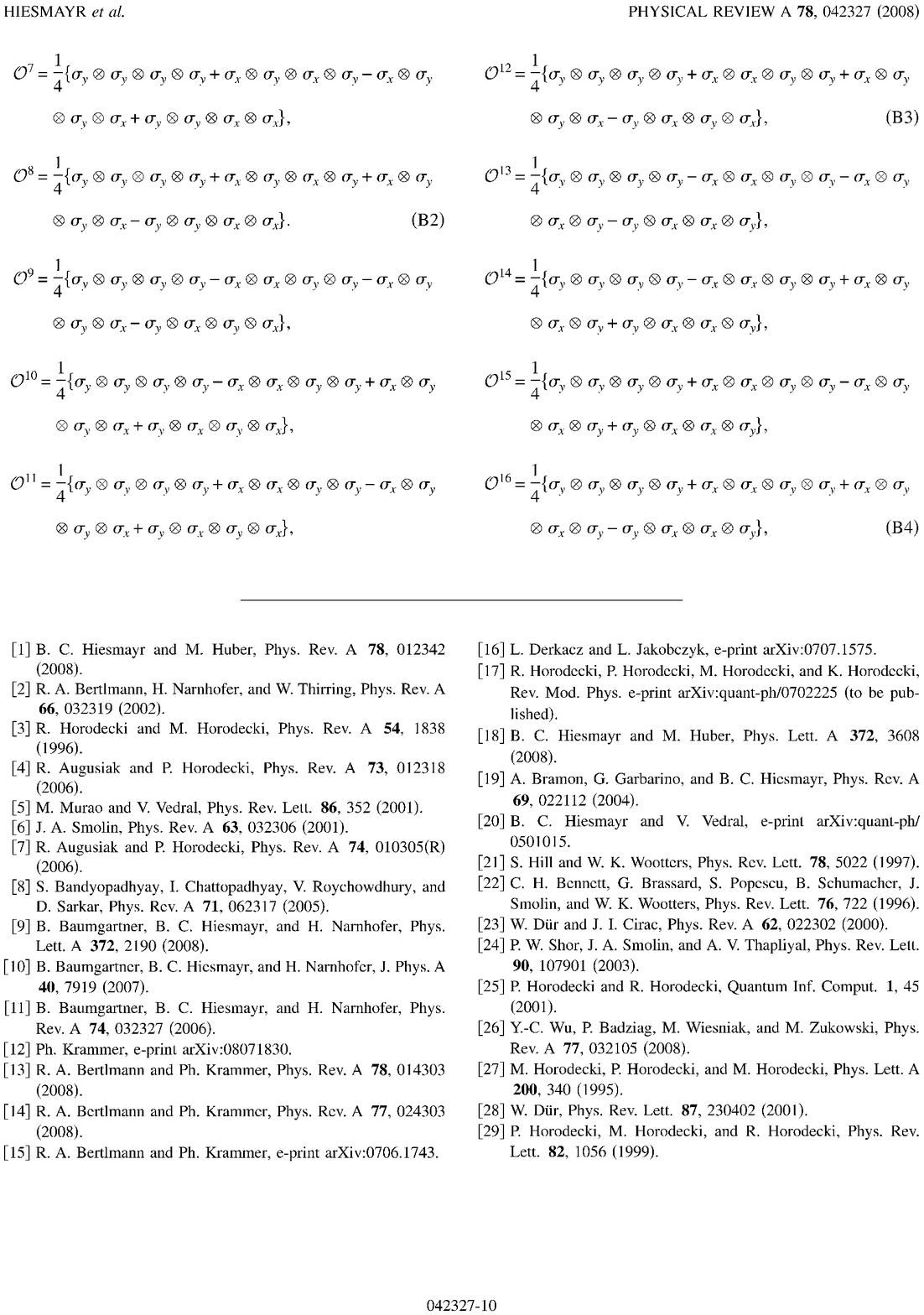}

\newpage
\section*{Summary and outlook}
\addcontentsline{toc}{section}{Summary and outlook} 
It is one of the most seminal discoveries that quantum physics contradicts local realism. This characteristic can be ascribed to quantum entanglement and manifests itself through the violation of a Bell inequality. Closer considerations within recent years have raised the question of whether there is a discrepancy between entanglement and nonlocality. The clarification of this question is of great importance for our understanding of the theory, and also might have consequences for future applications. The aim of this diploma thesis was to give an overview on the latest state of knowledge in this field, and to confront entanglement with the violation of the CGLMP inequality for a certain set of states that is called the magic simplex.\\
\\
We began our study in Chapter 1 with an introduction to the mathematical framework including important operator bases such as the Pauli operator basis for qubits and its generalisations for d-dimensional systems, that is the Gell-Mann and the Weyl operator basis. Afterwards, we discussed several issues concerning entanglement. We showed that it is very demanding to establish whether a mixed bipartite qudit state with $d>2$ is separable or entangled, because of the fact that all known practical criteria (e.g. PPT and matrix realignment) are only necessary but not sufficient for separability. We reviewed the requirements for entanglement measures and presented possible candidates based on distance and convex roof. In addition, we provided the proof that entangled PPT states cannot be distilled via "local operations and classical communication".\\
\\
In Chapter 2 we made clear the contradiction of quantum physics with local realistic theories. In particular, we explicitly derived that any local realistic description of a bipartite d-dimensional system obeys the CGLMP inequality, whereas quantum physics predicts a violation for the maximally entangled state. We investigated the problem of determining if a given state $\rho$ is able to violate the CGLMP inequality and therefore non-local. This is a high-dimensional nonlinear optimisation problem where all possible measurement settings of both parties have to be taken into account. An analytical solution has been found only for the case $d=2$. In order to study the nonlocality of higher dimensional systems, we have developed a numerical optimisation algorithm which utilises the generalised Euler angle parameterisation of $SU(N)$ and the Nelder-Mead method. The advantages of our method are its robustness against local maxima and its adaptability to other Bell inequalities. The chapter concluded with an overview of recent approaches for clarifying the possible discrepancy between nonlocality and entanglement, such as a discussion on hidden-nonlocality, tight Bell inequalities, the maximal violation of the CGLMP inequality through non-maximally entangled states and the non-local machine.\\
\\
In Chapter 3 we studied the state space of bipartite qubits with the aim of finding simple representatives of locally unitarily equivalent states. We showed that all locally maximally mixed states can be represented by elements of a tetrahedron spanned by the Bell states. We discussed the difficulties with regard to finding an extension for qudits. We introduced a possible generalisation in form of the magic simplex, which is a mixture of maximally entangled two-qudit states. Afterwards we considered the problem of separability for this set of states, and showed how it can be simplified by exploiting symmetries. Considering two- and three-parameter families of qutrit states $(d=3)$, we used these concepts to determine PPT states and to construct optimal entanglement witnesses. In order to reveal the non-local states of this families, we applied our numerical optimisation algorithm. As expected, we found that there is a large region of entangled states that do not violate the CGLMP inequality. We made a supposition about the exact form of the boundary of CGLMP violation for the three-parameter family of states on a line, which coincides with the numerical data up to the order $10^{-5}$. Comparing two three-parameter families, we revealed that for the mixtures off a line, the region of nonlocality is smaller while the region of entanglement is larger. This has demonstrated that entanglement and nonlocality do not behave conformably in any case. In addition we presented a publication of our group, where we investigated a generalisation of the tetrahedron for multipartite qubit systems. Investigating separability, we realised that all entangled states in this tetrahedron are bound entangled. We showed that there is a relation between distillability and the available type of entanglement. In the multipartite case, entanglement can be shared in many different ways. In our publication we argued that bipartite Bell states cannot be distilled from the occuring entangled states, due to the fact that they only possess n-partite $(n\neq2)$ entanglement.\\
\\
We conclude with some remarks regarding future research. As we have seen in this thesis, there is an abundance of open problems that make further considerations desirable. In order to solve them, progress is needed in the theory of nonlocality and entanglement. This means that Bell inequalities have to be improved, or alternatively it has to be shown that this is impossible. In addition, it is necessary to find more advanced techniques to solve the separability problem. Without developments in this field high-dimensional and/or multipartite systems are almost impossible to study. Once this has been achieved, we might be able to fully understand the relation between nonlocality and entanglement.

\appendix
\newpage
\section{MATHEMATICA notebook: Partial transposition of multipartite density matrices}
\includegraphics[scale=0.9]{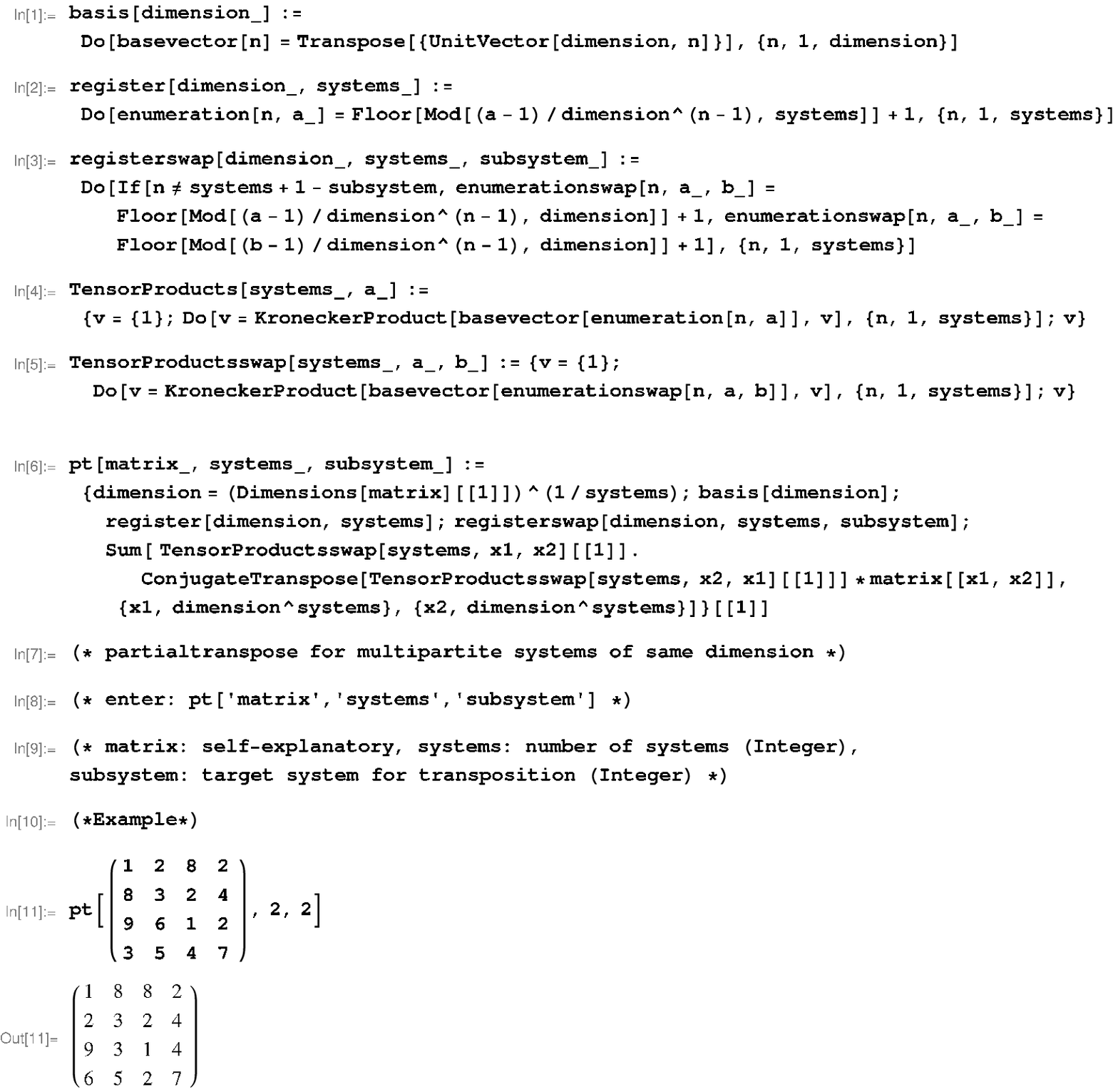}

\newpage
\section{MATHEMATICA notebook: Optimisation of $\mathcal{B}_{I_3}$} \label{OPBI3}
\includegraphics[scale=0.9]{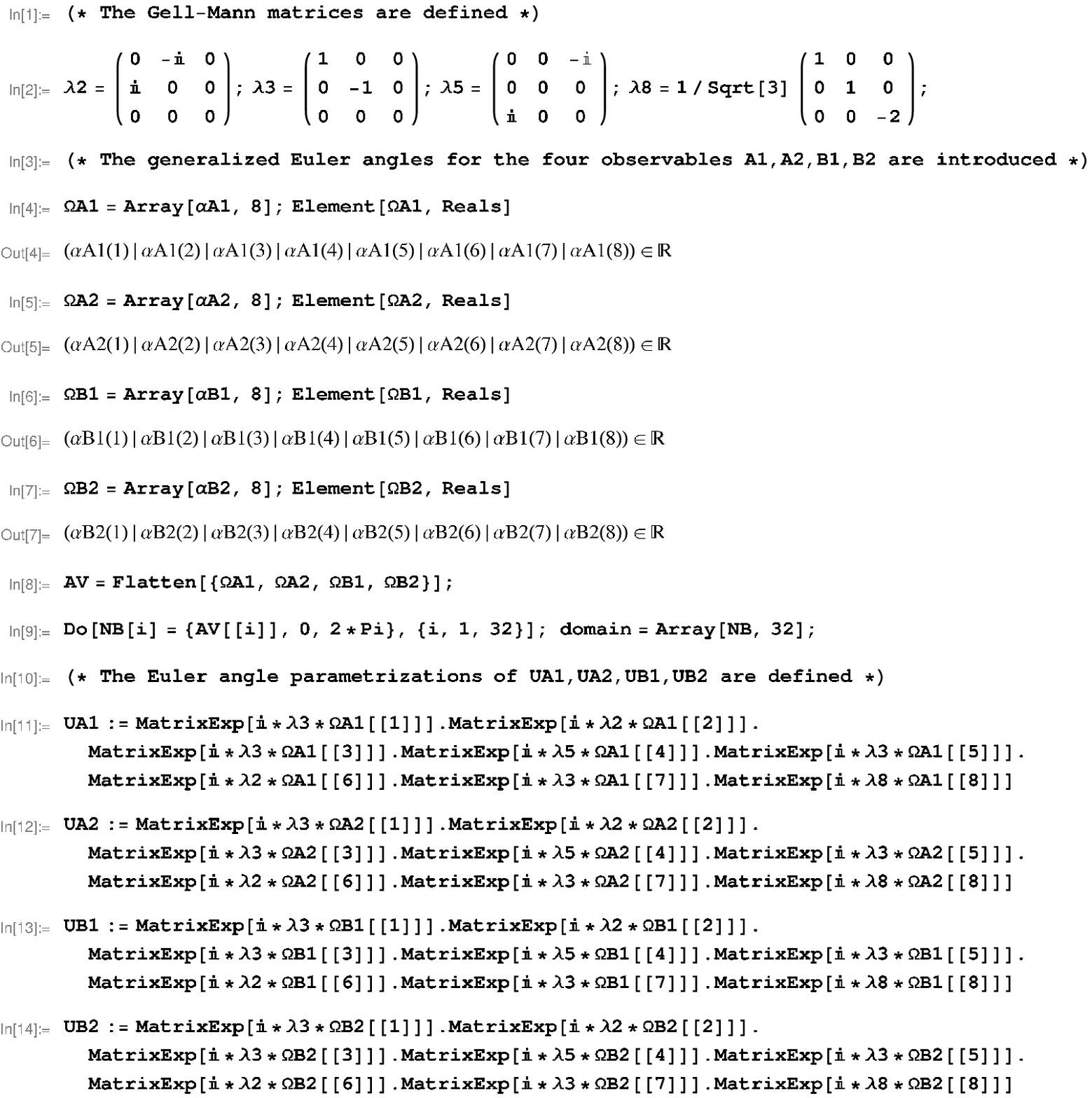}
\newpage
\includegraphics[scale=0.9]{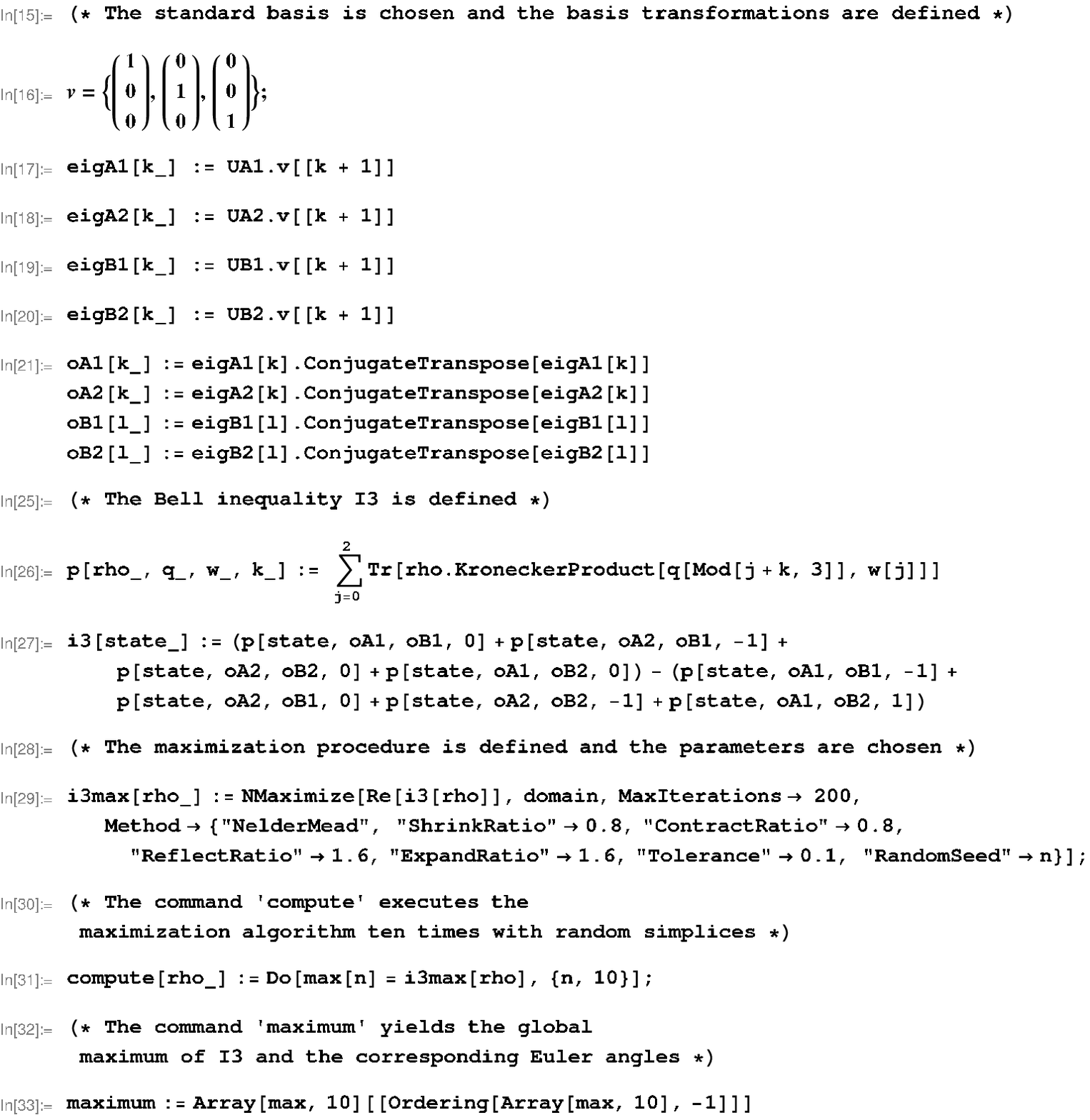}
\newpage
\includegraphics[scale=0.9]{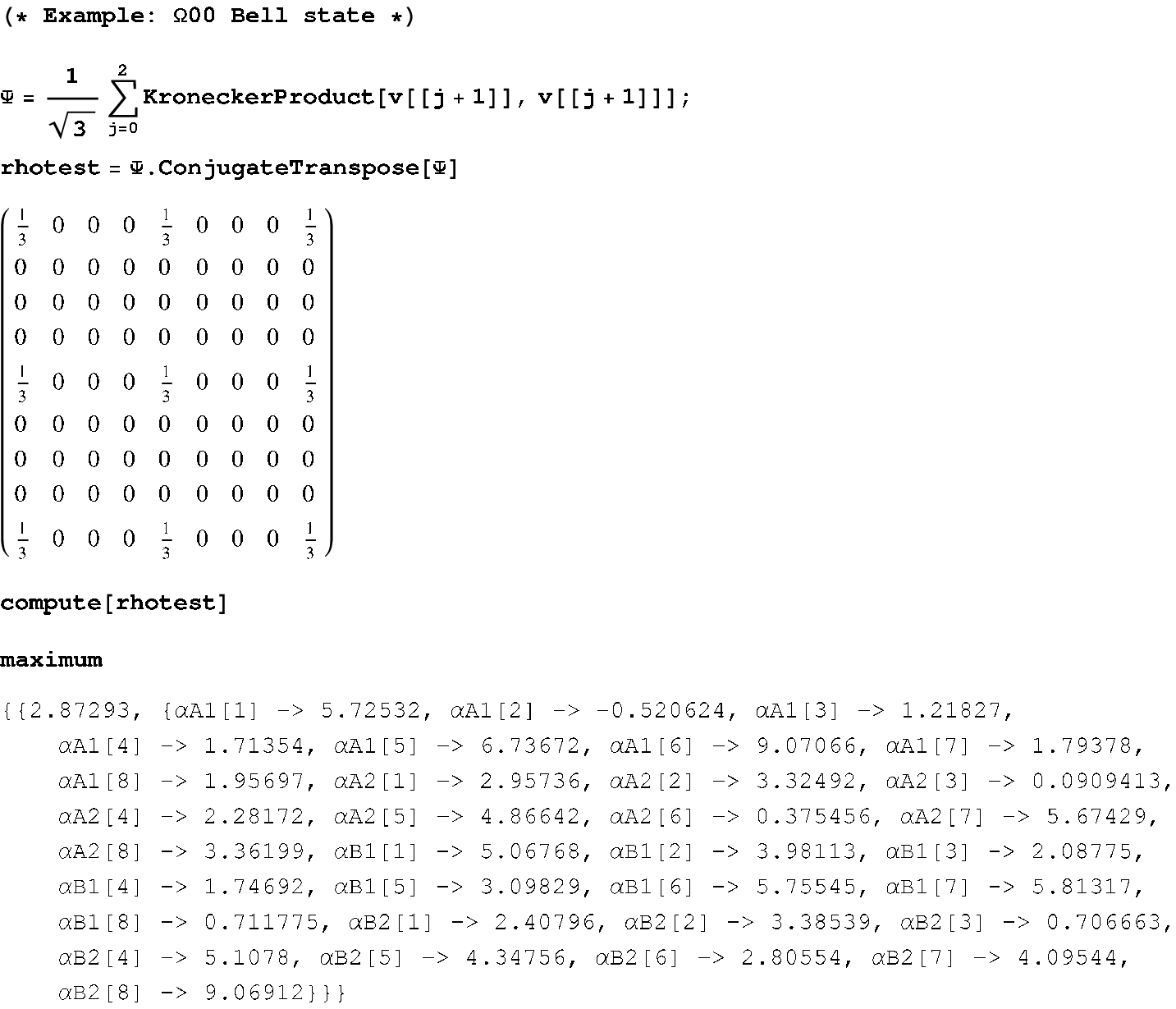}

\newpage
\section{MATHEMATICA notebook: Optimisation of $\mathcal{B}_{I_4}$} \label{OPBI4}
\includegraphics[scale=0.9]{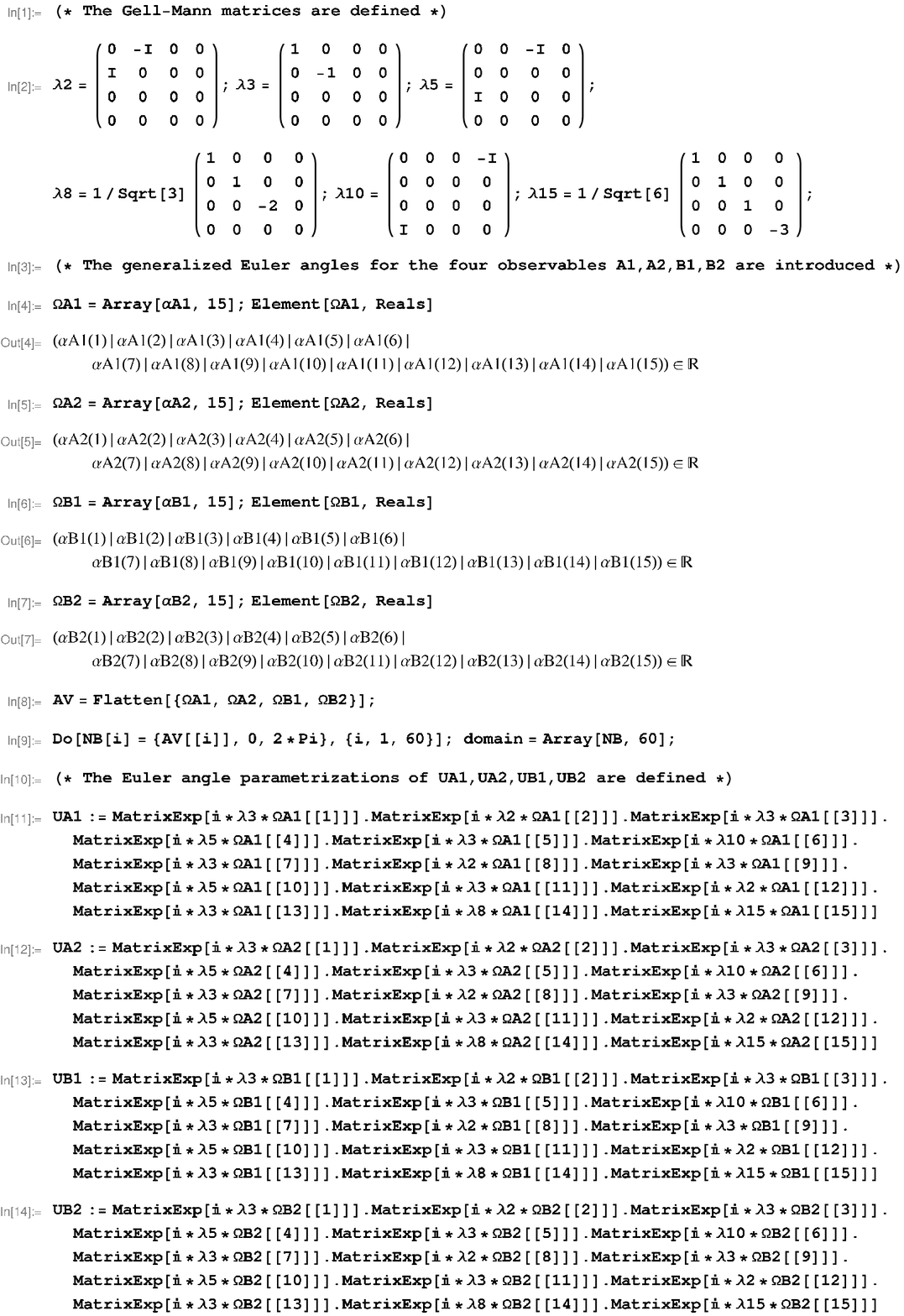}
\newpage
\includegraphics[scale=0.9]{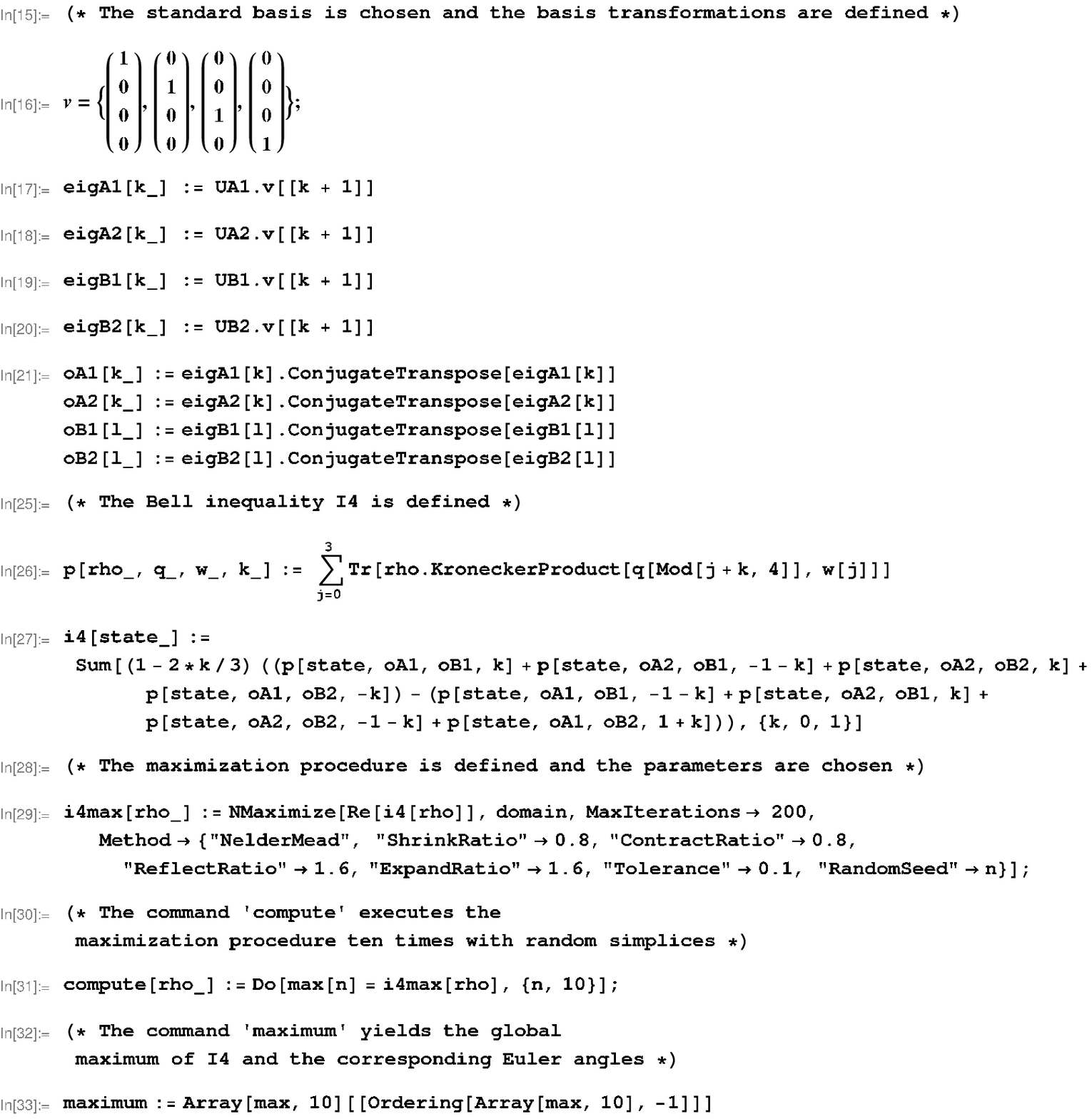}
\newpage
\includegraphics[scale=0.9]{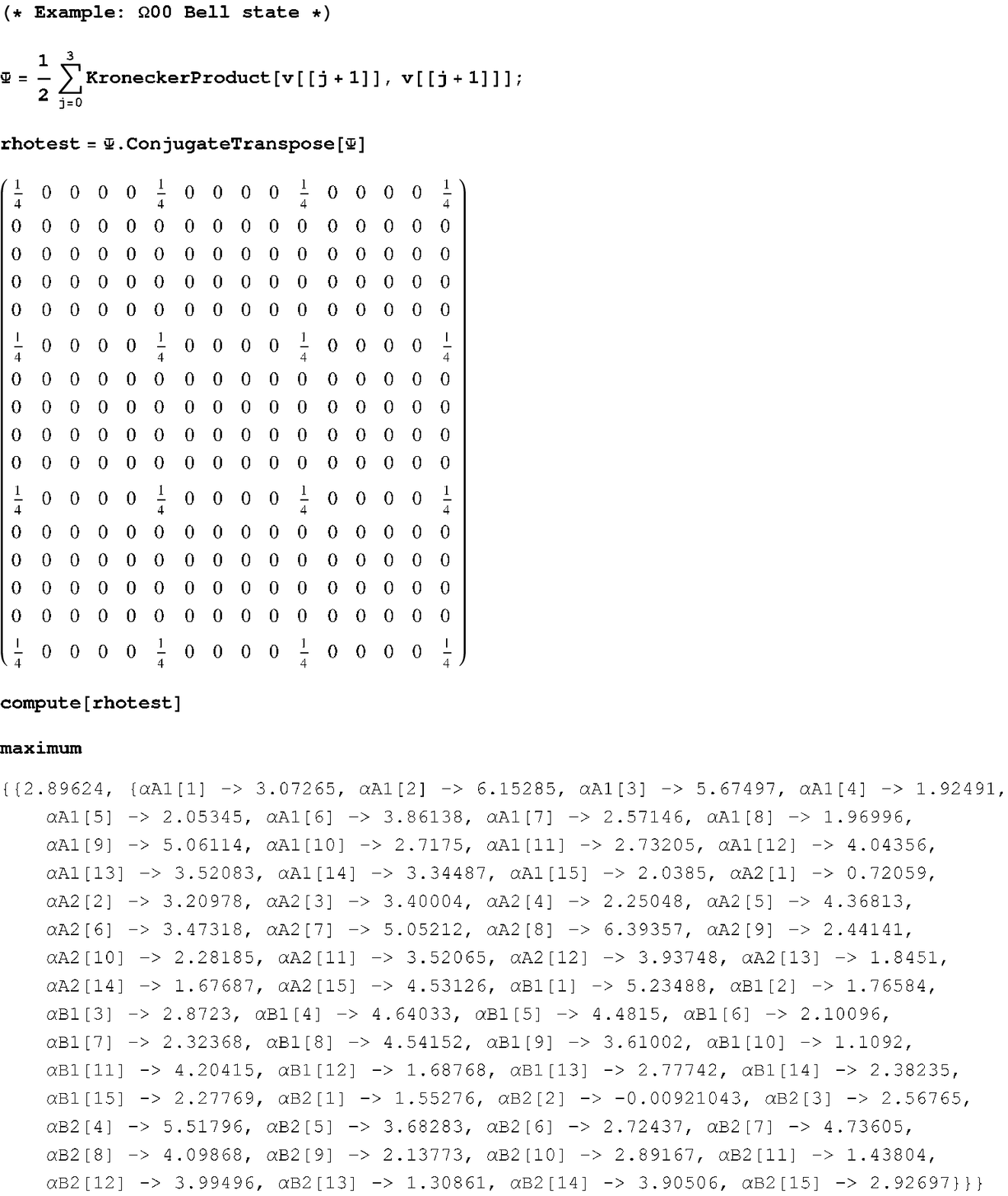}

\newpage
\section{Computation details on $\rho=\frac{1-\alpha-\beta}{9}\mathbbm{1}+\alpha P_{0,0}+\beta P_{1,0}$} \label{compdet1}
PPT:
\begin{align*}
w&=e^{2\pi i /3}\\
c_{0,0}&=\frac{1-\alpha-\beta}{9}+\alpha\\
c_{1,0}&=\frac{1-\alpha-\beta}{9}+\beta\\
c_{k,l}&=\frac{1-\alpha-\beta}{9} \hspace{2cm} \forall \ (k,l)\notin \{(0,0),(1,0)\} \\
\left(B_0\right)_{s,t}&=\frac{1}{3}\sum_{k=0}^{2}c_{k,s+t} w^{k(s-t)}\\
\Rightarrow B_0&=\left(
\begin{array}{lll}
 \frac{1}{9} (2 \alpha+2 \beta+1) & 0 & 0 \\
 0 & \frac{1}{9} (-\alpha-\beta+1) & \frac{1}{3} \left(\alpha+\beta e^{-\frac{2  \pi i }{3}}\right) \\
 0 & \frac{1}{3} \left(\alpha+\beta e^{\frac{2 \pi i }{3}}\right) & \frac{1}{9} (-\alpha-\beta+1)
\end{array}
\right)\\
\Rightarrow detB_0&=-\frac{1}{729} (2 \alpha + 2 \beta +1) (8 \alpha^2+8 \beta^2 -11 \alpha \beta +2 \alpha +2 \beta-1)
\end{align*}\\

\newpage
\section{Computation details on $\rho=\frac{1-\alpha-\beta-\gamma}{9}\mathbbm{1}+\alpha P_{0,0}+\beta P_{1,0}+\gamma P_{2,0}$} \label{compdet2}
PPT:
\begin{align*}
w&=e^{2\pi i /3}\\
c_{0,0}&=\frac{1-\alpha-\beta-\gamma}{9}+\alpha\\
c_{1,0}&=\frac{1-\alpha-\beta-\gamma}{9}+\beta\\
c_{2,0}&=\frac{1-\alpha-\beta-\gamma}{9}+\gamma\\
c_{k,l}&=\frac{1-\alpha-\beta-\gamma}{9} \hspace{2cm} \forall \ (k,l)\notin \{(0,0),(1,0),(2,0)\} \\
\left(B_0\right)_{s,t}&=\frac{1}{3}\sum_{k=0}^{2}c_{k,s+t} w^{k(s-t)}\\
\Rightarrow B_0&=\left(
\begin{array}{lll}
 \frac{1}{9} (2 \alpha+2 \beta+2 \gamma+1) & 0 & 0 \\
 0 & \frac{1}{9} (-\alpha-\beta-\gamma+1) & \frac{1}{3} \left(\alpha+\beta e^{-\frac{2  \pi i }{3}}+\gamma e^{\frac{2  \pi i }{3}}\right) \\
 0 & \frac{1}{3} \left(\alpha+\beta e^{\frac{2  \pi i }{3}}+\gamma e^{-\frac{2  \pi i }{3}}\right) & \frac{1}{9} (-\alpha-\beta-\gamma+1)
\end{array}
\right)\\
\Rightarrow detB_0&=-\frac{1}{729} (2 \alpha + 2 \beta + 2 \gamma + 1) \\
& \hspace{0.6cm} \cdot (8 \alpha^2 + 8 \beta^2 + 8 \gamma^2 + 2 \alpha  + 2 \beta + 2 \gamma - 11 \beta \alpha  -    11 \alpha \gamma - 11 \beta \gamma  - 1)
\end{align*}\\
\vspace{1cm}

\newpage
\section{Computation details on $\rho=\frac{1-\alpha-\beta-\gamma}{9}\mathbbm{1}+\alpha P_{0,0}+\beta P_{1,0}+\gamma P_{0,1}$} \label{compdet3}
PPT:
\begin{align*}
w&=e^{2\pi i /3}\\
c_{0,0}&=\frac{1-\alpha-\beta-\gamma}{9}+\alpha\\
c_{1,0}&=\frac{1-\alpha-\beta-\gamma}{9}+\beta\\
c_{0,1}&=\frac{1-\alpha-\beta-\gamma}{9}+\gamma\\
c_{k,l}&=\frac{1-\alpha-\beta-\gamma}{9} \hspace{2cm} \forall \ (k,l)\notin \{(0,0),(1,0),(0,1)\} \\
\left(B_0\right)_{s,t}&=\frac{1}{3}\sum_{k=0}^{2}c_{k,s+t} w^{k(s-t)}\\
\Rightarrow B_0&=\left(
\begin{array}{lll}
 \frac{1}{9} (2 \alpha+2 \beta- \gamma+1) & \frac{\gamma}{3} & 0 \\
 \frac{\gamma}{3} & \frac{1}{9} (-\alpha-\beta-\gamma+1) & \frac{1}{3} \left(\alpha+\beta e^{-\frac{2  \pi i }{3}}\right) \\
 0 & \frac{1}{3} \left(\alpha+\beta e^{\frac{2  \pi i }{3}}\right) & \frac{1}{9} (-\alpha-\beta+2\gamma+1)
\end{array}
\right)\\
\Rightarrow detB_0&=\frac{1}{729} (-16 \alpha^3-16 \beta^3-16 \gamma^3
+6 \beta \alpha^2 + 6 \gamma \alpha^2+6 \gamma^2 \alpha+6 \beta^2 \alpha+6 \beta^2 \gamma \\ & \hspace{1cm}+6 \beta \gamma^2
-12 \alpha^2-12 \beta^2-12 \gamma^2
+3 \beta \alpha+3 \gamma \alpha+3 \beta \gamma
-15 \beta \gamma \alpha
+1)
\end{align*}

\newpage
\vspace*{4cm}
\addcontentsline{toc}{section}{Acknowledgements}
\begin{center}
\Large{\textbf{Acknowledgements}}
\end{center}
\vspace*{1cm}
Ich m\"ochte mich bei allen bedanken, die zu dieser Arbeit beigetragen haben:\\
\\
Insbesondere danke ich Dora Kopf, Ingrid Pintaritsch und Marcel Meyer f\"ur die Englisch-Korrekturen. Ohne ihre Hilfe w\"urde der meiste Teil der Arbeit, anstatt aus S\"atzen aus einem Sammelsurium an W\"ortern bestehen.\\
\\
Ausserdem besonders zu erw\"ahnen ist David Rottensteiner, bei dem ich einen grossen Teil der Arbeit verfasst habe. Bei ihm bedanke ich mich f\"ur das produktive Beisammensein, das gute Essen und die wertvollen Tipps in allen Bereichen.\\
\\
Des weiteren bedanke ich mich bei allen Wiener Physikern, besonders f\"ur die gute Atmosph\"are am Institut. Am meisten danke ich hierbei nat\"urlich meiner Betreuerin Beatrix Hiesmayr, f\"ur ihre gute Betreuung, guten Ratschl\"age und ihre stets freund- liche Art. Zudem m\"ochte ich mich auch noch f\"ur die vielen Freiheiten im Bezug auf den Inhalt bedanken.\\
\\
Zuletzt m\"ochte ich noch ein ganz grosses "Danke" an meine Eltern, Hans-Joachim Ari und Margit Spengler richten, die mich die ganzen Jahre \"uber unterst\"utzt haben.\\
\\
Nochmals vielen Dank!

\newpage
\hfill
\vfill
\newpage

\addcontentsline{toc}{section}{Curriculum Vitae} 
\begin{center}
\Large{\textbf{Curriculum Vitae}}
\end{center}

\vspace{0.2cm}

\begin{flushleft}
 \begin{tabbing}
    \hspace{3cm} \= \\
    \textbf{Pers\"onliche Angaben:}\\
    \\
    \>Name: Christoph Ari Spengler\\
    \>Kontakt: ch.spengler@gmx.de\\
    \>Geburtsort: Augsburg \\
    \>Geburtstag: 18.02.1982 \\
    \>Staatsangeh\"origkeit: Deutschland \\
    \>Familienstand: ledig, keine Kinder\\
    
    \\
    
   \textbf{Schulbildung:}\\
   \\
   \> 09/1988 - \=07/1994 Volksschule Fischach\\
\>09/1994 - 07/1998 Staatliche Realschule Neus\"ass\\
\>09/1998 - 07/2000 Fachoberschule Augsburg\\
\\

\textbf{Zivildienst:}\\
\\
\> 10/2000 - 08/2001 Salesianum M\"unchen\\
\\
\textbf{Hochschulstudium:}\\
\\
\>10/2001 - 09/2002 FH M\"unchen\\
\> Studienrichtung: Physikalische Technik\\
\> 09/2002 Diplom-Vorpr\"ufung (gut)\\
\\
\>10/2002 - 02/2006 Universit\"at Augsburg\\
\> Studienrichtung: Physik\\
\> 10/2004 Diplom-Vorpr\"ufung (gut)\\
\\
\>seit 03/2006 Universit\"at Wien\\
\> Studienrichtung: Physik\\
\\
\textbf{Konferenzen/Seminare:}\\
\\
\> seit 12/2007 Non-local-Seminar Vienna-Bratislava\\

\> 09/2008 \"OPG-FAKT Tagung in Aflenz, Vortrag:\\
\> \textit{"Bell-Ungleichungen und Geometrie in der Quantenphysik"}\\

\> seit 10/2008 Vienna-Theory-Lunch-Club-Seminar\\
\\
\textbf{Publikationen:}\\
\\
 \> B.C.Hiesmayr, F.Hipp, M.Huber, P.Krammer and Ch.Spengler\\
\> \textit{"A simplex of bound entangled multipartite qubit states"}\\
\> Phys. Rev. A 78, 042327 (2008) 
    
\end{tabbing}
\end{flushleft}

\end{document}